\documentclass[10pt,aps,prd,onecolumn,showpacs,amsmath,amssymb,nofootinbib,eqsecnum,preprintnumbers,superscriptaddress]{revtex4-2}

\usepackage{mathtools}					
\usepackage[colorlinks=true,
            citecolor=red,
            linkcolor=blue,
            urlcolor=violet,
            filecolor=cyan,
            backref=false]{hyperref}
\usepackage{xcolor}
\usepackage{nicefrac}
\usepackage{empheq}
\usepackage{stmaryrd}
\usepackage[shortlabels]{enumitem}

\usepackage{pst-node}

\newcommand{\appsection}[1]{\section{\MakeUppercase{#1}}}
\newcommand\bs[1]{\boldsymbol{#1}} 
\newcommand\dd{\mathrm{d}} 
\newcommand\pp{\partial} 
\newcommand\feq{\mathrel{\phantom{=}}} 
\newcommand{\erf}{\operatorname{erf}}
\newcommand{\erfc}{\operatorname{erfc}}
\renewcommand{\cosh}{\operatorname{ch}}
\renewcommand{\sinh}{\operatorname{sh}}
\renewcommand{\tanh}{\operatorname{th}}
\renewcommand{\coth}{\operatorname{cth}}
\newcommand{\csch}{\operatorname{csch}}
\newcommand{\sech}{\operatorname{sech}}
\newcommand{\arccosh}{\operatorname{arcch}}
\newcommand{\Li}{\operatorname{Li}}

\renewcommand{\Re}{\operatorname{Re}}

\newcommand\equivn[1]{\overset{\scriptscriptstyle #1}{\simeq}}

\begin{document}
\title{
Propagators in AdS for higher-derivative and nonlocal gravity: Heat kernel approach
}

\author{Ivan Kol\'a\v{r}}
\email{ivan.kolar@matfyz.cuni.cz}
\affiliation{Institute of Theoretical Physics, Faculty of Mathematics and Physics, Charles University,
V Hole\v{s}ovi\v{c}k\'ach 2, Prague 180 00, Czech Republic}
\affiliation{Van Swinderen Institute, University of Groningen, 9747 AG, Groningen, Netherlands}

\author{Tom\'a\v{s} M\'alek}
\email{malek@math.cas.cz}
\affiliation{Institute of Mathematics of the Czech Academy of Sciences, \v{Z}itn\'a 25, 115 67 Prague 1, Czech Republic}

\date{\today}

\begin{abstract}
We present a new covariant method of construction of the (position space) propagators in the $N$-dimensional (Euclidean) anti-de Sitter background for any gravitational theory with the Lagrangian that is an analytic expression in the metric, curvature, and covariant derivative. We show that the propagators (in Landau gauge) for all such theories can be expressed using the heat kernels for scalars and symmetric transverse-traceless rank-2 tensors on the hyperbolic $N$-space. The latter heat kernels are constructed explicitly and shown to be directly related to the former if an improved bi-scalar representation is used. Our heat kernel approach is first tested on general relativity, where we find equivalent forms of the propagators. Then it is used to obtain explicit expressions for propagators for various higher-derivative as well as infinite-derivative/nonlocal theories of gravity. As a by-product, we also provide a new derivation of the equivalent action (correcting a mistake in the original derivation) and an extension of the quadratic action to arbitrary ${N\geq 3}$ dimensions. 
\end{abstract}

\maketitle

\section{Introduction}

\textit{Anti-de Sitter space (AdS)} is a \textit{maximally symmetric (MS)} spacetime of negative constant curvature, which plays a crucial role in the AdS/CFT correspondence \cite{Maldacena:1997re,Gubser:1998bc,Witten:1998qj}. An important aspect of AdS/CFT is the calculation of the correlation functions, for which the (position space) bulk-to-bulk propagators are required \cite{Liu:1998ty, Liu:1998th, Freedman:1998bj,Burgess:1984ti,DHoker:1998bqu,Liu:1998th,DHoker:1999kzh,DHoker:1999mqo,Costa:2014kfa}. The propagators in MS spacetime for \textit{general relativity (GR)} have been studied by many authors \cite{Allen:1986tt,Turyn:1988af,Antoniadis:1986sb,DHoker:1999bve,Leonhardt:2003qu,Leonhardt:2003sn,Faizal:2011sa,Miao:2011fc,Mora:2012zi,Costa:2014kfa,Glavan:2019msf,Kleppe:1994ga}. Although there have been considerable research on propagation in AdS for modified theories of gravity \cite{Gullu:2009vy,Myung:2010rj,Tekin:2016vli,Biswas:2016egy,Biswas:2016etb,Mazumdar:2018xjz}, to the best of our knowledge, no explicit position-space formulas for propagators in AdS beyond GR have been presented in the literature.

In this paper, we try to fill in this gap by presenting a very generic covariant method of construction of propagators in AdS that is applicable to a wide class of gravitational theories --- the theories with the Lagrangian that is an analytic expression in the metric, curvature, and covariant derivative. This class contains majority of the popular metric theories of gravity considered in the literature such as any \textit{higher-derivative gravity (HDG)} \cite{Stelle:1976gc,Stelle:1977ry,Asorey:1996hz,Bergshoeff:2009hq,Bergshoeff:2009aq,Ohta:2011rv} as well as \textit{infinite-derivative/nonlocal gravity (IDG)} \cite{Krasnikov1987,Tomboulis:1997gg,Biswas:2011ar,Modesto:2011kw,Frolov:2015bia,Buoninfante:2018lnh,Buoninfante:2020ctr,Koshelev:2018}. Terms with higher finite and infinite orders of derivatives are very well motivated by their appearance in many effective
descriptions of quantum gravity. They tend to ameliorate issues
related to ultraviolet incompleteness of GR such as the existence of curvature singularities and quantum non-renormalizability.\footnote{On the downside, HDG typically suffer from the Ostrogradsky instability and ghosts. IDG can be rendered ghost-free at the expense of becoming nonlocal, which leads to issues with causality and initial value problem.} HDG has also attracted significant attention in the context of AdS/CFT \cite{Fukuma:2001uf,Nojiri:1999mh,Bobev:2021qxx}.

Our method of constructing propagators for gravitational theories in AdS combines the following mathematical tools and approaches:

\begin{description}
    \item[Euclidean AdS] 
    First and foremost let us stress that we work in the Euclidean version of AdS, i.e., the hyperbolic space, where we impose the fastest-falloff boundary conditions for all quantities involved. This significantly simplifies our problem since the hyperbolic space has many mathematically important properties, which are absolutely essential for our derivations. Although, the analytic continuation to the Lorentzian signature as well as the discussion of various boundary conditions in AdS are very interesting topics, they go well beyond the scope of this paper. Hence, our study should be viewed as a generalization of \cite{DHoker:1999bve}. In fact, the propagators calculated in the Euclidean signature are of their own interest because amplitudes in many higher-derivative \cite{Aglietti:2016pwz,Anselmi:2017lia,Anselmi:2017yux} and infinite-derivative/nonlocal \cite{Buoninfante:2018mre,Buoninfante:2022krn} quantum field theories can be sensibly defined only in the Euclidean signature.
    
    \item[Quadratic action] 
    We derive the equations for the propagators for the generic class of gravitational theories by first constructing the equivalent action possessing the same linearized gravity around MS backgrounds \cite{Biswas:2016egy,Biswas:2016etb}, and then calculating quadratic action by taking the second variation of the equivalent action \cite{Biswas:2016egy,Biswas:2016etb,Mazumdar:2018xjz} (see also \cite{Koshelev:2016xqb,SravanKumar:2019eqt}). Since the original derivation of the equivalent action contains a mistake, we re-derive it using an alternative method. We also re-derive the quadratic action in order to extend it from dimensions ${N=3,4}$ to arbitrary dimensions ${N\geq 3}$ (still for generic MS backgrounds). The equations for the propagators are written in the Landau gauge, as it anihilates the gauge-dependent terms that are not present in the quadratic action. This means that one only needs to find the purely scalar and transverse-traceless (TT) tensor parts of the propagators.
    
    \item[MS bi-tensors]
    Due to the Euclidean signature and assumption on the invariance of the propagator under all isometries of the hyperbolic space, we can conveniently use the standard approaches from the older literature \cite{Allen:1986tt,Turyn:1988af,Antoniadis:1986sb,DHoker:1999bve,Faizal:2011sa} (see also \cite{Allen:1985wd}). These approaches are fully covariant (i.e., they are independent of coordinates in AdS) and make use of various properties of MS bi-tensors in the hyperbolic space. A very crucial property here is that one can fully characterize an arbitrary MS TT bi-tensor using a single MS bi-scalar. We further improve this trick by introducing an alternative MS bi-scalar, which renders the action of the Laplace operator even simpler. In total, the problem is reduced to finding two MS bi-scalars, i.e., two functions of the geodesic distance only.
    
    \item[Heat kernels] 
    Finally, our approach takes the inspiration from the calculations in HDG and IDG that are linearized around Minkowski spacetime \cite{Frolov:2015bta,Frolov:2015usa,Frolov:2015bia}. Here, the authors used the relation between the Green's function of the scalar operators that are functions of the Laplace operator and the heat kernels. Although this trick was used in the flat space, it can be easily generalized to the curved Riemannian background\footnote{For its application to the nonlocal scalar fields in curved backgrounds, see \cite{Kolar:2022kgx}.} and also extended to MS TT bi-tensors. As a consequence, we can express the TT bi-tensor and bi-scalar parts of the propagators using the heat kernels for scalars and symmetric TT rank-2 tensors. Furthermore, we show that these heat kernels are actually related to each other. Let us remark that the scalar heat kernels in hyperbolic spaces are very well understood \cite{McKean:1970,Debiard:1976,Chavel:1984,DaviesMandouvalos:1988,Davies:1989,Grigoryan:1998,Matsumoto:2001,Grigor:weight,Avramidi:2015}. On the other hand, as far as we are aware, the `full' heat kernels on symmetric rank-2 TT tensors have not been presented so explicitly before.\footnote{Only the `traced' heat kernels on symmetric TT tensors have been studied before \cite{David:2009xg,Gopakumar:2011qs,Lal:2012ax,Shahidi:2018smw,Lal:2012aku} while the `full' heat kernels have been calculated only for vectors \cite{David:2009xg,Hatzinikitas:2017vgl} and symmetric traceless (but not TT) rank-2 tensors in ${N=3}$ \cite{Giombi:2008vd} (recently applied in \cite{Suzuki:2021pyw}).}
\end{description}

The paper is organized as follows: In Sec.~\ref{seclinearization} we derive the equations for the propagator in AdS for a generic class of gravitational theories via the equivalent and quadratic actions. In Sec.~\ref{sec:heatkernelapproach} we relate the propagators for any such gravitational theory to the heat kernels, for which we also find the explicit expressions. In Sec.~\ref{sec:evaluationofgravitonpropagators}, we evaluate the propagators for a variety of gravitational theories, such as GR, HDG, and IDG. The paper is concluded with a summary and discussion of our results in Sec.~\ref{sec:conclusions}. The text is supplemented by Appx.~\ref{apx:varmetcur}--\ref{apx:powersofbox} involving lengthy calculations such as the derivation of equivalent and quadratic actions and the formulas for powers of the Laplace operator on MS TT bi-tensors.

\section{Linearization of generic gravity in AdS}\label{seclinearization}

Main goal of this section is to derive the equations for the propagator in AdS for any gravitational theory with the Lagrangian that is an analytic scalar function of the metric, curvature, and covariant derivative. In particular, we first introduce the Euclidean AdS and review the covariant tensor decomposition of symmetric rank-2 tensors. Then, we derive the equivalent gravitational action which leads to the same linearized theory as the generic gravitational theory. Next, we derive the quadratic action and show that it involves only the TT part and the scalar part of the metric perturbations. Finally, taking the variation of the quadratic action and imposing the Landau gauge (in which only these parts are non-zero), we arrive at the simple form of the equations for propagators.

\subsection{Euclidean AdS}

The $N$-dimensional \textit{AdS} of radius $\alpha>0$ is an MS spacetime described by the Lorentzian manifold $(\mathbb{S}^1\times\mathbb{R}^{N-1}$,$\tilde{\bs{g}})$, where $\tilde{\bs{g}}$ is the metric tensor $\tilde{g}_{ab}$ of the constant negative curvature. Here, the indices $a,b,\dots$ are the \textit{abstract tensor indices} \cite{penrose_rindler_1984}; the \textbf{boldface} denotes tensorial character of the quantity when the indices are suppressed. It can be characterized by global embedding
\begin{equation}
    -X_1^2-X_2^2+X_3^2+\dots+X_{N+1}^2=-\alpha^2
\end{equation}
in the flat space with metric $-\bs{\dd}X_1^2-\bs{\dd}X_2^2+\bs{\dd}X_3^2+\dots +\bs{\dd}X_{N+1}^2$. The space can be covered by \textit{static coordinates} of \textit{type I}, \textit{type II}, and \textit{Poincar\'e type (P)}, which are associated to timelike Killing vectors of uniform acceleration that is less than $1/\alpha$, larger than $1/\alpha$, and equal to $1/\alpha$, respectively \cite{Krtous:2005ej,Krtous:2014pva}. The prototypical examples of these coordinates are
\begin{equation}
\begin{aligned}
    \tilde{\bs{g}}
    &=-\cosh^2\left(\tfrac{p}{\alpha}\right)\bs{\dd}t_{\textrm{I}}^2+\bs{\dd}p^2+\alpha^2\sinh^2\left(\tfrac{p}{\alpha}\right) \bs{g}_{\mathbb{S}^{N-2}} & &\textrm{(I)}
    \\
    &=\tfrac{\alpha^2}{q^2}\sec^2\zeta\Big[-\big(1-\tfrac{q^2}{\alpha^2}\big)\bs{\dd}t_{\textrm{II}}^2+\big(1-\tfrac{q^2}{\alpha^2}\big)^{-1}\bs{\dd}q^2+q^2\big(\bs{\dd}\zeta^2+\sin^2\zeta\bs{g}_{\mathbb{S}^{N-3}}\big)\Big] & &\textrm{(II)}
    \\
    &=\tfrac{\alpha^2}{z^2}\big(-\bs{\dd}t_{\textrm{P}}^2+\bs{\dd}z^2+\bs{g}_{\mathbb{R}^{N-2}}\big)\;.  & &\textrm{(P)}
\end{aligned}
\end{equation}
Out of these three, only the type I coordinates are actually globally static because the Killing vector $\bs{\pp}_{t_{\textrm{I}}}$ is timelike and smooth everywhere while the Killing vectors $\bs{\pp}_{t_{\textrm{II}}}$ and $\bs{\pp}_{t_{\textrm{P}}}$ become spacelike or null in some domains. The coordinate $t_{\textrm{I}}\in(0,2\pi)$ is periodic but it is common to unwrap it and define the \textit{universal covering space of AdS} instead given by the Lorentzian manifold $(\mathbb{R}^{N}$,$\tilde{\bs{g}})$.

The `Euclidean version' of AdS is the \textit{hyperbolic space} $\mathbb{H}^N$ or sometimes called the \textit{Euclidean AdS}. It is the $N$-dimensional Riemannian manifold of constant negative curvature (of radius $\alpha>0$), which we denote simply by $(M,\bs{g})$. The metric $\bs{g}$ of $\mathbb{H}^N$ can be obtained by analytic continuation in any of the three temporal coordinates mentioned above, $t_{\textrm{I}}\to i t_{\textrm{I}}^{\textrm{E}}$, $t_{\textrm{II}}\to i t_{\textrm{II}}^{\textrm{E}}$, or $t_{\textrm{P}}\to i t_{\textrm{P}}^{\textrm{E}}$, each of the resulting coordinates then cover the entire manifold ${M:=\mathbb{R}^N}$.\footnote{The same result also follows from a more general definition of the Wick rotation \cite{Visser:2017atf}, ${g_{ab} \to g_{ab}-2v_a v_b/(v^c v_c)}$, with the timelike Killing vector $\bs{v}$ being chosen either as $\bs{\pp}_{t_{\textrm{I}}}$, $\bs{\pp}_{t_{\textrm{II}}}$, or $\bs{\pp}_{t_{\textrm{P}}}$.} However, only the analytic continuation with respect to $t_{\textrm{I}}$ defines a global bijective map between the complete AdS and $\mathbb{H}^N$, because it actually corresponds to ${X_1\to i X_1^{\textrm{E}}}$ in the embedding space. Let us note that $\mathbb{H}^N$ also describes geometries on spacelike sections of $(N+1)$-dimensional AdS of constant $t_{\textrm{I}}$, $t_{\textrm{II}}$, or $t_{\textrm{P}}$, providing us with yet another global coordinates on $\mathbb{H}^N$. For example, setting $t_{\textrm{I}}=\textrm{constant}$, gives us coordinates with manifest spherical symmetry,
\begin{equation}
    \bs{g}
    =\bs{\dd}r^2+\alpha^2\sinh^2(\tfrac{r}{\alpha}) \bs{g}_{\mathbb{S}^{N-1}}\;,
\end{equation}
which we will refer to simply as the \textit{spherical coordinates}. They can be related back to Euclidean type I coordinates via
\begin{equation}\label{eq:sphcoor}
\begin{aligned}
    r &=\alpha \sinh ^{-1}\left(\cosh \left({t_{\textrm{I}}^{\textrm{E}}}/{\alpha}\right) \sqrt{\sinh ^2\left({p}/{\alpha}\right)+\tanh ^2\left({t_{\textrm{I}}^{\textrm{E}}}/{\alpha}\right)}\right)\;,
    \\
    \vartheta_1 &=\tan ^{-1}\left(\tanh \left({p}/{\alpha}\right) \csch\left({t_{\textrm{I}}^{\textrm{E}}}/{\alpha}\right)\right)\;.
\end{aligned}
\end{equation}
Since $\mathbb{H}^N$ is connected and geodesically complete, any two points $\mathrm{x},\mathrm{x}'\in M$ can be joined by a geodesic with the \textit{geodesic distance} $\rho(\mathrm{x},\mathrm{x}')$. It can be written explicitly in spherical coordinates \cite{Cohl:2012},
\begin{equation}\label{eq:geodinsph}
\begin{aligned}
    \rho(\mathrm{x},\mathrm{x}') &= \alpha \arccosh\big[\cosh\tfrac{r}{\alpha}\cosh\tfrac{r'}{\alpha}-\sinh\tfrac{r}{\alpha}\sinh\tfrac{r'}{\alpha}
    \\
    &\feq\times\left(\cos\vartheta_1\cos\vartheta_1'+\sin\vartheta_1\sin\vartheta_1'\left(\dots(\cos\vartheta_{N-2}\cos\vartheta_{N-2}'+\sin\vartheta_{N-2}\sin\vartheta_{N-2}'\cos(\varphi-\varphi'))\dots\right)\right)\big]\;.
\end{aligned}
\end{equation}
Due to homogeneity of $\mathbb{H}^N$, we can move the origin of the spherical coordinates to an arbitrary point $\mathrm{x}\in M$ such that the coordinate $r$ then coincides with the geodesic distance to an arbitrary point $\mathrm{x}'\in M$, $r=\rho(\mathrm{x},\mathrm{x}')$.

\subsection{Covariant tensor decomposition}

Let us consider an arbitrary symmetric rank-2 tensor $\bs{\phi}$ in $\mathbb{H}^N$. It can be covariantly decomposed into TT tensor $\bs{\phi}^{\perp}$, transverse covector $\bs{\phi}^{\asymp}$, and two scalars $\phi^{\bowtie}$ and $\phi^{\top}$ as follows \cite{Antoniadis:1986sb,DHoker:1999bve}:
\begin{equation}\label{eq:decomp}
\begin{aligned}
    \phi_{ab} &=\phi_{ab}^{\perp}+2\nabla_{(a} \phi_{b)}^{\asymp}+\nabla_{a} \nabla_{b}\phi^{\bowtie}+\tfrac{1}{N} {g}_{ab} \phi^{\top}\;,
    \\
     &\phi^{\top} :=\phi-\Box \phi^{\bowtie}\;, \quad \phi:={g}^{ab}\phi_{ab}\;, \quad \nabla^{a} \phi_{ab}^{\perp}={g}^{ab} \phi_{ab}^{\perp}=0\;, \quad \nabla^{a} \phi_{a}^{\asymp}=0\;,
\end{aligned}
\end{equation}
where $\bs{\nabla}$ is the covariant derivative of $(M,\bs{g})$ and ${\Box={g}^{ab}{\nabla}_a{\nabla}_b}$ is the Laplace operator (Euclideanized wave operator). Notice that the scalar $\phi^{\top}$ is actually different from the pure trace $\phi$. A nice property of $\mathbb{H}^N$ is that the decomposition \eqref{eq:decomp} is actually unique if one considers the fastest-falloff boundary conditions of all quantities involved \cite{DHoker:1999bve}. Its inversion can be characterized by the \textit{projection bi-tensors} $\bs{P}^{\perp}(\mathrm{x},\mathrm{x}')$, $\bs{P}^{\asymp}(\mathrm{x},\mathrm{x}')$, $\bs{P}^{\bowtie}(\mathrm{x},\mathrm{x}')$, and $\bs{P}^{\top}(\mathrm{x},\mathrm{x}')$,\footnote{For the explicit expressions involving scalar and vector Green's functions of $\Box$, we refer the reader to \cite{DHoker:1999bve,Antoniadis:1986sb}.} 
\begin{equation}\label{eq:projections}
\begin{aligned}
    \phi^{\perp}_{ab}(\mathrm{x}) &=\int_{\mathrm{x}'\in{M}} {\mathfrak{g}}^{\nicefrac{1}{2}}(\mathrm{x}') P^{\perp}_{ab}{}^{a'b'}(\mathrm{x},\mathrm{x}')\phi_{a'b'}(\mathrm{x}')\;,
    & \quad
    \phi^{\asymp}_{a}(\mathrm{x}) &=\int_{\mathrm{x}'\in{M}} {\mathfrak{g}}^{\nicefrac{1}{2}}(\mathrm{x}') P^{\asymp}_{a}{}^{a'b'}(\mathrm{x},\mathrm{x}')\phi_{a'b'}(\mathrm{x}')\;,
    \\
    \phi^{\bowtie}(\mathrm{x}) &=\int_{\mathrm{x}'\in{M}} {\mathfrak{g}}^{\nicefrac{1}{2}}(\mathrm{x}') P^{\bowtie}{}^{a'b'}(\mathrm{x},\mathrm{x}')\phi_{a'b'}(\mathrm{x}')\;,
    & \quad
    \phi^{\top}(\mathrm{x}) &=\int_{\mathrm{x}'\in{M}} {\mathfrak{g}}^{\nicefrac{1}{2}}(\mathrm{x}') P^{\top}{}^{a'b'}(\mathrm{x},\mathrm{x}')\phi_{a'b'}(\mathrm{x}')\;.
\end{aligned}
\end{equation}
The symbol ${\mathfrak{g}}^{\nicefrac{1}{2}}:=\sqrt{\det {g}_{ab}}dx^1\cdots dx^N$ stands for the standard metric density. We also introduced a common bi-tensor convention \cite{Allen:1985wd,Allen:1986tt,Poisson:2011nh} in which indices ${a,b,\dots}$ and ${a',b',\dots}$ are associated with the tangent spaces at points ${\mathrm{x}\in{M}}$ and ${\mathrm{x}'\in{M}}$, respectively. 

It is natural to demand that Green's functions of covariant operators on $\mathbb{H}^N$, which also depend on two points $\mathrm{x},\mathrm{x}'\in M$, are invariant under all isometries of $\mathbb{H}^N$. The scalars with such properties will be referred to as  the \textit{MS bi-scalars}; they depend on $\mathrm{x}$ and $\mathrm{x}'$ only through the geodesic distance $\rho(\mathrm{x},\mathrm{x}')$. Since we are interested in gravitational perturbations, we also need the \textit{MS bi-tensors} $\bs{\Phi}(\mathrm{x},\mathrm{x}')$ of rank-2 in tangent-spaces of both points $\mathrm{x},\mathrm{x}'\in M$. Apart from requiring the invariance under all isometries of $\mathbb{H}^N$, we also implicitly demand the following symmetries $\Phi_{ab}{}^{a'b'}=\Phi_{ba}{}^{a'b'}=\Phi_{ab}{}^{b'a'}$ and $\Phi_{aba'b'}(\mathrm{x},\mathrm{x}')=\Phi_{a'b'ab}(\mathrm{x}',\mathrm{x})$. Naturally, they can be decomposed in the sense of \eqref{eq:decomp} in both pairs of indices. For example, the MS TT bi-tensors $\bs{\Phi}^{\perp}$, which will be of special interest for us, satisfy
\begin{equation}\label{eq:TTbitensors}
\begin{aligned}
    {\nabla^a}\Phi^{\perp}_{ab}{}^{a'b'}&=0\;, & {\nabla'_{a'}}\Phi^{\perp}_{ab}{}^{a'b'}&=0\;, 
    \\ 
    g^{ab}\Phi^{\perp}_{ab}{}^{a'b'}&=0\;, & g_{a'b'}\Phi^{\perp}_{ab}{}^{a'b'}&=0\;.
\end{aligned}
\end{equation}

\subsection{Equivalent and quadratic actions}

We will start our discussion with a general $N$-dimensional Riemannian manifold denoted by $(\hat{M},\hat{\bs{g}})$ and consider a (Euclidean) gravitational action $S^{\textrm{E}}$ of the generic form
\begin{equation}\label{eq:analyticlagrangian}
    S^{\textrm{E}}[\hat{\bs{g}}]:=-\int_{\hat{M}}\hat{\mathfrak{g}}^{\nicefrac{1}{2}}L(\hat{\bs{g}},\hat{\bs{R}},\hat{\bs{\nabla}})\;,
\end{equation}
where the Lagrangian $L$ is an arbitrary analytic expression in the metric tensor $\hat{\bs{g}}$, the Riemann tensor $\hat{\bs{R}}$, and the covariant derivative $\hat{\bs{\nabla}}$ that are organized into a scalar function $L$ on $\hat{M}$. In other words, $L$ may contain arbitrary products of covariant derivatives of curvature of any (possibly infinite) orders in curvature and covariant derivatives. We also assume that $\mathbb{H}^N$ of radius $\alpha$ (i.e., the Ricci scalar ${R=-{(N-1)N}/{\alpha^2}}$) is the vacuum solution of the theory and denote it by $({M},{\bs{g}})$, i.e., ${\delta S^{\textrm{E}}/\delta \hat{\bs{g}}|_{\hat{\bs{g}}={\bs{g}}}=0}$. Let us now consider $(\hat{M},\hat{\bs{g}})$ that differs from $({M},{\bs{g}})$ only by a small metric perturbation $\bs{h}$, $(\hat{M},\hat{\bs{g}})=(M,\bs{g}+\varepsilon\bs{h})$, $\varepsilon>0$. As far as the linearized theory around AdS (or any MS background) is concerned, one can work with a simpler \textit{equivalent action} $S_{\textrm{equiv}}^{\textrm{E}}$ satisfying ${S^{\textrm{E}}-S_{\textrm{equiv}}^{\textrm{E}}=O(\varepsilon^3)}$ \cite{Biswas:2016egy,Biswas:2016etb}. The equivalent action is re-derived using a new approach (for arbitrary MS backgrounds) in Appx.~\ref{apx:equivaction}, where we also fixed a mistake in the original derivation. It takes the form
\begin{equation}\label{eq:IDG}
    S_{\textrm{equiv}}^{\textrm{E}}[\hat{\bs{g}}]:=-\tfrac{1}{2}\int_{\hat{M}}\hat{\mathfrak{g}}^{\nicefrac{1}{2}}\big[\varkappa^{-1}(\hat{R}-2\Lambda)+\hat{R}\mathcal{F}_1(\hat{\Box})\hat{R}+\hat{S}^{ab}\mathcal{F}_2(\hat{\Box})\hat{S}_{ab}+\hat{C}^{abcd}\mathcal{F}_3(\hat{\Box})\hat{C}_{abcd}\big]\;,
\end{equation}
where $\hat{R}$, $\hat{\bs{S}}$, and $\hat{\bs{C}}$ denote the Ricci scalar, the traceless Ricci tensor, and the Weyl tensor of $(\hat{M},\hat{\bs{g}})$, respectively. Here, the operators $\mathcal{F}_i(\hat{\Box})$ are given by some analytic functions $\mathcal{F}_i$ of the wave operator ${\hat{\Box}}$. Analyticity of $\mathcal{F}_i$ allows us to expand the operators $\mathcal{F}_i(\hat{\Box})$ in $\hat{\Box}$,
\begin{equation}
    \mathcal{F}_i(\hat{\Box})=\sum_{n=0}^{\infty}f_{i,n} \hat{\Box}^n\;,
\end{equation}
with $f_{i,k}$ being arbitrary constant coefficients. The constants $\varkappa$ and $\Lambda$ have been chosen so that $({M},{\bs{g}})$ is the vacuum solution and the theory reduces to GR for $\mathcal{F}_i(\hat{\Box})=0$. Specifically, the cosmological constant $\Lambda$ is related to the radius $\alpha$ by means of
\begin{equation}
    \Lambda=\tfrac{(N-4)(N-1)^2 N}{2\alpha^4}\mathcal{F}_1(0)-\tfrac{(N-2)(N-1)}{2\alpha^2}\;,
\end{equation}
which is derived in Appx.~\ref{apx:quadraticaction} [cf. \eqref{eq:bgFEs}]. The equivalent action \eqref{eq:IDG} can be actually further simplified depending on $N$. Indeed, if $N=3$, we can take $\mathcal{F}_3(\hat{\Box})=0$ because the Weyl tensor identically vanishes in $N=3$. On the other hand, if $N\geq4$, we can use the identity \cite{Frolov:2015usa,Koshelev:2017tvv}\footnote{This identity is a direct consequence of commutators of covariant derivatives and (contracted) Bianchi identities.}
\begin{equation}\label{eq:GBtypeidentity}
    \hat{C}_{a b c d} \hat{\Box}^n \hat{C}^{a b c d} - 4 \tfrac{N-3}{N-2} \hat{S}_{a b} \hat{\Box}^n \hat{S}^{a b}+\tfrac{(N-2)(N-3)}{N(N-1)} \hat{R} \hat{\Box}^n \hat{R} = O\left(\hat{\bs{R}}^3\right) +\textrm{div}\;, \quad n\geq 1\;,
\end{equation}
to set $\mathcal{F}_2(\hat{\Box})=\mathcal{F}_2(0)$. Moreover, if $N=4$ and $k=0$, the expression above turns into the Gauss--Bonnet term,
\begin{equation}
    \hat{C}_{a b c d} \hat{C}^{a b c d} - 2 \hat{S}_{a b} \hat{S}^{a b}+\tfrac{1}{6} \hat{R}^2 = \textrm{div}\;,
\end{equation}
which allows us to further put $\mathcal{F}_2(0)=0$ in $N=4$. Thus, without loss of generality, we may consider the following simplifications in the equivalent action $S_{\textrm{equiv}}^{\textrm{E}}$:
\begin{equation}\label{eq:simplificationofF}
\begin{aligned}
    \mathcal{F}_3(\hat{\Box}) &=0 \;, & N &=3\;,
    \\
    \mathcal{F}_2(\hat{\Box}) &=0 \;, & N &=4\;,
    \\
    \mathcal{F}_2(\hat{\Box}) &=\mathcal{F}_2(0) \;, & N &\geq5\;.
\end{aligned}
\end{equation}

The \textit{quadratic action} $S_{\textrm{quad}}^{\textrm{E}}$ corresponding to the linearized gravity of $S^{\textrm{E}}$ is then obtained by taking the second variation of $S_{\textrm{equiv}}^{\textrm{E}}$  \cite{ChristensenDuff:1980},
\begin{equation}\label{eq:quadactiondef}
    S_{\textrm{quad}}^{\textrm{E}}[\bs{h}]:=\pp_{\varepsilon}^2S_{\textrm{equiv}}^{\textrm{E}}[{\bs{g}}+\varepsilon \bs{h}]\big|_{\varepsilon=0}- \int_{{M}} {\mathfrak{g}}^{\nicefrac{1}{2}}\tau^{ab}h_{ab}\;,
\end{equation}
where we also included the term corresponding to the standard coupling to the energy-momentum tensor $\bs{\tau}$. One can recast $S_{\textrm{quad}}^{\textrm{E}}$ in a relatively simple form with the help of the covariant tensor decomposition \eqref{eq:decomp}. The important aspect of this decomposition, when applied to the metric perturbation $\bs{h}$, is that it isolates the physical perturbations from the gauge-dependent ones. Recall that tensor fields $h_{ab}$ and $h_{ab}+2{\nabla}_{(a}\xi_{b)}$ with an arbitrary covector field $\bs{\xi}$ represent the same physical perturbations because $({M},{\bs{g}})$ is physically equivalent to $({M},\chi^*{\bs{g}})$ for any diffeomorphism $\chi$ on ${M}$ \cite{wald:1984}.\footnote{We consider the active viewpoint, where $\chi$ denotes an actual map on ${M}$ rather than a coordinate transformation.} The TT tensor $\bs{h}^{\perp}$ and the scalar $h^{\top}$ encode the full gauge-independent information about the perturbation [$N(N-1)/2$ independent components] while the transverse vector  $\bs{h}^{\asymp}$ and the scalar $h^{\bowtie}$ are gauge-dependent ($N$ independent components) and should drop out of all physical quantities \cite{Antoniadis:1986sb}. Employing the decomposition \eqref{eq:decomp} for $\bs{h}$ and $\bs{\tau}$, the quadratic action $S_{\textrm{quad}}^{\textrm{E}}$ can be expressed as
\begin{equation}\label{eq:quadaction}
    S_{\textrm{quad}}^{\textrm{E}}[\bs{h}]  = -\tfrac{1}{2\varkappa}\int_{{M}} {\mathfrak{g}}^{\nicefrac{1}{2}}\left[\tfrac12 h^{\perp}{}^{ab}\left(\Box+\tfrac{2}{\alpha^2}\right) \mathcal{E}(\Box)h_{ab}^{\perp} -\tfrac{(N-2)(N-1)}{2N^2}h^{\top}\left(\Box-\tfrac{N}{\alpha^2}\right) \mathcal{I}(\Box)h^{\top}+ 2\varkappa \tau^{\perp}{}^{ab}h_{ab}^{\perp}+\tfrac{2\varkappa}{N} \tau h^{\top}\right]\;,
\end{equation}
where we assumed that $\bs{\tau}$ is conserved, ${{\nabla}_{a}\tau^{ab}=0}$. Here, the operators $\mathcal{E}(\Box)$ and $\mathcal{I}(\Box)$ are given by analytic functions of $\Box$ and can be expressed in terms of operators $\mathcal{F}_i(\Box)$,
\begin{equation}\label{eq:EIop}
\begin{aligned}
    \mathcal{E}({\Box}) &=\begin{cases}
    1-\varkappa\big[\tfrac{12}{\alpha^2} \mathcal{F}_1(0) - \big(\Box+\tfrac{{2}}{\alpha^2}\big)\mathcal{F}_2(\Box)\big]\;, & N=3\;,
    \\
    1-\varkappa\big[\tfrac{24}{\alpha^2}\mathcal{F}_1(0)-2\big({\Box}+\tfrac{4}{\alpha^2}\big) \mathcal{F}_3\big({\Box}-\tfrac{4}{\alpha^2}\big)\big]\;, & N=4\;,
    \\
    1-\varkappa\big[\tfrac{2N(N-1)}{\alpha^2}\mathcal{F}_1(0) - \mathcal{F}_2(0)\big(\Box+\tfrac{2}{\alpha^2}\big)  - 4 \frac{N-3}{N-2} \big({\Box}+\tfrac{N}{\alpha^2}\big) \mathcal{F}_3\big({\Box}-\tfrac{2(N-2)}{\alpha^2}\big)\big]\;, & N\geq5\;,
    \end{cases}
    \\
    \mathcal{I}({\Box}) &=
    \begin{cases}
    1-\varkappa \big[\tfrac{12}{\alpha^2} \mathcal{F}_1(0) +8\big(\Box-\tfrac{3}{\alpha^2})\mathcal{F}_1(\Box)+\tfrac{1}{3} \Box \mathcal{F}_2\big(\Box-\tfrac{6}{\alpha^2}\big) \big]\;, & N=3\;,
    \\
    1-\varkappa\big[\tfrac{24}{\alpha^2}\mathcal{F}_1(0) +6\big({\Box}-\tfrac{4}{\alpha^2}\big) \mathcal{F}_1({\Box})\big]\;,  & N=4\;,
    \\
    1-\varkappa\big[\tfrac{2N(N-1)}{\alpha^2}\mathcal{F}_1(0) + \tfrac{4(N-1)}{N-2} \big({\Box}-\tfrac{N}{\alpha^2}\big) \mathcal{F}_1({\Box}) + \tfrac{N-2}{N} \mathcal{F}_2(0){\Box} \big]\;, & N\geq5\;.
    \end{cases}
\end{aligned}
\end{equation}
This generalize the known formulas in \cite{Mazumdar:2018xjz,Biswas:2016egy,Biswas:2016etb} from ${N=3,4}$ to an arbitrary number of dimensions, ${N\geq3}$. The complete derivation (valid for arbitrary MS backgrounds) is performed in the Appx.~\ref{apx:quadraticaction} with preliminaries in Appx.~\ref{apx:varmetcur}-\ref{apx:commutators}.

The functional equations \eqref{eq:EIop} can be easily inverted by expressing functons $\mathcal{F}_i$ in terms of functions $\mathcal{E}$ and $\mathcal{I}$. In this process, we have to keep in mind that $\mathcal{F}_i$ are analytic, which imposes certain conditions on $\mathcal{E}$ and $\mathcal{I}$. Particularly, one can find that the analyticity of $\mathcal{F}_i$ leads to
\begin{equation}\label{eq:EIcond}
\begin{aligned}
    &\mathcal{E}\left(-\tfrac{3}{\alpha^2}\right)=\mathcal{I}\left(\tfrac{3}{\alpha^2}\right)\;, \quad \mathcal{I}(0)=2-\mathcal{E}\left(-\tfrac{2}{\alpha^2}\right)\;, & N&=3\;,
    \\
    &\mathcal{E}\left(-\tfrac{4}{\alpha^2}\right)=\mathcal{I}\left(\tfrac{4}{\alpha^2}\right) \;,\quad \mathcal{I}(0)=1\;, & N &=4\;,
    \\
    &\mathcal{E}\left(-\tfrac{N}{\alpha^2}\right)=\mathcal{I}\left(\tfrac{N}{\alpha^2}\right)\;, & N &\geq5\;,
\end{aligned}
\end{equation}
and the inverse of \eqref{eq:EIop} reads
\begin{equation}\label{eq:inverseF}
\begin{aligned}
    \mathcal{F}_1(\Box) &=\tfrac{\mathcal{I}(\Box){-}\mathcal{I}\left(\frac{3}{\alpha^2}\right)}{-8\varkappa\left(\Box-\frac{3}{\alpha^2}\right)}+\tfrac{\left[\mathcal{E}\left(\Box{-}\frac{6}{\alpha^2}\right){-}\mathcal{E}\left(-\frac{2}{\alpha^2}\right)\right]\Box+3\left[\mathcal{E}\left(-\frac{3}{\alpha^2}\right){-}\mathcal{E}\left(-\frac{2}{\alpha^2}\right)\right]\left(\Box{-}\frac{4}{\alpha^2}\right)}{-24\varkappa\left(\Box-\frac{3}{\alpha^2}\right)\left(\Box-\frac{4}{\alpha^2}\right)}\;, \quad \mathcal{F}_2(\Box) =\tfrac{\mathcal{E}(\Box)-\mathcal{E}\left(-\frac{2}{\alpha^2}\right)}{\varkappa\left(\Box+\frac{2}{\alpha^2}\right)}\;, & N&=3\;,
    \\
    \mathcal{F}_1({\Box}) &=\tfrac{\mathcal{I}(\Box)-\mathcal{I}\left(\frac{4}{\alpha^2}\right)}{-6\varkappa\left({\Box}-\frac{4}{\alpha^2}\right)}\;, \quad \mathcal{F}_3({\Box})=\tfrac{\mathcal{E}\left(\Box+\frac{4}{\alpha^2}\right)-\mathcal{E}\left(-\frac{4}{\alpha^2}\right)}{2\varkappa\left({\Box}+\frac{8}{\alpha^2}\right)}\;, & N&=4\;, 
    \\
    \mathcal{F}_{1}(\Box) &=\tfrac{\mathcal{I}(\Box)-\mathcal{I}\left(\frac{N}{\alpha ^2}\right)}{-\frac{4   (N-1) }{N-2}\varkappa\left(\Box-\frac{N}{\alpha ^2}\right)}-\tfrac{(N-2)^2}{4 (N-1) N}\mathcal{F}_2(0)\;, \quad
    \mathcal{F}_3\left(\Box\right) =\tfrac{\mathcal{E}\left(\Box+\frac{2 (N-2)}{\alpha ^2}\right)-\mathcal{E}\left(-\frac{N}{\alpha ^2}\right)}{\frac{4 \varkappa  (N-3) }{N-2}\left(\Box+\frac{3 N-4}{\alpha ^2}\right)}-\tfrac{N-2}{4 (N-3)}\mathcal{F}_2(0)\;,
    \\
    \mathcal{F}_2(0) &=\tfrac{\alpha ^2\mathcal{I}(0)}{  (N-4)\varkappa}-\tfrac{\alpha ^2\mathcal{I}\left(\frac{N}{\alpha ^2}\right)}{ (N-2)\varkappa } -\tfrac{\alpha ^2}{\frac{1}{2}   (N-4) (N-2)\varkappa}\;, & N &\geq5\;.
\end{aligned}
\end{equation}
The common condition ${\mathcal{E}\left(-{N}/{\alpha^2}\right)=\mathcal{I}\left({N}/{\alpha^2}\right)}$, $N\geq3$, arises in the inversion of \eqref{eq:EIop} [i.e., in the derivation of \eqref{eq:inverseF}]. However, in ${N=3,4}$ the inversion formulas \eqref{eq:inverseF} require additional consistency conditions \eqref{eq:EIcond} to guarantee that, when inserted in \eqref{eq:EIop}, they return the identity. We will refer to all conditions in \eqref{eq:EIcond} as the \textit{$\mathcal{F}_i$-analyticity conditions}. Notice that the linearization of an arbitrary gravitational theory \eqref{eq:analyticlagrangian} is now completely characterized by arbitrary functions $\mathcal{E}$ and $\mathcal{I}$ provided that they are analytic and satisfy the $\mathcal{F}_i$-analyticity conditions \eqref{eq:EIcond}.

\subsection{Propagator in Landau gauge}

The field equations for $\bs{h}^{\perp}$ and $h^{\top}$ are obtained by taking the corresponding functional derivatives of $S_{\textrm{quad}}^{\textrm{E}}$ given by \eqref{eq:quadaction},
\begin{equation}
\begin{aligned}
   \frac{\delta S_{\textrm{quad}}^{\textrm{E}}}{\delta \bs{h}^{\perp}} &=-\tfrac{1}{2\varkappa}\Big[\left(\Box+\tfrac{2}{\alpha^2}\right)\mathcal{E}(\Box)\bs{h}^{\perp}+2\varkappa \bs{\tau}^{\perp}\Big]\;,
    \\
    \frac{\delta S_{\textrm{quad}}^{\textrm{E}}}{\delta h^{\top}} &=-\tfrac{1}{2\varkappa }\Big[-\tfrac{(N-2)(N-1)}{N^2}\left(\Box-\tfrac{N}{\alpha^2}\right)\mathcal{I}(\Box) h^{\top}+\tfrac{2\varkappa}{N} \tau\Big]\;.
\end{aligned}
\end{equation}
Notice that these formulas do not actually determine the gauge-dependent terms $\bs{h}^{\asymp}$ and $h^{\bowtie}$ because these terms disappeared in the calculation of the quadratic action. Thus, it is natural to impose the following covariant gauge conditions
\begin{equation}
    \nabla_a h^a_b=\tfrac{1}{N}\nabla_b h\;,
\end{equation}
known as the \textit{Landau gauge}, which implies $\bs{h}^{\asymp}=0$ and $h^{\bowtie}=0$ \cite{DHoker:1999bve}. Furthermore, the scalar $h^{\top}$ reduces to the pure trace, $h^{\top}=h$, so the full metric perturbation is now given by $\bs{h}=\bs{h}^{\perp}+\frac{h}{N}{\bs{g}}$. Thus, the linearized field equations take the form
\begin{equation}\label{eq:LFE}
\begin{aligned}
    -\left(\Box+\tfrac{2}{\alpha^2}\right)\mathcal{E}(\Box)\bs{h}^{\perp} &=2\varkappa \bs{\tau}^{\perp}\;,
    \\
    -\left(\Box-\tfrac{N}{\alpha^2}\right)\mathcal{I}(\Box) h &=-\tfrac{2\varkappa N}{(N-2)(N-1)} \tau\;.
\end{aligned}
\end{equation}

The equations \eqref{eq:LFE} can be solved using the Green's function methods. Following \cite{Antoniadis:1986sb,DHoker:1999bve}, we introduce the \textit{propagator (in Landau gauge)}  $\bs{G}(\mathrm{x},\mathrm{x}')$ as an MS bi-tensor (i.e., $G_{ab}{}^{a'b'}(\mathrm{x},\mathrm{x}')$) of the form
\begin{equation}\label{eq:gravprop}
    \bs{G}:=\bs{G}^{\perp}-\tfrac{1}{(N-2)(N-1)}G\bs{O}_1\;,
\end{equation}
where ${G(\rho)}$ is an MS bi-scalar and $\bs{G}^{\perp}(\mathrm{x},\mathrm{x}')$ is an MS TT bi-tensor satisfying\footnote{Let us stress that the equations \eqref{eq:GF} are valid only in the Landau gauge. The approach through the quadratic action (which allowed us to linearize all gravitational theories at the same time) becomes less useful in other covariant gauges, where the vector part is non-zero and contains very complicated terms already in GR \cite{Faizal:2011sa}. Derivation of the equations for propagator and finding their solutions may then require a very different procedure compared to what we will present in the next section.}
\begin{equation}\label{eq:GF}\boxed{
\begin{aligned}
    -\left(\Box+\tfrac{2}{\alpha^2}\right)\mathcal{E}(\Box)\bs{G}^{\perp} &= \bs{P}^{\perp}\;,
    \\
    -\left(\Box-\tfrac{N}{\alpha^2}\right)\mathcal{I}(\Box) G &= \delta\;.
\end{aligned}}
\end{equation}
The boundary conditions are fixed by demanding the fastest possible falloff at long geodesic distances and matching the flat space results ($\alpha\to\infty$) at short geodesic distances. Remark that the TT projection bi-tensor $\bs{P}^{\perp}(\mathrm{x},\mathrm{x'})$ plays a role of the \textit{Dirac-delta bi-scalar} $\delta(\mathrm{x},\mathrm{x}')$ on the space of symmetric TT rank-2 tensors. 

Given $\bs{G}^{\perp}$ and $G$, the solution of \eqref{eq:LFE} can be written as a convolution with $\bs{\tau}^{\perp}$ and $\tau$,
\begin{equation}\label{eq:conv}
\begin{aligned}
    h^{\perp}_{ab} &=2\varkappa\int_{\mathrm{x}'}G^{\perp}_{ab}{}^{a'b'}\tau^{\perp}_{a'b'}\;, 
    \\
    h &=-2\varkappa\tfrac{N}{(N-2)(N-1)}\int_{\mathrm{x}'}G\tau\;,
\end{aligned}
\end{equation}
where we suppressed dependencies on points $\mathrm{x},\mathrm{x}'\in{M}$ and introduced the shorthand notation $\int_{\mathrm{x}'}:=\int_{\mathrm{x}'\in {M}}{\mathfrak{g}}^{\nicefrac{1}{2}}(\mathrm{x}')$. By adding these two contributions together, ${\bs{h}=\bs{h}^{\perp}+\frac{h}{N}\bs{g}}$, and writing  $\bs{\tau}^{\perp}$ and $\tau$ as projections of $\bs{\tau}$, we can express $\bs{h}$ as a convolution of the propagator $\bs{G}(\mathrm{x},\mathrm{x}')$ with the full energy-momentum tensor $\bs{\tau}$,
\begin{equation}
        h_{ab}=2\varkappa\int_{\mathrm{x}'}G{}_{ab}{}^{a'b'}\tau_{a'b'}\;.
\end{equation}
Alternatively, the last equation can be also verified by inserting its projections directly in \eqref{eq:LFE}.

\section{Heat kernel approach}\label{sec:heatkernelapproach}

In this section, we show that the propagators for generic theories (introduced in the previous section) can be expressed by means of the heat kernels for scalars and symmetric TT rank-2 tensors in $\mathbb{H}^N$. We start by recasting the action of operators $\mathcal{D}(\Box)$ using the convolution with the heat kernels for scalars as well as symmetric TT rank-2 tensors. With the help of this heat kernel representation, we invert the kinetic operators and find the desired relations for the propagators involving integrals of the heat kernels. Upon realizing that every MS TT bi-tensor is fully characterized by a single MS bi-scalar, we further reduce the bi-tensor formulas to bi-scalar expressions and provide explicit formulas for the heat kernels.

\subsection{Heat kernel representation of \texorpdfstring{$\mathcal{D}(\Box)$}{D(box)}}

There exist various representations of the operators of the form $\mathcal{D}(\Box)$. The most straightforward definition views the operator $\mathcal{D}(\Box)$ as a (finite or infinite) sum of differential operators $\Box^n$,
\begin{equation}\label{eq:infsum}
    \mathcal{D}(\Box):=\sum_{n=0}^{\infty} \frac{\mathcal{D}^{(n)}(0)}{n!} \Box^n\;.
\end{equation}
The obvious drawback is that it holds only for analytic functions $\mathcal{D}$. More importantly, a direct calculation of higher derivatives of tensor fields can be very tedious and lead to convoluted expressions. It is especially difficult to write finite or infinite sums of these expressions in closed forms that would be useful for further analysis.

An alternative representation, which is typically more under control, uses the results of harmonic analysis on $\mathbb{H}^N$. This allows us to characterize the action of operators $\mathcal{D}(\Box)$ on the tensor fields by the spectral decomposition of $-\Box$. In particular, we can decompose the symmetric TT rank-2 tensors $\bs{\phi}^{\perp}$ and scalars $\phi$ in $\mathbb{H}^N$ into complete sets of eigenmodes, ${-\Box\bs{\psi}^{\perp}_{\lambda,\omega} =\left(\lambda+\frac{2}{\alpha^2}\right)\bs{\psi}^{\perp}_{\lambda,\omega}}$ and ${-\Box{\psi}_{\lambda,\omega} =\lambda{\psi}_{\lambda,\omega}}$, respectively. The explicit expressions for $\bs{\psi}^{\perp}_{\lambda,\omega}$ and $\psi_{\lambda,\omega}$ on $\mathbb{H}^N$ are well known and can be found, e.g., in \cite{Camporesi:1994ga,Camporesi:1995fb}. Then, the operators $\mathcal{D}(\Box)$ can be defined through the spectral integrals,
\begin{equation}\label{eq:spectralresol}
    \mathcal{D}(\Box):=\int_{\lambda_{\textrm{min}}}^{\infty} d E_{\lambda}\; \mathcal{D}(-\lambda-\lambda_{\textrm{spin}})\;, \quad \lambda_{\textrm{min}}:=\tfrac{(N-1)^2}{4\alpha^2}\;, \quad \lambda_{\textrm{spin}}:=\begin{cases}
        \tfrac{2}{\alpha^2}\;, &\text{for $\bs{\phi}^{\perp}$}\;,
        \\
        0\;, &\text{for $\phi$}\;,
    \end{cases}
\end{equation}
where $dE_{\lambda}$ is the spectral measure of $-\Box$, which encodes the information about the projections to the eigenmodes $\bs{\psi}^{\perp}_{\lambda,\omega}$ and ${\psi}_{\lambda,\omega}$. Here, the spectrum is continuous because $\mathbb{H}^N$ is a non-compact manifold. In contrast to \eqref{eq:infsum}, the definition \eqref{eq:spectralresol} extends also to non-analytic functions $\mathcal{D}$ such as the meromophic functions in $\mathbb{C}$. Hence, it enables us to solve equations \eqref{eq:GF} by inverting the kinetic operators on the left-hand side in the spirit of the Fourier-transform method for solving differential equations. Although technically doable, the evaluation of spectral integrals \eqref{eq:spectralresol} is still quite difficult as it requires detailed knowledge of the complete sets of eigenmodes $\bs{\psi}^{\perp}_{\lambda,\omega}$ and $\psi_{\lambda,\omega}$.

Interestingly, one can skip this step by relating $\mathcal{D}(\Box)$ to the \textit{evolution operators} $e^{s(\Box+\lambda_{\textrm{spin}})}$, $s\geq0$, or $e^{i\sigma(\Box+\lambda_{\textrm{spin}})}$, $\sigma\in\mathbb{R}$. Then, in the view of the spectral representation ${\mathcal{D}(\Box):=\int_{\lambda_{\textrm{min}}}^{\infty} d E_{\lambda'}\mathcal{D}({-}\lambda'{-}\lambda_{\textrm{spin}})}$ and using the integral representations of Dirac-delta in terms of the
inverse Laplace or Fourier transforms ${\delta(\lambda-\lambda')=\mathsf{L}^{-1}[e^{-\lambda's}](\lambda)=\mathsf{F}^{-1}[e^{-i\lambda'\sigma}](\lambda)}$,\footnote{We use the following conventions for the Laplace and Fourier transforms
\begin{equation}
\begin{aligned}
    \mathsf{L}[f](y) &=\int_0^{\infty} dx f(x) e^{-xy}, & \mathsf{L}^{-1}[f](x) &=\frac{1}{2 \pi i} \int_{\upsilon-i\infty}^{\upsilon+i\infty}dy f(y) e^{xy}\;,
    \\
    \mathsf{F}[f](y) &=\int_{-\infty}^{\infty} d x f(x) e^{-i xy}, & \mathsf{F}^{-1}[f](x) &=\frac{1}{2 \pi} \int_{-\infty}^{\infty} dy f(y) e^{ixy} \;,
\end{aligned}
\end{equation}
where the parameter $\upsilon$ must be chosen so that the integration path lies in the domain of analyticity of $f$.} we can re-cast the spectral integral \eqref{eq:spectralresol} to a more concrete form
\begin{equation}\label{eq:DboxHK1}
\begin{aligned}
    {\mathcal{D}(\Box)} &= \int_{\lambda_{\textrm{min}}}^{\infty} d \lambda\, \mathcal{D}(-\lambda-\lambda_{\textrm{spin}})
    \times\begin{cases}
    \mathsf{L}^{-1}\big[e^{s(\Box+\lambda_{\textrm{spin}})}\big](\lambda)\;,
    \\
    \mathsf{F}^{-1}\big[e^{i\sigma(\Box+\lambda_{\textrm{spin}})}\big](\lambda)\;.
    \end{cases}
\end{aligned}
\end{equation}
This way we can avoid the explicit use of $\bs{\psi}^{\perp}_{\lambda,\omega}$ and $\psi_{\lambda,\omega}$ (hidden in the spectral measure $d E_{\lambda}$) because the actions of $e^{s(\Box+\lambda_{\textrm{spin}})}$ and $e^{i\sigma(\Box+\lambda_{\textrm{spin}})}$ can be also characterized by the evolution dictated by the heat and the Schr\"odinger equations. Furthermore, by taking the advantage of the inversion theorems for the Laplace and Fourier transforms, ${\mathsf{L}[\mathsf{L}^{-1}[f]](\lambda')=\mathsf{F}[\mathsf{F}^{-1}[f]](\lambda')=f(\lambda')}$, we can derive another integral representations of ${\mathcal{D}(\Box)}$ that is related even more directly to the evolution operators,
\begin{equation}\label{eq:DboxHK2}
\begin{aligned}
    {\mathcal{D}(\Box)} &= \int_0^{\infty} ds\, \mathsf{L}^{-1}[\mathcal{D}(-\lambda-\lambda_{\textrm{spin}})](s) e^{s(\Box+\lambda_{\textrm{spin}})}
    \\
    &=\int_{-\infty}^{\infty} d\sigma\, \mathsf{F}^{-1}[\mathcal{D}(-\lambda-\lambda_{\textrm{spin}})](\sigma) e^{i\sigma(\Box+\lambda_{\textrm{spin}})}\;.
\end{aligned}
\end{equation}
Note that the presence of $\lambda_{\textrm{spin}}$ in above equations is purely conventional; it can be removed by the shift properties of $\mathsf{L}^{-1}$ and $\mathsf{F}^{-1}$. More importantly, the equations involving $e^{i\sigma(\Box+\lambda_{\textrm{spin}})}$ and $\mathsf{F}^{-1}$ are related to those involving $e^{s(\Box+\lambda_{\textrm{spin}})}$ and $\mathsf{L}^{-1}$ by an analytic continuation ${s\to i\sigma+0^+}$. For the ease of notation, we will explicitly write most formulas only for the latter; the formulas for the former are usually implicit. For brevity, we will sometimes omit the small positive part $0^{+}$; it can be always reintroduced by a formal replacement ${\sigma\to\sigma-i0^+}$.

The action of the evolution operator $e^{s(\Box+\lambda_{\textrm{spin}})}$ on symmetric TT rank-2 tensors $\bs{\phi}^{\perp}$ and scalars $\phi$ in $\mathbb{H}^N$ describes the solution of the heat equation on $M\times\mathbb{R}_+$ with the initial conditions (at ${s=0}$) corresponding to $\bs{\phi}^{\perp}$ and $\phi$, respectively. Thus, the resulting tensor field is determined by the convolution with the \textit{heat kernels} $\bs{K}^{\perp}(\mathrm{x},\mathrm{x}';s)$ and $K(\mathrm{x},\mathrm{x}';s)$,
\begin{equation}\label{eq:etoboxisK}
\begin{aligned}
    e^{s\left(\Box+\frac{2}{\alpha^2}\right)}\phi_{ab}^{\perp} &=\int_{\mathrm{x}'}K^{\perp}_{ab}{}^{a'b'}ˇ(s)\phi_{a'b'}^{\perp}\;,
    \\
    e^{s\Box}\phi &=\int_{\mathrm{x}'}K(s)\phi\;,
\end{aligned}
\end{equation}
which are solutions of the heat equations with the initial conditions given by $\bs{P}^{\perp}$ and $\delta$,
\begin{equation}\label{eq:heatequations}\boxed{
\begin{aligned}
    \left(\Box+\tfrac{2}{\alpha^2}\right) \bs{K}^{\perp} &=\pp_s \bs{K}^{\perp}\;, &  \bs{K}^{\perp}(0) &=\bs{P}^{\perp}\;,
    \\
    \Box K &=\pp_s K\;,  & K(0) &=\delta\;,
\end{aligned}}
\end{equation}
In other words, the heat kernel on symmetric TT rank-2 tensors $\bs{K}^{\perp}$ and the heat kernel on scalars $K$ are results of the action of the evolution operators $e^{s(\Box+\lambda_{\textrm{spin}})}$ on the TT projection bi-tensor $\bs{P}^{\perp}$ and the Dirac-delta bi-scalar $\delta$,
\begin{equation}\label{eq:eboxdelta}
\begin{aligned}
    e^{s\left(\Box+\frac{2}{\alpha^2}\right)}\bs{P}^{\perp} &=\bs{K}^{\perp}(s)\;,
    \\ 
    e^{s\Box}\delta &=K(s)\;.
\end{aligned}
\end{equation}
Moreover, for every value of ${s\geq0}$, $K(\rho;s)$ is the MS bi-scalar and $\bs{K}^{\perp}(\mathrm{x},\mathrm{x}';s)$ is the MS TT bi-tensor. Naturally, the \textit{Schr\"odinger kernels} $\bs{K}^{\perp}(\mathrm{x},\mathrm{x}';i\sigma)$ and $K(\mathrm{x},\mathrm{x}';i\sigma)$ can be obtained by analytic continuation mentioned above.

\subsection{Relation between propagator and heat kernel}

With this heat kernel representation of $\mathcal{D}(\Box)$, we can now return to the propagator $\bs{G}(\mathrm{x},\mathrm{x}')$ of any gravitational theory characterized by $\mathcal{E}(\Box)$ and $\mathcal{I}(\Box)$. Inverting the operators on the left-hand side of \eqref{eq:GF} in the sense of spectral integrals \eqref{eq:spectralresol}, we can write
\begin{equation}\label{eq:Gusinginverse}
\begin{aligned}
    \bs{G}^{\perp} &=\left(-\left(\Box+\tfrac{2}{\alpha^2}\right)\mathcal{E}(\Box)\right)^{-1}\bs{P}^{\perp} & &:=\int_{\lambda_{\textrm{min}}}^{\infty} d E_{\lambda} \left({\lambda\,\mathcal{E}\big(-\lambda-\tfrac{2}{\alpha^2}\big)}\right)^{-1}\bs{P}^{\perp}\;, 
    \\
    G &=\left(-\left(\Box-\tfrac{N}{\alpha^2}\right)\mathcal{I}(\Box)\right)^{-1}\delta & &:=\int_{\lambda_{\textrm{min}}}^{\infty} d E_{\lambda}\left(\left(\lambda+\tfrac{N}{\alpha^2}\right)\mathcal{I}(-\lambda)\right)^{-1}\delta\;.
\end{aligned}
\end{equation}
Now, applying \eqref{eq:DboxHK1} and \eqref{eq:DboxHK2} to the inverted operators in \eqref{eq:Gusinginverse} and using \eqref{eq:eboxdelta}, we can find explicit formulas for $\bs{G}^{\perp}$ and $G$ in terms of the heat kernels $\bs{K}^{\perp}(s)$ and $K(s)$ as well as the analogous formulas in terms of the Schr\"odinger kernels $\bs{K}^{\perp}(i\sigma)$ and $K(i\sigma)$. Specifically, the bi-scalar $G$ can be written as
\begin{equation}\label{eq:GK}
\boxed{\begin{aligned}
    G &=\int_{\lambda_{\textrm{min}}}^{\infty} d\lambda\,\left(\left(\lambda+\tfrac{N}{\alpha^2}\right)\mathcal{I}(-\lambda)\right)^{-1}\mathsf{L}^{-1}\left[K(s)\right](\lambda) 
    \\
     &= \int_{0}^{\infty} ds\,\mathsf{L}^{-1}\left[\left(\left(\lambda+\tfrac{N}{\alpha^2}\right)\mathcal{I}(-\lambda)\right)^{-1}\right](s) K(s)\;,
\end{aligned}} 
\end{equation}
while the bi-tensor $\bs{G}^{\perp}$ as
\begin{equation}\label{eq:GKbs}
\boxed{\begin{aligned}
    \bs{G}^{\perp} &= \int_{\lambda_{\textrm{min}}}^{\infty} d\lambda\, \left({\lambda\,\mathcal{E}\big(-\lambda-\tfrac{2}{\alpha^2}\big)}\right)^{-1} \mathsf{L}^{-1}\left[\bs{K}^{\perp}(s)\right](\lambda) 
    \\
    &=\int_{0}^{\infty} ds\, \mathsf{L}^{-1}\left[\left({\lambda\,\mathcal{E}\big(-\lambda-\tfrac{2}{\alpha^2}\big)}\right)^{-1}\right](s) \bs{K}^{\perp}(s)\;.
\end{aligned}}
\end{equation}
The integrals in first lines of \eqref{eq:GK} and \eqref{eq:GKbs} actually hide the well-known \textit{inverse spherical transform on $\mathbb{H}^N$} of $G$ \cite{bray1994fourier,Helgason:2022,Awonusika:2016} and its TT bi-tensor version applied to $\bs{G}^{\perp}$. On the other hand, the integrals in the second lines of \eqref{eq:GK} and \eqref{eq:GKbs}, involving the inverse Laplace transform of the inverse of the kinetic operators from \eqref{eq:GF}, utilize the heat kernel approach similar to the one used in \cite{Frolov:2015bta,Frolov:2015usa,Frolov:2015bia} but in the case of the hyperbolic instead of the flat spaces. The relevant integrals are typically simpler to evaluate in many cases of physical interests. Naturally, relations \eqref{eq:GK} and \eqref{eq:GKbs} also exist in the versions with Schr\"odinger kernels $K(i\sigma)$ and $\bs{K}^{\perp}(i\sigma)$ (and with $\mathsf{L}^{-1}$ replaced by $\mathsf{F}^{-1}$). Of course, the usefulness of all these formulas depends on the convergence properties and the complexity of the respective integrals and the inverse Laplace/Fourier transforms for the theories in question. Recall that \eqref{eq:GK} and \eqref{eq:GKbs} are valid for an arbitrary gravitational theory of the form \eqref{eq:analyticlagrangian}, which is fully characterized by analytic functions $\mathcal{E}$ and $\mathcal{I}$ satisfying the $\mathcal{F}_i$-analyticity conditions \eqref{eq:EIcond}. In what follows, we will use the properties of MS TT bi-tensors to reduce \eqref{eq:GKbs} to a single bi-scalar relation [similar to \eqref{eq:GK}].

\subsection{MS TT bi-tensors via MS bi-scalars}

It is well known \cite{Allen:1986tt,Turyn:1988af,Antoniadis:1986sb,DHoker:1999bve} (see also \cite{Allen:1985wd}) that every MS bi-tensor (with the assumed symmetries) can be written as a linear combination of five MS bi-tensors $\bs{O}_{k}(\mathrm{x},\mathrm{x}')$ with the MS bi-scalar coefficients $\Phi_{k}(\rho)$,
\begin{equation}\label{eq:bidecomposition}
     \bs{\Phi}(\mathrm{x},\mathrm{x}')=\sum_{k=1}^{5} \Phi_{k}(\rho) \bs{O}_{k}(\mathrm{x},\mathrm{x}')\;.
\end{equation} 
The basis MS bi-tensors $\bs{O}_{k}$ are expressible using the following geometric objects \cite{Allen:1986tt,Turyn:1988af}: i) the metric tensor $\bs{g}(\mathrm{x})$, ii) the \textit{parallel transporter} ${\bs{\delta}}_{\parallel}(\mathrm{x},\mathrm{x}')$, which parallel-transport vectors $\bs{v}$ as $v^{a'}(\mathrm{x}') =v^{a}(\mathrm{x}){\delta}_{\parallel}{}_{a}{}^{a'}(\mathrm{x},\mathrm{x}')$, and iii) the \textit{gradients of geodesic distance} $\bs{n}(\mathrm{x},\mathrm{x}'):=\bs{\dd}\rho(\mathrm{x},\mathrm{x}')$, $\bs{n}'(\mathrm{x},\mathrm{x}'):=\bs{\dd}'\rho(\mathrm{x},\mathrm{x}')$, which have unit norms and point away from each other, ${n_a=-n'_{a'} {\delta}_{\parallel}{}_{a}{}^{a'}}$. Namely, $\bs{O}_{k}$ can be defined as \cite{Allen:1986tt,Turyn:1988af}
\begin{equation}
\begin{aligned}
    O_{1}{}_{ab}{}^{a'b'} &:=g_{ab}g^{a'b'}\;,
    \\
    O_{2}{}_{ab}{}^{a'b'} &:=n_a n_b n'^{a'}n'^{b'}\;,
    \\
    O_{3}{}_{ab}{}^{a'b'} &:=2\delta_{\parallel}{}_a{}^{(a'}\delta_{\parallel}{}_b{}^{b')}\;,
    \\
    O_{4}{}_{ab}{}^{a'b'} &:=g_{ab}n'^{a'}n'^{b'}+g^{a'b'}n_{a}n_{b}\;,
    \\
    O_{5}{}_{ab}{}^{a'b'} &:=4{\delta}_{\parallel}{}_{(a}{}^{(a'}n_{b)}n'^{b')}\;.
\end{aligned}
\end{equation}
The above MS bi-tensor/bi-scalar relation gets further simplified in the case of MS TT bi-tensors $\bs{\Phi}^{\perp}(\mathrm{x},\mathrm{x}')$. Introducing two auxiliary bi-scalars $\breve{\Phi}:=\Phi_3-\Phi_5$ and $\bar{\Phi}:=2\Phi_3-(N-1)\Phi_4$, the conditions on tracelesness and trasversality \eqref{eq:TTbitensors} can be recast to the form
\begin{equation}
\begin{aligned}
    0&= \Phi_2+N\Phi_4-4\Phi_5 \;,
    \\
    0&=N \Phi_1+2\Phi_3+\Phi_4\;,
    \\
    0&=\pp_{\rho}\breve{\Phi}+\tfrac{N}{\alpha} \coth\tfrac{\rho}{\alpha}\, \breve{\Phi}-\tfrac{N}{2\alpha}\csch\tfrac{\rho}{\alpha}\, \bar{\Phi} -\tfrac{N(N-1)-2}{2\alpha}\csch\tfrac{\rho}{\alpha}\, \Phi_4\;,
    \\
    0&=\pp_{\rho}\bar{\Phi}+\tfrac{N}{\alpha}\coth\tfrac{\rho}{\alpha}\, \bar{\Phi}-\tfrac{2N}{\alpha}\csch\tfrac{\rho}{\alpha}\, \breve{\Phi}\;.
\end{aligned}
\end{equation}
Interestingly, it was first noticed in \cite{Allen:1986tt,Turyn:1988af} that these equations can be solved for all $\Phi_k$ (and $\breve{\Phi}$) in terms of $\bar{\Phi}$, $\pp_{\rho}\bar{\Phi}$, $\pp_{\rho}^2\bar{\Phi}$. As a consequence of this, any MS TT bi-tensor $\bs{\Phi}^{\perp}$ is fully and uniquely characterized by a single MS bi-scalar $\bar{\Phi}(\rho)$. Namely, once $\bar{\Phi}$ is known, we can easily calculate all the components of \eqref{eq:bidecomposition} as
\begin{equation}\label{eq:barexplicit}
\begin{aligned}
    \Phi_1 &=-\tfrac{1}{(N-2) N (N+1)}\left[\alpha \sinh \tfrac{\rho }{\alpha} \left(\alpha \sinh \tfrac{\rho }{\alpha} \pp_{\rho}^2\bar{\Phi}+(2 N+1) \cosh \tfrac{\rho }{\alpha} \pp_{\rho}\bar{\Phi}\right)+(N+1)  \left(N \cosh ^2\tfrac{\rho }{\alpha}-2\right)\bar{\Phi}\right]\;,
    \\
    \Phi_2 &=\tfrac{1}{N (N+1)}\left[\alpha \sinh \tfrac{\rho }{\alpha} \left(\alpha \sinh \tfrac{\rho }{\alpha} \,\pp_{\rho}^2\bar{\Phi}+ \left((2 N+1) \cosh \tfrac{\rho }{\alpha}-2 (N+1)\right)\pp_{\rho}\bar{\Phi}\right)\right]+4  \sinh ^4\tfrac{\rho}{2\alpha}\,\bar{\Phi}\;,
    \\
    \Phi_3 &=\tfrac{1}{4 (N-2) N (N+1)}\left[\alpha (N-1) \left(2 \alpha \sinh ^2\tfrac{\rho }{\alpha}\, \pp_{\rho}^2\bar{\Phi}+(2 N+1) \sinh \tfrac{2 \rho }{\alpha}\, \pp_{\rho}\bar{\Phi}\right)+N (N+1)  \left((N-1) \cosh \tfrac{2 \rho }{\alpha}+N-3\right)\bar{\Phi}\right]\;,
    \\
    \Phi_4 &=\tfrac{1}{(N-2) N (N+1)}\left[\sinh \tfrac{\rho }{\alpha} \left(\sinh \tfrac{\rho }{\alpha} \left(\alpha^2 \pp_{\rho}^2\bar{\Phi}+N (N+1) \bar{\Phi}\right)+\alpha (2 N+1) \cosh \tfrac{\rho }{\alpha} \,\pp_{\rho}\bar{\Phi}\right)\right]\;,
    \\
    \Phi_5 &=\tfrac{1}{2 (N-2) N (N+1)}\left[\alpha \sinh \tfrac{\rho }{\alpha} \left(\alpha (N-1) \sinh \left(\tfrac{\rho }{\alpha}\right) \pp_{\rho}^2\bar{\Phi}+ \left((N-1) (2 N+1) \cosh \tfrac{\rho }{\alpha}-N^2+N+2\right)\pp_{\rho}\bar{\Phi}\right)\right.
    \\
    &\feq\left.+2 N (N+1)  \sinh ^2\tfrac{\rho}{2\alpha} \left((N-1) \cosh \tfrac{\rho }{\alpha}+1\right)\bar{\Phi}\right]\;.
\end{aligned}
\end{equation}
Furthermore, the bi-scalar $\bar{\Phi}$ can be also easily extracted directly from $\bs{\Phi}^{\perp}$ if we contract it with the gradients of the geodesic distance $\bs{n}$ and $\bs{n}'$ \cite{Antoniadis:1986sb,DHoker:1999bve},
\begin{equation}\label{eq:extraction}
    \bar{\Phi}=\tfrac{N}{N-1}n^{a}n^{b}n'_{a'}n'_{b'}\Phi^{\perp}_{ab}{}^{a'b'}\;.
\end{equation}

As we will see below, it actually proves useful to further rewrite $\bar{\Phi}(\rho)$ using yet another bi-scalar $\mathring{\Phi}(\rho)$, which is defined as a solution of
\begin{equation}\label{eq:ringdef}
    \bar{\Phi}=\left(\frac{1}{\alpha\sinh \tfrac{\rho}{\alpha}} \pp_{\rho}\right)^2\mathring{\Phi}\;.
\end{equation}
This definition is motivated by the properties of the Laplace operator, which will be discussed in the next subsection. In addition to that, $\mathring{\Phi}$ is typically simpler to calculate and takes a simpler form than $\bar{\Phi}$, which then follows from it by a straightforward differentiation. Note that the relation between $\bar{\Phi}$ and $\mathring{\Phi}$ can be inverted up to the freedom in the homogeneous part,
\begin{equation}\label{eq:hom}
    \left(\frac{1}{\alpha\sinh \tfrac{\rho}{\alpha}} \pp_{\rho}\right)^2\mathring{\Phi}_{\textrm{hom}}=0 \quad \iff \quad \mathring{\Phi}_{\textrm{hom}}= c_1 \alpha \cosh \left(\tfrac{\rho }{\alpha }\right)+c_2\;.
\end{equation}
Since $\bar{\Phi}$ for the quantities $\bs{\Phi}^{\perp}$ of our interest tend to decay faster than $e^{-\kappa{\rho}/{\alpha}}$ with $\kappa>2$ (typically as $e^{-N{\rho}/{\alpha}}$), it is possible to fix $c_1$ and $c_2$ by demanding
\begin{equation}\label{eq:ringcond}
    \lim_{\rho\to\infty}\mathring{\Phi}=0\;,
\end{equation}
which then determines $\mathring{\Phi}$ uniquely. For future reference, we denote the above map from MS TT bi-tensors $\bs{\Phi}^{\perp}$ to MS bi-scalars $\bar{\Phi}$ and $\mathring{\Phi}$ by $\bar{\Pi}$ and $\mathring{\Pi}$, respectively, i.e., ${\bar{\Pi}(\bs{\Phi}^{\perp})=\bar{\Phi}}$ and ${\mathring{\Pi}(\bs{\Phi}^{\perp})=\mathring{\Phi}}$. We will also use the symbol $\Upsilon$ for the transition map from $\mathring{\Phi}$ to $\bar{\Phi}$, ${\Upsilon:=\bar{\Pi}\circ\mathring{\Pi}^{-1}}$.

\subsection{Bi-scalar reduction of \texorpdfstring{$\bs{G}^{\perp}$ and $\bs{K}^{\perp}$}{Gperp and Kperp}}

Since $\bs{G}^{\perp}$ as well as $\bs{K}^{\perp}(s)$ for every $s\geq0$ are MS TT bi-tensors, they can be characterized by MS bi-scalars $\bar{G}=\bar{\Pi}(\bs{G}^{\perp})$ and $\bar{K}(s)=\bar{\Pi}(\bs{K}^{\perp}(s))$ through the equations \eqref{eq:bidecomposition} and \eqref{eq:barexplicit}. These MS bi-scalars encode the only independent pieces of information that are contained in the first lines of \eqref{eq:GF} and \eqref{eq:heatequations}. By applying \eqref{eq:extraction} to \eqref{eq:GKbs}, we can reduce the bi-tensor relations between $\bs{G}^{\perp}$ and $\bs{K}^{\perp}$ to the bi-scalar relations between $\bar{G}$ and $\bar{K}$,
\begin{equation}\label{eq:GKbar}
\begin{aligned}
    \bar{G} &= \int_{\lambda_{\textrm{min}}}^{\infty} d\lambda\, \left({\lambda\,\mathcal{E}\big(-\lambda-\tfrac{2}{\alpha^2}\big)}\right)^{-1} \mathsf{L}^{-1}\left[\bar{K}(s)\right](\lambda)
    \\
    &= \int_{0}^{\infty} ds\, \mathsf{L}^{-1}\left[\left({\lambda\,\mathcal{E}\big(-\lambda-\tfrac{2}{\alpha^2}\big)}\right)^{-1}\right](s) \bar{K}(s)\;.
\end{aligned}
\end{equation}
We can go one step further and introduce ${\mathring{G}=\mathring{\Pi}(\bs{G}^{\perp})}$ and ${\mathring{K}(s)=\mathring{\Pi}(\bs{K}^{\perp}(s))}$ by means of \eqref{eq:ringdef} with the condition \eqref{eq:ringcond}. The previous relations then take the form
\begin{equation}\label{eq:GKring}
\begin{aligned}
    \mathring{G} &=\int_{\lambda_{\textrm{min}}}^{\infty} d\lambda\,\left({\lambda\,\mathcal{E}\big(-\lambda-\tfrac{2}{\alpha^2}\big)}\right)^{-1} \mathsf{L}^{-1}\big[\mathring{K}(s)\big]\big(\lambda\big) 
    \\    
    &=\int_{0}^{\infty} ds\,\mathsf{L}^{-1}\left[\left({\lambda\,\mathcal{E}\big(-\lambda-\tfrac{2}{\alpha^2}\big)}\right)^{-1}\right](s)  \mathring{K}(s)\;.
\end{aligned}
\end{equation}
Hence, the bi-scalars $\bar{K}$ and $\mathring{K}$ in \eqref{eq:GKbar} and \eqref{eq:GKring} play a similar role as the heat kernel $K$ in \eqref{eq:GK}. As we will see below, the formulas in \eqref{eq:GKring} can be also extracted from equations \eqref{eq:GF} and \eqref{eq:heatequations}. This exercise will also reveal an important relation between $\mathring{K}$ and $K$.

To proceed further, we have to find the bi-scalar representations of the action of $\Box$ on MS TT bi-tensors. First, we can realize that $\Box$ acts on MS bi-scalars $\Phi$ simply like it does on functions of $r$ in spherical coordinates,
\begin{equation}\label{eq:boxonscalars}
    \Box\Phi=\left(\pp_{\rho}^2 +\tfrac{N-1}{\alpha}\coth\tfrac{\rho}{\alpha}\,\pp_{\rho}\right)\Phi\;.
\end{equation}
Since the result of this operation is still a function of $\rho$, i.e., the MS bi-scalar, we can easily iterate the last formula to find $\Box^n$. Hence, for an arbitrary analytic function $\mathcal{D}$ we get [see \eqref{eq:infsum}]
\begin{equation}
    \mathcal{D}(\Box)\Phi=\mathcal{D}\left(\pp_{\rho}^2 +\tfrac{N-1}{\alpha}\coth\tfrac{\rho}{\alpha}\,\pp_{\rho}\right)\Phi\;.
\end{equation}
Moving on to the MS TT bi-tensors $\bs{\Phi}^{\perp}$, one can realize that $\Box\bs{\Phi}^{\perp}$ is again an MS bi-tensor. Hence, it is fully characterized by a single MS bi-scalar via \eqref{eq:bidecomposition} and \eqref{eq:barexplicit}, which takes the form \cite{Allen:1986tt,Turyn:1988af}
\begin{equation}\label{eq:box}
    \bar{\Pi}(\Box\bs{\Phi}^{\perp}) = \left(\pp_{\rho}^2 +\tfrac{N+3}{\alpha}\coth\tfrac{\rho}{\alpha}\,\pp_{\rho}+\tfrac{2N}{\alpha^2}\right)\bar{\Phi}\;,
\end{equation}
where we should recall that ${\bar{\Phi}=\bar{\Pi}(\bs{\Phi}^{\perp})}$. For the reader's convenience, the derivation of this formula is reviewed in Appx.~\ref{apx:powersofbox}. Starting with  $\bar{\Pi}(\Box^{n}\bs{\Phi}^{\perp})$ we can now apply \eqref{eq:box} repeatedly (with $\bs{\Phi}^{\perp}$ replaced by $\Box^{n-1}\bs{\Phi}^{\perp}$, $\Box^{n-2}\bs{\Phi}^{\perp}$, etc.) to arrive at ${\left(\pp_{\rho}^2 +\tfrac{N+3}{\alpha}\coth\tfrac{\rho}{\alpha}\,\pp_{\rho}+\tfrac{2N}{\alpha^2}\right)^n\bar{\Phi}}$. Hence, for an arbitrary analytic $\mathcal{D}$ we obtain
\begin{equation}
    \bar{\Pi}(\mathcal{D}(\Box)\bs{\Phi}^{\perp}) = \mathcal{D}\left(\pp_{\rho}^2 +\tfrac{N+3}{\alpha}\coth\tfrac{\rho}{\alpha}\,\pp_{\rho}+\tfrac{2N}{\alpha^2}\right)\bar{\Phi}\;.
\end{equation}
Notice that the operator ${\pp_{\rho}^2 +\tfrac{N+3}{\alpha}\coth\tfrac{\rho}{\alpha}\,\pp_{\rho}+\tfrac{2N}{\alpha^2}}$ resembles $\Box$ acting on bi-scalars in $\mathbb{H}^{N+4}$ [cf. \eqref{eq:boxonscalars}] with an additional constant term $-{2(N+1)}/{\alpha^2}$. Inspired by the recursive dimensional relations for the heat kernels $K$ (see \eqref{eq:recur} below), one can recast above operator to the dimension $N$ by two applications of ${(\alpha\sinh \tfrac{\rho}{\alpha})^{-1}} \pp_{\rho}$. Concretely, if we consider an arbitrary MS bi-scalar $\Phi$, we can show that
\begin{equation}
    \left(\pp_{\rho}^2 +\tfrac{N+3}{\alpha}\coth\tfrac{\rho}{\alpha}\,\pp_{\rho}+\tfrac{2N}{\alpha^2}\right)\left(\frac{1}{\alpha\sinh \tfrac{\rho}{\alpha}} \pp_{\rho}\right)^2\Phi=\left(\frac{1}{\alpha\sinh \tfrac{\rho}{\alpha}} \pp_{\rho}\right)^2\left(\Box-\tfrac{2}{\alpha^2}\right)\Phi\;.
\end{equation}
Hence, by passing from $\bar{\Phi}$ to $\mathring{\Phi}$ via \eqref{eq:ringdef}, i.e., by using $\pp_{\rho}^2 +\tfrac{N+3}{\alpha}\coth\tfrac{\rho}{\alpha}\,\pp_{\rho}+\tfrac{2N}{\alpha^2}=\Upsilon\circ\left(\Box-\tfrac{2}{\alpha^2}\right)\circ\Upsilon^{-1}$, we can recast \eqref{eq:box} to the relation between $\Box$ on MS TT bi-tensors and $\Box$ on MS bi-scalars,
\begin{equation}
    \mathring{\Pi}(\Box\bs{\Phi}^{\perp}) = \left(\Box-\tfrac{2}{\alpha^2}\right)\mathring{\Phi}\;.
\end{equation}
Iterating this relation, we find
\begin{equation}\label{eq:ringDbox}
    \mathring{\Pi}(\mathcal{D}(\Box)\bs{\Phi}^{\perp}) = \mathcal{D}\left(\Box-\tfrac{2}{\alpha^2}\right)\mathring{\Phi}\;,
\end{equation}
for an arbitrary analytic $\mathcal{D}$.

With the help of \eqref{eq:ringDbox}, we can now go back to the equation for $\bs{G}^{\perp}$ [the first line in \eqref{eq:GF}], and recast it to the form
\begin{equation}\label{eq:Gringequation}
     -\Box\mathcal{E}\left(\Box-\tfrac{2}{\alpha^2}\right)\mathring{G} = \mathring{P}\;,
\end{equation}
where we introduced $\mathring{P}=\mathring{\Pi}(\bs{P}^{\perp})$. Similarly, we can also revisit the equations for $\bs{K}^{\perp}$ [the first line in \eqref{eq:heatequations}] and map the heat equation as well as the initial condition to
\begin{equation}
    \Box \mathring{K} =\pp_s \mathring{K}\;, \quad  \mathring{K}(0) =\mathring{P}\;.
\end{equation}
We see that $\mathring{K}$ is a solution of the scalar heat equation \eqref{eq:heatequations} with the initial condition given by $\mathring{P}$; hence, $\mathring{K}$ must be obtainable as the convolution of $K$ and $\mathring{P}$,
\begin{equation}\label{eq:KPconvolution}
    \boxed{\mathring{K}=\int_{\mathrm{x}'} \mathring{P} K\;.}
\end{equation}
Inverting the operators on the left-hand side of \eqref{eq:Gringequation} in the sense of spectral integrals \eqref{eq:spectralresol}, we get
\begin{equation}\label{eq:mathringGspect}
        \mathring{G} =\left(-\Box\mathcal{E}\left(\Box-\tfrac{2}{\alpha^2}\right)\right)^{-1}\mathring{P} :=\int_{\lambda_{\textrm{min}}}^{\infty} d E_{\lambda} \left({\lambda\,\mathcal{E}\big(-\lambda-\tfrac{2}{\alpha^2}\big)}\right)^{-1}\mathring{P}\;.
\end{equation}
Upon application of \eqref{eq:DboxHK1} and \eqref{eq:DboxHK2} to the inverted operator in \eqref{eq:mathringGspect} and employing \eqref{eq:etoboxisK} together with \eqref{eq:KPconvolution}, we recover the previous relations between $\mathring{G}$ and $\mathring{K}$ in \eqref{eq:GKring}. Furthermore, motivated by \eqref{eq:GKring} with \eqref{eq:KPconvolution}, we may also introduce an auxiliary MS bi-scalar $\check{G}$, which is the Green's function of $-\Box\mathcal{E}\left(\Box-{2}/{\alpha^2}\right)$, 
\begin{equation}\
        \check{G} =\left(-\Box\mathcal{E}\left(\Box-\tfrac{2}{\alpha^2}\right)\right)^{-1}\delta :=\int_{\lambda_{\textrm{min}}}^{\infty} d E_{\lambda} \left({\lambda\,\mathcal{E}\big(-\lambda-\tfrac{2}{\alpha^2}\big)}\right)^{-1}\delta\;,
\end{equation}
and we get instead
\begin{equation}\label{eq:GKcheck}
\begin{aligned}
    \check{G} &=\int_{\lambda_{\textrm{min}}}^{\infty} d\lambda\,\left({\lambda\,\mathcal{E}\big(-\lambda-\tfrac{2}{\alpha^2}\big)}\right)^{-1} \mathsf{L}^{-1}\big[K(s)\big]\big(\lambda\big) 
    \\    
    &=\int_{0}^{\infty} ds\,\mathsf{L}^{-1}\left[\left({\lambda\,\mathcal{E}\big(-\lambda-\tfrac{2}{\alpha^2}\big)}\right)^{-1}\right](s)  K(s)\;,
\end{aligned}
\end{equation}
which allows us to write $\mathring{G}$ as the convolution of $\mathring{P}$ with $\check{G}$,
\begin{equation}\label{eq:Gcheckconv}
    \mathring{G}=\int_{\mathrm{x}'}\mathring{P}\check{G}\;.
\end{equation}
Of course, \eqref{eq:GKbar}, \eqref{eq:GKring}, and \eqref{eq:GKcheck} still have Schr\"odinger kernels counterparts with $\bar{K}(i\sigma)$, $\mathring{K}(i\sigma)$, and $K(i\sigma)$ (and with $\mathsf{L}^{-1}$ replaced by $\mathsf{F}^{-1}$).

The MS bi-scalar $\mathring{P}$ appearing in various expressions above can be found explicitly in various dimensions $N$. Since ${\bar{P}=\bar{\Pi}(\bs{P}^{\perp})}$ was calculated in \cite{DHoker:1999bve},
\begin{equation}\label{eq:Pbar}
    \bar{P}=\frac{(N+1)  \Gamma \left(\frac{N}{2}+1\right) }{\pi ^{\frac{N}{2}}}\frac{  \cosh\tfrac{\rho}{\alpha}-\frac{1}{N-1}}{\alpha^N\sinh ^{N}\tfrac{\rho}{\alpha}\left(\cosh\tfrac{\rho}{\alpha}+1\right)}\;, \quad N\geq3\;,
\end{equation}
all we need is to solve the equation ${\Upsilon(\mathring{P})=\bar{P}}$ for $\mathring{P}$. In the dimensions of the typical physical interests, we obtain
\begin{equation}
    \mathring{P}=\begin{dcases}
    \tfrac{3 \alpha}{ \pi}\tfrac{1}{e^{\rho /\alpha}+1 }\;, & N=3\;,
    \\
    -\tfrac{5}{6 \pi ^2} \left(\sech^2\tfrac{\rho }{2 \alpha }+2 \log \left(\tanh \tfrac{\rho }{2 \alpha }\right)\right)\;, &N=4\;,
    \\
    \tfrac{15}{4 \pi ^2 \alpha  }\tfrac{2 e^{\rho /\alpha }+1}{\left(e^{\rho /\alpha }-1\right) \left(e^{\rho /\alpha }+1\right)^3}\;, &N=5\;.
    \end{dcases}
\end{equation}
More generally, for arbitrary odd ${N\geq3}$, we can also express $\mathring{P}$ in terms of the hypergeometric function,
\begin{equation}
\begin{aligned}
    \mathring{P} &=-\frac{1}{4 k^2 \pi ^{k+\frac{1}{2}} \alpha ^{2 k-3}\sinh ^{2k+1}\tfrac{\rho }{\alpha }} \left[\tfrac{4 k (k+1)}{2 k-1} \Gamma \left(k+\tfrac{3}{2}\right) \sinh ^{2}\tfrac{\rho }{\alpha } \left(\, _2F_1\left(\tfrac{1}{2},\tfrac{1}{2}-k;\tfrac{3}{2}-k;-\sinh ^2\tfrac{\rho }{\alpha }\right)-1\right)\right.
    \\
    &\feq\left.+2^{-2k-1} \Gamma \left(\tfrac{1}{2}-k\right) \Gamma (2 k+3)\sinh ^{2k+1}\tfrac{\rho }{\alpha }\right]\;, \quad N=2k+1\;,
\end{aligned}
\end{equation}
where ${k\geq1}$.

\subsection{Explicit heat kernels}\label{sec:explicitheatkernels}

The heat kernel $K$ for scalars in $\mathbb{H}^{N}$ is well known in the literature \cite{Debiard:1976,DaviesMandouvalos:1988,Grigoryan:1998}. It is given by the following expressions:\footnote{There also exists an alternative exact expression that does not distinguish between odd and even dimensions \cite{Gruet:1996}, or a precise two-sided global estimate \cite{DaviesMandouvalos:1988},
\begin{equation*}
\begin{aligned}
    K =\frac{\Gamma\left(\frac{N+1}{2}\right)\alpha^{N+1} e^{-\frac{(N-1)^2s}{8\alpha^2} }}{\pi(2 \pi)^{N / 2} \sqrt{s}}  \int_0^{\infty}\!d v\, \frac{e^{\alpha^2\frac{\pi^2-v^2}{2 s}} \sinh v\, \sin (\frac{\pi \alpha^2 v}{s})}{\left(\cosh  v+\cosh (\tfrac{\rho}{\alpha})\right)^{\frac{N+1}{2}}}
    \simeq \frac{(1+\frac{\rho}{\alpha}+\frac{s}{\alpha^2})^{\frac{N-3}{2}}(1+\frac{\rho}{\alpha^2})}{s^{N / 2}} e^{-\tfrac{(N-1)^2}{4\alpha^2} s-\frac{\rho^2}{4 s}-\tfrac{(N-1)}{2\alpha} \rho}\;,
\end{aligned}
\end{equation*}
where $f\simeq g$ is a shorthand for $\exists c>0:c^{-1}\leq f/g\leq c$.}
\begin{equation}\label{eq:scalarK}
        K= \begin{dcases}
            \frac{(-1)^k}{2^k \pi^k} \frac{1}{(4 \pi s)^{\frac{1}{2}}}\left(\frac{1}{\alpha\sinh \tfrac{\rho}{\alpha}} \pp_{\rho}\right)^k e^{-\tfrac{k^2s}{\alpha^2} -\frac{\rho^2}{4 s}}\;, & N=2k+1\;,
            \\
            \frac{(-1)^{k}}{2^{k+\frac{5}{2}} \pi^{k+\frac{3}{2}}}  \frac{\alpha e^{-\frac{\left(k+\frac12\right)^2s}{\alpha^2} }}{s^{\frac{3}{2}}}\left(\frac{1}{\alpha\sinh \tfrac{\rho}{\alpha}} \pp_{\rho}\right)^k \int_{\frac{\rho}{\alpha}}^{\infty}\!d u \,\frac{u e^{-\frac{\alpha^2u^2}{4 s}}}{\sqrt{\cosh u-\cosh \tfrac{\rho }{\alpha }}}\;, & N=2k+2\;,
        \end{dcases}
\end{equation}
where $k\geq1$. In particular, for ${N=3,4,5}$, we get
\begin{equation}
 K= \begin{dcases}
    \frac{\rho  \csch\frac{\rho }{\alpha} e^{-\frac{s}{\alpha^2}-\frac{\rho ^2}{4 s}}}{8 \pi ^{\frac32} \alpha s^{3/2}}\;, & N=3\;,
            \\
    -\frac{\alpha   e^{-\frac{9 s}{4 \alpha ^2}}}{8 \sqrt{2} \pi ^{\frac52} s^{\frac32} }\left(\frac{1}{\alpha\sinh \tfrac{\rho}{\alpha}} \pp_{\rho}\right) \int_{\frac{\rho}{\alpha}}^{\infty}\!d u \,\frac{u e^{-\frac{\alpha^2u^2}{4 s}}}{\sqrt{\cosh u-\cosh \tfrac{\rho }{\alpha }}}\;, & N=4\;,
    \\
    \frac{\csch^2 \frac{\rho }{\alpha }}{32 \pi ^{\frac52} \alpha ^3 s^{\frac52}}   \left(\alpha  \rho ^2+2 \rho  s \coth \tfrac{\rho }{\alpha }-2 \alpha  s\right)e^{-\frac{4 s}{\alpha ^2}-\frac{\rho ^2}{4 s}}\;, & N=5\;.  
    \end{dcases}
\end{equation}
Using \eqref{eq:boxonscalars}, it can be shown that these formulas satisfy the second line in \eqref{eq:heatequations}. Furthermore, the heat kernels \eqref{eq:scalarK} can be obtained inductively through the recurrence formula,
\begin{equation}\label{eq:recur}
    K_{N+2}=-\frac{e^{-\frac{N}{\alpha^2}s}}{2\pi}\frac{1}{ \alpha\sinh \tfrac{\rho}{\alpha}}\pp_{\rho}K_{N}\;,
\end{equation}
where the subscript denotes the dimension of the hyperbolic space.

Recall that the heat kernels for symmetric TT rank-2 tensors $\bs{K}^{\perp}$ are characterized the MS bi-scalars $\mathring{K}$ which are given by the convolution with $\mathring{P}$, see \eqref{eq:KPconvolution}. Interestingly, any convolutional integral of this type can be always reduced to a 2-dimensional integral. Indeed, if we consider a convolution of two arbitrary MS bi-scalars $\Phi(\rho)$ and $\Psi(\rho)$ that results in a third MS bi-scalar $\Omega(\rho)$, we can employ the spherical coordinates \eqref{eq:sphcoor} [with the geodesic distance given by \eqref{eq:geodinsph}] and adjust the coordinate systems such that ${r=\rho_{\mathrm{x}\mathrm{x}''}}$, $\vartheta_k=0$, $k=1,\dots,N-1$, and ${r'=\rho_{\mathrm{x}\mathrm{x}'}}$, where we introduced the shorthand ${\rho_{\mathrm{x}\mathrm{x}'}:=\rho(\mathrm{x},\mathrm{x}')}$. The resulting integral then takes the form
\begin{equation}\label{eq:convolution}
\begin{aligned}
    \Omega(\rho_{\mathrm{x}\mathrm{x}''}) &= \int_{\mathrm{x}'} \Phi\left(\rho_{\mathrm{x}\mathrm{x'}}\right) \Psi\left(\rho_{\mathrm{x}'\mathrm{x}''}\right)=\int_{\mathrm{x}'} \Psi\left(\rho_{\mathrm{x}\mathrm{x'}}\right) \Phi\left(\rho_{\mathrm{x}'\mathrm{x}''}\right)
    \\
    &=\tfrac{2 \pi ^{\frac{N-1}{2}} \alpha^{N-1}}{\Gamma \left(\frac{N-1}{2}\right)}\int_0^{\infty} dr' \sinh ^{N-1}\tfrac{r'}{\alpha}\Phi(r')\int_0^{\pi} d\vartheta'_1   \sin ^{N-2}\vartheta'_1\,\Psi\left(\alpha \arccosh\left(\cosh\tfrac{\rho_{\mathrm{x}\mathrm{x}''}}{\alpha}\cosh\tfrac{r'}{\alpha}-\sinh\tfrac{\rho_{\mathrm{x}\mathrm{x}''}}{\alpha}\sinh\tfrac{r'}{\alpha}
    \cos\vartheta_1'\right)\right)\;.
\end{aligned}
\end{equation}
Although we employed specific coordinates, the result is actually coordinate independent as it only depends on the geodesic distance between points ${\mathrm{x},\mathrm{x}''\in M}$. Therefore, in order to calculate $\mathring{K}$ from \eqref{eq:KPconvolution}, we need to evaluate the 2-dimensional integral of the form \eqref{eq:convolution}. In ${N=3}$, this can be done analytically; the resulting $\mathring{K}$ is given by the closed-form expression,
\begin{equation}\label{eq:Kring3}
    \mathring{K}=\tfrac{3 \alpha}{4 \pi } \left(\coth \tfrac{\rho}{\alpha}-1\right) \left[2 e^{\frac{\alpha \rho-s}{\alpha ^2}} \erf\tfrac{\rho }{2 \sqrt{s}}- \erf\tfrac{\alpha  \rho -2 s}{2 \alpha  \sqrt{s}}+e^{\frac{2 \rho }{\alpha }} \erfc\tfrac{\alpha  \rho +2 s}{2 \alpha  \sqrt{s}}-1\right]\;, \quad N=3\;,
\end{equation}
from which we can also easily find $\bar{K}$ by means of \eqref{eq:ringdef},
\begin{equation}
    \bar{K}=\frac{3 \csch^5\tfrac{\rho}{\alpha}}{4 \pi  \alpha ^3} \left[2 e^{-\frac{s}{\alpha ^2}} \left(\cosh \tfrac{2 \rho }{\alpha }+2\right) \erf\tfrac{\rho }{2 \sqrt{s}}+3 \cosh \tfrac{\rho}{\alpha} \left(\erf\tfrac{2 s-\alpha  \rho }{2 \alpha  \sqrt{s}}-\erf\tfrac{2 s+\alpha  \rho }{2 \alpha  \sqrt{s}}\right)\right]\;, \quad N=3\;.
\end{equation}

Since \eqref{eq:GK} and \eqref{eq:GKcheck} contain the inverse Laplace transform of the scalar heat kernels $\mathsf{L}^{-1}[K(s)]$, it is useful to study these MS bi-scalar expressions in more detail. It turns out that $\mathsf{L}^{-1}[K(s)]$ are closely related to the \textit{spherical functions} in the inverse spherical transform \cite{bray1994fourier,Awonusika:2016}. Since there are no singularities in $K(s)$ for ${\Re{s}>0}$, the integration contour of $\mathsf{L}^{-1}$ can be taken arbitrarily close to the imaginary axis, ${s=i\sigma+0^+}$, ${\sigma\in\mathbb{R}}$. Hence, $\mathsf{L}^{-1}[K(s)]$ can be calculated by taking the inverse Fourier transform of the Schr\"odinger kernel, $\mathsf{F}^{-1}[K(i\sigma)]$. Assuming ${\lambda\geq\lambda_{\textrm{min}}}$, one can find the explicit expressions in an arbitrary dimension ${N\geq3}$,
\begin{equation}\label{eq:inverseofSK}
    \mathsf{L}^{-1}[K(s)]= \begin{dcases}
        \frac{(-1)^k}{2^{k+1} \pi^{k+1}} \frac{1}{\sqrt{\lambda{-}\lambda_{\textrm{min}}}}\left(\frac{1}{\alpha\sinh \tfrac{\rho}{\alpha}} \pp_{\rho}\right)^k \cos\left(  \sqrt{\lambda{-}\lambda_{\textrm{min}}}\, \rho\right)\;, & N=2k+1\;,
        \\
        \frac{(-1)^k}{2^{k+2} \pi ^{k+1}} \tanh \left(\pi   \alpha\sqrt{\lambda{-}\lambda_{\textrm{min}}}\right)\left(\frac{1}{\alpha\sinh \tfrac{\rho}{\alpha}} \pp_{\rho}\right)^k  P_{i  \alpha\sqrt{\lambda{-}\lambda_{\textrm{min}}}-\frac{1}{2}}\left(\cosh \tfrac{\rho }{\alpha}\right)\;, & N=2k+2\;,
    \end{dcases}
\end{equation}
where we used the integral representation of the Legendre function
\begin{equation}
       P_{i w-\frac{1}{2}}\left(\cosh \tfrac{\rho }{\alpha}\right)=\tfrac{\sqrt{2}}{\pi}\coth (\pi  w)\int_{\frac{\rho}{\alpha}}^{\infty}\!d u \,\frac{\sin \left( w u \right)}{\sqrt{\cosh u-\cosh \frac{\rho }{\alpha}}}\;.
\end{equation}
In ${N=3,4,5}$, we have
\begin{equation}
    \mathsf{L}^{-1}[K(s)]= \begin{dcases}
    \frac{\csch\tfrac{\rho }{\alpha} }{4 \pi ^2 \alpha}\sin \left(\sqrt{\lambda{-}\tfrac{1}{\alpha ^2}}\,\rho\right)\;, &N=3\;,
    \\
    -\frac{i \csch\tfrac{\rho }{\alpha}}{8 \pi ^2 \alpha^2} \tanh \left(\pi\alpha  \sqrt{\lambda{-}\tfrac{9}{4\alpha ^2}}\right)  P_{i \alpha\sqrt{\lambda{-}\tfrac{9}{4\alpha ^2}}-\frac{1}{2}}^1\left(\cosh \tfrac{\rho }{\alpha}\right)\;, &N=4\;,
    \\
    \frac{\csch^2\frac{\rho }{\alpha }}{8 \pi ^3 \alpha ^3}\left(\coth \tfrac{\rho }{\alpha } \sin \left(\sqrt{\lambda{-}\tfrac{4}{\alpha ^2}}\,\rho\right)-\sqrt{\lambda{-}\tfrac{4}{\alpha ^2}} \cos \left(\sqrt{\lambda{-}\tfrac{4}{\alpha ^2}}\,\rho\right)\right)\;, &N=5\;,
    \end{dcases}
\end{equation}
where
\begin{equation}
    \quad P_{i w-\frac{1}{2}}^1\left(\cosh \tfrac{\rho }{\alpha}\right)=-i\alpha\pp_\rho P_{i w-\frac{1}{2}}\left(\cosh \tfrac{\rho }{\alpha}\right)\;.
\end{equation}
Similarly, we should also discuss the inverse Laplace transforms $\mathsf{L}^{-1}[\mathring{K}(s)]$, which appears in the first line of \eqref{eq:GKring}. The MS bi-scalars $\mathsf{L}^{-1}[\mathring{K}(s)]$ are essentially bi-scalar representations of the TT bi-tensor version of the spherical functions $\mathsf{L}^{-1}[\bs{K}^{\perp}(s)]$. Assuming ${\lambda\geq\lambda_{\textrm{min}}}$, these functions can be found either by direct calculation (e.g., using the relation to $\mathsf{F}^{-1}[\mathring{K}(i\sigma)]$) or by calculating the 2-dimensional convolutional integral $\mathsf{L}^{-1}[\mathring{K}(s)]=\int_{\mathrm{x}'} \mathring{P} \mathsf{L}^{-1}[K(s)]$ with the help of \eqref{eq:convolution}, where $\mathsf{L}^{-1}[K(s)]$ is given by \eqref{eq:inverseofSK}. In ${N=3}$, the result is surprisingly simple,
\begin{equation}\label{eq:inverseofSKring3}
    \mathsf{L}^{-1}[\mathring{K}(s)]=\frac{3   \csch\tfrac{\rho }{\alpha } }{2 \pi ^2 \alpha}\frac{\sin   \left(\sqrt{\lambda -\frac{1}{\alpha ^2 }}\rho\right)}{\lambda  \left( \lambda -\frac{1}{\alpha^2}\right)} \;, \quad N=3\;.
\end{equation}

\section{Evaluation of propagators}\label{sec:evaluationofgravitonpropagators} 

Having found the expressions for the heat kernels $K$ and $\bs{K}^{\perp}$ [in the form of ${\mathring{K}=\mathring{\Pi}(\bs{K}^{\perp})}$], we have all necessary ingredients to calculate the propagators in AdS for various gravitational theories using the \textit{heat kernel approach} introduced in the previous sections. Recall that the full propagator $\bs{G}$ is described by the MS bi-scalar $G$ and the MS TT bi-tensor $\bs{G}^{\perp}$ [see \eqref{eq:gravprop}], which is completely encoded in another MS bi-scalar ${\mathring{G}=\mathring{\Pi}(\bs{G}^{\perp})}$. After $\mathring{G}$ is known, finding $\bar{G}$ as well as all components of $\bs{G}^{\perp}$ [in the bi-tensor frame \eqref{eq:bidecomposition}] is just a matter of straightforward application of \eqref{eq:barexplicit} and \eqref{eq:ringdef}, which only involves differentiation and algebraic operations.\footnote{Alternatively, it is also possible to calculate $\bar{G}$ directly using \eqref{eq:GKbar}, however, the expressions for $\mathring{G}$ are typically more compact than for $\bar{G}$ and relevant integrals are simpler.} Since the evaluation of $\mathring{G}$ is complicated by the presence of extra convolution with $\mathring{P}$, it is not always possible to find a closed-form expression. In such cases, we present the results in the form of $\check{G}$, which is then related to $\mathring{G}$ by the 2-dimensional convolutional integral \eqref{eq:Gcheckconv} of the form \eqref{eq:convolution}.

\subsection{General relativity} \label{ssc:GR}

As a consistency check of our heat kernel approach, we first derive the propagator for GR and compare it to the previous results in the literature, namely \cite{DHoker:1999bve}. Recall that GR corresponds to the choice $\mathcal{F}_i(\Box)=0$, i.e., ${\mathcal{E}(\Box)=\mathcal{I}(\Box)=1}$. The evaluation of $G_{\textrm{GR}}$ is relatively simple in arbitrary dimension $N\geq3$ if we use the second formula in \eqref{eq:GK}. Using
\begin{equation}
    \mathsf{L}^{-1}\left[\left(\lambda+\tfrac{N}{\alpha^2}\right)^{-1}\right](s) =e^{-\frac{N s}{\alpha^2}}\;,
\end{equation}
and the heat kernels $K$ from \eqref{eq:scalarK}, we may find
\begin{equation}
    G_{\textrm{GR}} =
    \begin{dcases}
    \frac{ (-1)^{k}  }{2^{k+1}\pi^k}\left(\frac{1}{\alpha\sinh \tfrac{\rho}{\alpha}} \pp_{\rho}\right)^k\frac{e^{-\frac{(k+1) \rho }{\alpha}}}{\frac{k+1}{\alpha}}\;, & N=2k+1\;,
    \\
    \frac{(-1)^k }{2^{k+\frac{3}{2}} \pi ^{k+1}}\left(\frac{1}{\alpha\sinh \tfrac{\rho}{\alpha}} \pp_{\rho}\right)^k \int_{\frac{\rho}{\alpha}}^{\infty}\!d u \,\frac{ e^{-\left(k+\frac{3}{2}\right) u}}{\sqrt{\cosh u-\cosh \tfrac{\rho }{\alpha}}}\;, & N=2k+2\;.
    \end{dcases}
\end{equation}
Specifically, in ${N=3,4,5}$, after some differentiation and integration, one can arrive at the closed-form expressions
\begin{equation}
     G_{\textrm{GR}} =
     \begin{dcases}
    \frac{e^{-\frac{2 \rho }{\alpha}} \csch\frac{\rho }{\alpha}}{4 \pi  \alpha}\;, & N=3\;,
    \\
    \frac{\csch^2\frac{\rho }{\alpha}-6 \cosh \frac{\rho }{\alpha} \coth ^{-1}\left(e^{\frac{\rho }{\alpha }}\right)+3}{4 \pi ^2 \alpha^2}\;, & N=4\;,
    \\
    \frac{e^{-\frac{3 \rho }{\alpha }} \left(\coth \frac{\rho }{\alpha }+3\right) \csch^2\frac{\rho }{\alpha }}{8 \pi ^2 \alpha ^3}\;, & N=5\;.
    \end{dcases}
\end{equation}
These expressions match with the general formula obtained in \cite{DHoker:1999bve},\footnote{The formulas in \cite{DHoker:1999bve} are expressed using the \textit{chordal distance}, ${x=\cosh ^2\left({\rho }/{(2 \alpha)}\right)}$, and the authors set ${\alpha=1}$.}
\begin{equation}
    G_{\textrm{GR}} =\frac{ \Gamma \left(\frac{N}{2}+1\right) \Gamma (N)  }{2^{N} \pi ^{\frac{N}{2}} \alpha ^{N-2}\Gamma (N+2)\cosh ^{2N}\frac{\rho }{2 \alpha }}\, _2F_1\left(\tfrac{N}{2}+1,N;N+2;\sech^2\tfrac{\rho }{2 \alpha }\right)\;, \quad N\geq3\;.
\end{equation}

The evaluation of $\mathring{G}_{\textrm{GR}}$ is more difficult due to the convolution with $\mathring{P}$ which enters the calculation either through \eqref{eq:KPconvolution} or through \eqref{eq:Gcheckconv}. First, we realize that 
\begin{equation}\label{eq:GRlaplaceE}
    \mathsf{L}^{-1}\left[\lambda^{-1}\right](s)=1\;,
\end{equation}
meaning that $\mathring{G}_{\textrm{GR}}$ is just an integral of $\mathring{K}$ due to the second line of \eqref{eq:GKring}. Furthermore, in ${N=3}$, we know the closed-form expressions for $\mathring{K}$, see \eqref{eq:Kring3}. Interestingly, the resulting $\mathring{G}_{\textrm{GR}}$ can be also written in a closed form,
\begin{equation}\label{eq:Gring3}
    \mathring{G}_{\textrm{GR}} =\frac{3 \alpha ^3}{\pi  \big(e^{\frac{\rho}{\alpha} }+1\big) }-\frac{3 \alpha ^2 \rho  \left(\coth \tfrac{\rho}{\alpha}-1\right)}{4 \pi } \;, \quad N=3\;.
\end{equation}
Upon calculating  ${\bar{G}_{\textrm{GR}}=\Upsilon(\mathring{G}_{\textrm{GR}})}$, one can check that this result matches perfectly the expression from \cite{DHoker:1999bve}. Now, since $\mathring{K}$ is not known in the closed-form in ${N\geq4}$, it is more practical to express $\mathring{G}_{\textrm{GR}}$ using the auxiliary bi-scalar ${\check{G}_{\textrm{GR}}}$. In arbitrary ${N\geq3}$, the second line of \eqref{eq:GKcheck} together with \eqref{eq:GRlaplaceE} implies that ${\check{G}_{\textrm{GR}}}$ are directly integrals of heat kernels $K$ from \eqref{eq:scalarK}. Hence, we obtain
\begin{equation}
    \check{G}_{\textrm{GR}} =
    \begin{dcases}
    \frac{ (-1)^{k}  }{2^{k+1}\pi^k }\left(\frac{1}{\alpha\sinh \tfrac{\rho}{\alpha}} \pp_{\rho}\right)^k\frac{e^{-\frac{k \rho }{\alpha}}}{\frac{k}{\alpha}}\;, & N=2k+1\;,
    \\
    \frac{(-1)^k }{2^{k+\frac{3}{2}} \pi ^{k+1}}\left(\frac{1}{\alpha\sinh \tfrac{\rho}{\alpha}} \pp_{\rho}\right)^k \int_{\frac{\rho}{\alpha}}^{\infty}\!d u \,\frac{ e^{-\left(k+\frac{1}{2}\right) u}}{\sqrt{\cosh u-\cosh \tfrac{\rho }{\alpha}}}\;, & N=2k+2\;.
    \end{dcases}
\end{equation}
In particular, in $N=3,4,5$, we get
\begin{equation}
     \check{G}_{\textrm{GR}} =
     \begin{dcases}
    \frac{1}{2 \pi  \alpha  \big(e^{\frac{2 \rho }{\alpha }}-1\big) }\;, & N=3\;,
    \\
    \frac{\log \tanh \frac{\rho }{2 \alpha }+\coth \frac{\rho }{\alpha } \csch\frac{\rho }{\alpha }}{4 \pi ^2 \alpha ^2}\;, & N=4\;,
    \\
    \frac{\coth ^3\frac{\rho }{\alpha }-3 \coth \frac{\rho }{\alpha }+2}{8 \pi ^2 \alpha ^3}\;, & N=5\;.
    \end{dcases}
\end{equation}
In a general dimension $N\geq3$, $\mathring{G}_{\textrm{GR}}$ can be obtained by evaluating the 2-dimensional convolution integral \eqref{eq:Gcheckconv}. In ${N=3}$, one recovers our previous closed-form expression \eqref{eq:Gring3}. In ${N=4}$, we calculated the integral numerically for various $\rho$ and $\alpha$. Our numerical results perfectly match the following expression
\begin{equation}
\begin{aligned}
    \mathring{G}_{\textrm{GR}} = &-\tfrac{5 \alpha^2 }{108 \pi ^2}\big[12 \Li_2\left(\cosh ^2\tfrac{\rho}{2\alpha}\right)+9 \sech^2\tfrac{\rho}{2\alpha}+6 \left(\csch^2\tfrac{\rho}{2\alpha}+4 i \pi +1\right) \log \left(\cosh \tfrac{\rho}{2\alpha}\right)
    \\
    &+6 \log \left(\sinh \tfrac{\rho}{2\alpha}\right) \left(\sech^2\tfrac{\rho}{2\alpha}+4 \log \left(\cosh \tfrac{\rho}{2\alpha}\right)-1\right)-4 \pi ^2\big]\;, &N=4\;,
\end{aligned}
\end{equation}
which we obtained by taking the $\bar{G}_{\textrm{GR}}$ from \cite{DHoker:1999bve} and solving ${\bar{G}_{\textrm{GR}}=\Upsilon(\mathring{G}_{\textrm{GR}})}$ for $\mathring{G}_{\textrm{GR}}$.

From the closed-form expressions for ${N=3}$, one can notice the divergent behavior of both parts of the propagator for ${\rho\to0}$,
\begin{equation}
    G_{\textrm{GR}}\approx \frac{1}{4 \pi  \rho }\;, \quad \bar{G}_{\textrm{GR}}\approx \frac{3}{16 \pi  \rho }\;, \quad {\rho\to0}\;, \quad N=3\;,
\end{equation}
and their exponential suppression for ${\rho\to\infty}$,
\begin{equation}
    G_{\textrm{GR}} \approx \frac{1}{2 \pi \alpha} e^{-\frac{3\rho}{\alpha}}\;, \quad \bar{G}_{\textrm{GR}} \approx \frac{24}{\pi \alpha} e^{-\frac{3\rho}{\alpha}}\;, \quad \rho\to\infty\;, \quad N=3\;.
\end{equation}

\subsection{Higher-derivative gravity} \label{ssc:HDG}

The real power of the heat kernel approach comes with its application to modified theories of gravity. Let us first consider gravitational theories with higher derivatives of finite orders, i.e. the HDG. Instead of solving the complicated higher-derivative equations directly, we only have to find the inverse Laplace transforms of the inverse of the kinetic operators and then evaluate the relevant integrals, which is essentially quite similar to the derivation we just did for GR. To demonstrate this explicitly, let us assume that $\mathcal{E}(\Box)$ and $\mathcal{I}(\Box)$ are polynomials of degrees $m_1$ and $m_2$, respectively. We can write them in the form
\begin{equation}\label{eq:EIHDG}
\begin{aligned}
    \mathcal{E}(\Box) &=\mathcal{E}\left(-\tfrac{2}{\alpha^2}\right)\prod_{n=1}^{m_1}\left(1-\frac{\Box+\frac{2}{\alpha^2}}{\mu_n^2}\right)\;,
    \\
    \mathcal{I}(\Box) &=\mathcal{I}\left(0\right)\prod_{n=1}^{m_2}\left(1-\frac{\Box}{\nu_n^2}\right)\;.
\end{aligned}
\end{equation}
Here, $\mu_n$ and $\nu_n$ correspond to the \textit{masses} of the additional degrees of freedom \cite{Tekin:2016vli}, some of which are ghosts.\footnote{The tachyonic instabilities can be avoided if the mass squares $\mu_n^2$ and $\nu_n^2$ satisfy the Breitenlohner--Freedman bound \cite{Lu:2011qx}, $\mu_n^2,\nu_n^2> -{(N-1)^2}/{(4\alpha^2)}$, as can be easily deduced from the spectrum \eqref{eq:spectralresol}.} Remark that the constants $\mathcal{E}(-{2}/{\alpha^2})$, $\mathcal{I}\left(0\right)$, $\mu_1,\dots,\mu_{m_1}$, $\nu_1,\dots,\nu_{m_2}$ are not mutually independent. They have to satisfy the $\mathcal{F}_i$-analyticity conditions \eqref{eq:EIcond},
\begin{equation}\label{eq:HDGanacond}
\begin{aligned}
    \mathcal{E}\left(-\tfrac{2}{\alpha^2}\right) &=1-\tfrac{1-X}{1+X}\;, & \mathcal{I}(0) &=1+\tfrac{1-X}{1+X}\;, & N&=3\;, 
    \\
    \mathcal{E}\left(-\tfrac{2}{\alpha^2}\right) &=X\;, & \mathcal{I}(0) &=1\;, & N&=4\;,
    \\
    \mathcal{E}\left(-\tfrac{2}{\alpha^2}\right) &=\mathcal{I}\left(0\right) X\;, \quad & & & N &\geq5\;,
\end{aligned}
\end{equation}
where we denoted
\begin{equation}
    X :=\frac{ \prod\limits _{n=1}^{m_2} \left(1-\frac{N}{\alpha ^2 \nu_n^2}\right)}{\prod\limits _{n=1}^{m_1} \left(1+\frac{N-2}{\alpha ^2\mu_n^2}\right)}\;.
\end{equation}
We will also assume that the zeros of $\mathcal{E}$ and $\mathcal{I}$ are of multiplicity $1$, i.e., $\mu_{n}^2\neq\mu_{\tilde{n}}^2\neq0$ and $\nu_{n}^2\neq\nu_{\tilde{n}}^2\neq N/\alpha^2$. In other words, we focus on non-degenerated cases. 

Again, we start with the calculation of $G_{\textrm{HDG}}$ and proceed exactly like for GR. The inverse Laplace transforms in the second line of \eqref{eq:GK} for the polynomial $\mathcal{I}$ can be easily calculated using the \textit{Heaviside expansion theorem} (for simple poles) \cite{Poularikas2010-zl},
\begin{equation}\label{eq:Heaviside}
    \mathsf{L}^{-1}\left[\prod_{i=0}^{n}\frac{1}{\lambda-c_i}\right](s)=\sum_{i=0}^n \frac{e^{c_i s}}{\prod_{\substack{j=0\\ j\neq i}}^n(c_i-c_j)}\;.
\end{equation}
In particular, if we consider \eqref{eq:EIHDG}, we find
\begin{equation}
\begin{gathered}
    \mathsf{L}^{-1}\left[\left({\left(\lambda+\tfrac{N}{\alpha^2}\right)\prod_{n=1}^{m_2}\left(1+\frac{\lambda}{\nu_n^2}\right)}\right)^{-1}\right](s) =V_0 e^{-\frac{N}{\alpha^2} s}+\sum_{n=1}^{m_2} V_n e^{-\nu_n^2 s}\;, 
    \\
    V_0:=\frac{\prod_{j=1}^{m_2}\nu_j^2}{\prod_{j=1}^{m_2}\left(\nu_j^2-\frac{N}{\alpha^2}\right)}\;, \quad V_n:=\frac{\prod_{j=1}^{m_2}\nu_j^2}{\left(\frac{N}{\alpha^2}-\nu_n^2\right)\prod_{\substack{j=1\\j\neq n}}^{m_2}(\nu_j^2-\nu_n^2)}\;.
\end{gathered}
\end{equation}
Inserting this expression together with the formulas for the heat kernels $K$ from \eqref{eq:scalarK} into the second line of \eqref{eq:GK}, we arrive at
\begin{equation}
    G_{\textrm{HDG}} =
    \begin{dcases}
    \frac{(-1)^{k}}{2^{k+1}\pi^k\mathcal{I}\left(0\right)}\left(\frac{1}{\alpha\sinh \tfrac{\rho}{\alpha}} \pp_{\rho}\right)^k \left[ \frac{V_0 e^{-\tfrac{(k+1)\rho}{\alpha} }}{\tfrac{k+1}{\alpha}}+\sum_{n=1}^{m_2} \frac{V_n e^{-\sqrt{\nu_n^2+\tfrac{k^2}{\alpha^2}}\, \rho }}{\sqrt{\nu_n^2+\tfrac{k^2}{\alpha^2}}}\right]\;, & N=2k+1\;,
    \\
    \frac{(-1)^k}{2^{k+\frac{3}{2}} \pi ^{k+1}\mathcal{I}\left(0\right)} \left(\frac{1}{\alpha\sinh \tfrac{\rho}{\alpha}} \pp_{\rho}\right)^k \int_{\frac{\rho}{\alpha}}^{\infty}\!d u \,\frac{V_0 e^{-\left(k+\frac32\right) u}+\sum\limits_{n=1}^{m_2}V_n e^{-\alpha\sqrt{\nu_n ^2+\frac{\left(k+\frac12\right)^2}{\alpha^2}}\, u}}{\sqrt{\cosh u-\cosh \tfrac{\rho }{\alpha}}}\;, & N=2k+2\;.
    \end{dcases}
\end{equation}
For example, in $N=3,4,5$, the above expression reduces to
\begin{equation}\label{eq:GHDG345}
    G_{\textrm{HDG}} =
    \begin{dcases}
    \frac{1}{4 \pi  \alpha \mathcal{I}\left(0\right) }\csch\tfrac{\rho }{\alpha}\left[V_0 e^{-\tfrac{2\rho}{\alpha}}+\sum_{n=1}^{m_2}V_n e^{-\sqrt{\nu_n^2+\tfrac{1}{\alpha^2}}\, \rho}\right]\;, & N=3\;,
    \\
    \frac{-1}{2^{\frac52} \pi^2 \mathcal{I}\left(0\right)} \left(\frac{1}{\alpha\sinh \tfrac{\rho}{\alpha}} \pp_{\rho}\right) \int_{\frac{\rho}{\alpha}}^{\infty}\!d u \,\frac{V_0 e^{-\frac52 u} + \sum\limits_{n=1}^{m_2}V_n e^{-\alpha\sqrt{\nu_n ^2+\frac{9}{4\alpha^2}}\, u}}{\sqrt{\cosh u-\cosh \tfrac{\rho }{\alpha}}}\;, & N=4\;,
    \\
    \frac{1}{8\pi^2\alpha^3\mathcal{I}\left(0\right)} \csch^2\tfrac{\rho}{\alpha} \left[ V_0 (\coth\tfrac{\rho}{\alpha} + 3) e^{-\tfrac{3\rho}{\alpha}} + \sum_{n=1}^{m_2} V_n \left(\coth\tfrac{\rho}{\alpha} + \alpha \sqrt{\nu_n^2+\tfrac{4}{\alpha^2}} \right) e^{-\sqrt{\nu_n^2+\tfrac{4}{\alpha^2}}\,\rho} \right] \;, & N=5\;.
    \end{dcases}
\end{equation}

The calculation of $\mathring{G}_{\textrm{HDG}}$ is also similar to the GR case. Using the Heaviside expansion theorem \eqref{eq:Heaviside}, we can find 
\begin{equation}\label{eq:LaplaceHDGE}
\begin{gathered}
    \mathsf{L}^{-1}\left[\left({\lambda\prod_{n=1}^{m_1}\left(1+\frac{\lambda}{\mu_n^2}\right)}\right)^{-1}\right](s) = 1-\sum_{n=1}^{m_1} U_n e^{-\mu_n^2 s}\;, 
    \\
    U_n:=\frac{\prod_{\substack{j=1\\j\neq n}}^{m_1}\mu_j^2}{\prod_{\substack{j=1\\j\neq n}}^{m_1}(\mu_j^2-\mu_n^2)}\;.
\end{gathered}
\end{equation}
As before, in ${N=3}$, we can make use of the closed-form expression for $\mathring{K}$ from \eqref{eq:Kring3} to evaluate the integral in the second line of \eqref{eq:GKring}. Also, the result can still be written in the closed form,
\begin{equation}\label{eq:GHDGring3}
    \mathring{G}_{\textrm{HDG}} =
     \frac{3\alpha^3}{2\pi\mathcal{E}\left(-\tfrac{2}{\alpha^2}\right)}\left(\coth \tfrac{\rho}{\alpha}-1\right)\left[ e^{\tfrac{\rho}{\alpha} }{-}1-\tfrac{\rho }{2\alpha}+\tfrac{1}{\alpha^2}\sum_{n=1}^{m_1}\frac{  U_n  }{    \mu_n ^2+ \tfrac{1}{\alpha^2}}\left(1- e^{\tfrac{\rho}{\alpha}}+\tfrac{1{-}e^{\frac{\rho}{\alpha}-\sqrt{\mu_n^2+\frac{1}{\alpha^2}}\rho}}{\alpha^2\mu_n ^2}\right) \right]\;, \quad N=3\;.
\end{equation}
In general $N\geq3$, we can still calculate the auxiliary bi-scalar $\check{G}_{\textrm{HDG}}$ using the formula at the second line of \eqref{eq:GKcheck} together with \eqref{eq:LaplaceHDGE} and the heat kernels $K$ from \eqref{eq:scalarK},
\begin{equation}
    \check{G}_{\textrm{HDG}} =
    \begin{dcases}
    \frac{(-1)^{k}}{2^{k+1}\pi^k\mathcal{E}\left(-\tfrac{2}{\alpha^2}\right)}\left(\frac{1}{\alpha\sinh \tfrac{\rho}{\alpha}} \pp_{\rho}\right)^k \left[ \frac{ e^{-\tfrac{k\rho}{\alpha} }}{\tfrac{k}{\alpha}}-\sum_{n=1}^{m_1} \frac{U_n e^{-\sqrt{\mu_n^2+\tfrac{k^2}{\alpha^2}}\, \rho }}{\sqrt{\mu_n^2+\tfrac{k^2}{\alpha^2}}}\right]\;, & N=2k+1\;,
    \\
    \frac{(-1)^k}{2^{k+\frac{3}{2}} \pi ^{k+1}\mathcal{E}\left(-\tfrac{2}{\alpha^2}\right)} \left(\frac{1}{\alpha\sinh \tfrac{\rho}{\alpha}} \pp_{\rho}\right)^k \int_{\frac{\rho}{\alpha}}^{\infty}\!d u \,\frac{ e^{-\left(k+\frac12\right) u}-\sum\limits_{n=1}^{m_1}U_n e^{-\alpha\sqrt{\mu_n ^2+\frac{\left(k+\frac12\right)^2}{\alpha^2}}\, u}}{\sqrt{\cosh u-\cosh \tfrac{\rho }{\alpha}}}\;, & N=2k+2\;.
    \end{dcases}
\end{equation}
In $N=3,4,5$, we get
\begin{equation}
    \check{G}_{\textrm{HDG}} =
    \begin{dcases}
    \frac{1}{4\pi\alpha\mathcal{E}\left(-\tfrac{2}{\alpha^2}\right)} \csch\tfrac{\rho}{\alpha} \left[ e^{-\tfrac{\rho}{\alpha} } - \sum_{n=1}^{m_1} U_n e^{-\sqrt{\mu_n^2+\tfrac{1}{\alpha^2}}\, \rho } \right]\;, & N=3\;,
    \\
    \frac{-1}{2^{\frac52} \pi^2 \mathcal{E}\left(-\tfrac{2}{\alpha^2}\right)} \left(\frac{1}{\alpha\sinh \tfrac{\rho}{\alpha}} \pp_{\rho}\right) \int_{\frac{\rho}{\alpha}}^{\infty}\!d u \,\frac{ e^{-\frac32 u}-\sum\limits_{n=1}^{m_1}U_n e^{-\alpha\sqrt{\mu_n ^2+\frac{9}{4\alpha^2}}\, u}}{\sqrt{\cosh u-\cosh \tfrac{\rho }{\alpha}}}\;, & N=4\;,
    \\
    \frac{1}{8\pi^2\alpha^3\mathcal{E}\left(-\tfrac{2}{\alpha^2}\right)} \csch^2\tfrac{\rho}{\alpha} \left[ (\coth\tfrac{\rho}{\alpha} + 2) e^{-\tfrac{2\rho}{\alpha} } - \sum_{n=1}^{m_1} U_n \left(\coth\tfrac{\rho}{\alpha} + \alpha \sqrt{\mu_n^2+\tfrac{4}{\alpha^2}} \right) e^{-\sqrt{\mu_n^2+\tfrac{4}{\alpha^2}}\, \rho } \right]\;, & N=5\;.
    \end{dcases}
\end{equation}
Naturally, $\mathring{G}_{\textrm{HDG}}$ is still obtainable (at least numerically) by the 2-dimensional convolutional integral \eqref{eq:Gcheckconv} in arbitrary $N\geq3$. In $N=3$, this leads to the closed-form expression \eqref{eq:GHDGring3}.

By investigating the closed-form expressions for ${N=3}$, we can observe that both parts of the propagator are regular for ${\rho\to0}$,
\begin{equation}
\begin{aligned}
    G_{\textrm{HDG}} &\approx \frac{-1}{4\pi\alpha\mathcal{I}(0)} \left[ 2 V_0 + \sum_{n=1}^{m_2} V_n \sqrt{1 + \alpha^2 \nu_n^2} \right] \;, \\
    \bar{G}_{\textrm{HDG}} &\approx \frac{1}{10\pi\alpha\mathcal{E}\left(-\tfrac{2}{\alpha^2}\right)} \left[ 3 + \sum_{n=1}^{m_1}U_n \sqrt{1+\alpha^2\mu_n^2} \left( 1 -\frac{4}{1+ \alpha^2\mu_n^2} \right) \right] \;, \quad {\rho\to0}\;, \quad N=3\;,
\end{aligned}
\end{equation}
but they still decay exponentially for ${\rho\to\infty}$, 
\begin{equation}
\begin{aligned}
    G_{\textrm{HDG}} &\approx \frac{1}{2 \pi \alpha\mathcal{I}(0)} \left[V_0 e^{-\frac{3\rho}{\alpha}} + {\sum_{n=1}^{m_2} V_n} e^{-\frac{\rho}{\alpha}(1+\sqrt{1 + \alpha^2 \nu_n^2})}\right]\;, 
    \\
    \bar{G}_{\textrm{HDG}} &\approx \frac{24}{\pi \alpha \mathcal{E}\left(-\tfrac{2}{\alpha^2}\right)} \left[1 - \sum_{n=1}^{m_1} \frac{U_n}{1 + \alpha^2 \mu_n^2}\right]e^{-3\frac{\rho}{\alpha}}\;, \quad\rho\to\infty\;, \quad N=3\;.
\end{aligned}
\end{equation}
Nevertheless, the overall factor as well as the damping constant (for $G_{\textrm{HDG}}$) may be different, when compared to GR, as it depends on the constants of the theory.

A simple example of HDG is the quadratic gravity for which $\mathcal{F}_i(\Box)$ are given just by constants $\mathcal{F}_i(0)$, hence $\mathcal{E}(\Box)$ and $\mathcal{I}(\Box)$ are just linear functions, i.e., ${m_1=m_2=1}$,
\begin{equation}\label{eq:EIQCG}
\begin{aligned}
    \mathcal{E}(\Box) &=\mathcal{E}\left(-\tfrac{2}{\alpha^2}\right)\left(1-\frac{\Box+\frac{2}{\alpha^2}}{\mu^2}\right)\;,     
    \\
    \mathcal{I}(\Box) &=\mathcal{I}\left(0\right)\left(1-\frac{\Box}{\nu^2}\right)\;,  
\end{aligned}
\end{equation}
where we denoted $\mu:=\mu_1$ and $\nu:=\nu_1$. The constants in \eqref{eq:EIQCG} can be expressed in terms of the independent coupling constants $\mathcal{F}_i(0)$,
\begin{equation}
\begin{aligned}
    \mathcal{E}\left(-\tfrac{2}{\alpha^2}\right) &=1-\tfrac{12 \varkappa}{\alpha^2} \mathcal{F}_1(0)\;, &  \mu^2 &= -\tfrac{1-\frac{12 \varkappa}{\alpha^2} \mathcal{F}_1(0)}{\varkappa \mathcal{F}_2(0)}\;,
    \\
    \mathcal{I}(0) &=1+\tfrac{12 \varkappa}{\alpha^2} \mathcal{F}_1(0)\;,  & \nu^2 &= \tfrac{1+\frac{12 \varkappa}{\alpha^2} \mathcal{F}_1(0)}{\varkappa[8\mathcal{F}_1(0)+\frac13\mathcal{F}_2(0)]}\;, & N&=3\;,
    \\
    \mathcal{E}\left(-\tfrac{2}{\alpha^2}\right) &=1+\tfrac{4 \varkappa}{\alpha^2} \left[-6\mathcal{F}_1(0)+\mathcal{F}_3(0)\right]\;, & \mu^2 &= -\tfrac{1+\frac{4 \varkappa}{\alpha ^2}  \left[\mathcal{F}_3(0)-6 \mathcal{F}_1(0)\right]}{2 \varkappa\mathcal{F}_3(0)  }\;,
    \\
    \mathcal{I}(0) &=1\;, & \nu^2 &= \tfrac{1}{6\varkappa\mathcal{F}_1(0)}\;, & N &=4\;,
        \\
    \mathcal{E}\left(-\tfrac{2}{\alpha^2}\right) &=1+ \tfrac{2\varkappa}{\alpha^2}\left(-N(N{-}1) \mathcal{F}_1(0)   +2    (N{-}3)\mathcal{F}_3(0)\right)\;, & \mu^2 &= -\tfrac{1- \frac{2   (N-1) N\varkappa}{\alpha ^2}\mathcal{F}_1(0)+\frac{4(N-3)\varkappa}{\alpha^2}\mathcal{F}_3(0)}{ \varkappa  \left[\mathcal{F}_2(0)+\frac{4  (N-3)}{N-2}\mathcal{F}_3(0)\right]}\;,
    \\
    \mathcal{I}(0) &=1-\tfrac{2  (N-4) (N-1) N \varkappa }{\alpha ^2 (N-2)}\mathcal{F}_1(0)\;, & \nu^2 &= \tfrac{  1-2  \frac{\varkappa (N-4) (N-1) N}{(N-2)\alpha ^2} \mathcal{F}_1(0)}{\alpha ^2 \varkappa  \left[\frac{4  (N-1) }{N-2}\mathcal{F}_1(0)+ \frac{N-2}{N}\mathcal{F}_2(0)\right]}\;, & N &\geq5\;.
\end{aligned}
\end{equation}
Notice that $\mathcal{F}_i$-analyticity conditions \eqref{eq:HDGanacond} are automatically satisfied, cf. \eqref{eq:EIcond}. Of course, the expressions for the propagators are just special cases of the expressions for HDG where we only keep the first term, ${n=1}$.

\subsection{Infinite-derivative gravity} \label{ssc:IDG}

The heat kernel approach also allows us to find propagators for nonlocal gravitational theories with derivatives of an infinite order, i.e., the IDG. These theories are typically defined by the condition that $\mathcal{E}(\Box)$ and $\mathcal{I}(\Box)$ are exponentials of some entire functions, which guarantees that they possess only the usual GR degrees of freedom and avoid the unphysical ghosts. Various choices of entire functions have been proposed in the literature in the flat background ($\alpha\to\infty$) \cite{Krasnikov1987,Tomboulis:1997gg,Biswas:2011ar,Modesto:2011kw,Frolov:2015bia,Buoninfante:2018lnh,Buoninfante:2020ctr} as well as in the constant curvature backgrounds \cite{Biswas:2016egy,Biswas:2016etb,Mazumdar:2018xjz,Koshelev:2018}. 

For simplicity, we will focus on the polynomial entire functions and write the nonlocal operators $\mathcal{E}(\Box)$ and $\mathcal{I}(\Box)$ in the form
\begin{equation}\label{eq:nonlocalEI}
    \begin{aligned}\mathcal{E}(\Box) &=\mathcal{E}\left(-\tfrac{2}{\alpha^2}\right)\exp\left[-\ell^{2(q_1+1)}\left(\Box+\tfrac{2}{\alpha^2}\right)\prod_{n=1}^{q_1}\left(-\left(\Box+\tfrac{2}{\alpha^2}\right)+\tfrac{a_n}{\alpha^2}\right)\right]\;,
    \\
    \mathcal{I}(\Box) &=\mathcal{I}\left(0\right)\exp\left[-\jmath^{2(q_2+1)}\Box\prod_{n=1}^{q_2}\left(-\Box+\tfrac{b_n}{\alpha^2}\right)\right]\;.
    \end{aligned}
\end{equation}
Here, the conventional negative sign of the highest power of $\Box$ in \eqref{eq:nonlocalEI} indicates that one is typically interested in choices with the exponential suppression at short geodesic distances (i.e., large values of $\lambda$). The constants $\mathcal{E}\left(-\tfrac{2}{\alpha^2}\right)$, $\mathcal{I}\left(0\right)$, $\ell$, $\jmath$, $a_2,\dots,a_{q_1}$, $b_2,\dots,b_{q_2}$ are again constrained by the $\mathcal{F}_i$-analyticity conditions \eqref{eq:EIcond},
\begin{equation}
\begin{aligned}
    \mathcal{E}\left(-\tfrac{2}{\alpha^2}\right) &=1-\tanh \tfrac{Y}{2}\;, & \mathcal{I}(0) &=1+\tanh\tfrac{Y}{2}\;, & N&=3\;, 
    \\
    \mathcal{E}\left(-\tfrac{2}{\alpha^2}\right) &=\exp (-Y)\;, & \mathcal{I}(0) &=1\;, & N&=4\;,
    \\
    \mathcal{E}\left(-\tfrac{2}{\alpha^2}\right) &=\mathcal{I}\left(0\right)\exp (-Y)\;, \quad & & & N &\geq5\;,
\end{aligned}
\end{equation}
where we denoted
\begin{equation}
    Y :=(N-2)\left(\tfrac{\ell}{\alpha}\right)^{2(q_1+1)}\textstyle\prod\limits _{n=1}^{q_1} \left(a_n+N-2\right)+N \left(\tfrac{\jmath}{\alpha}\right)^{2(q_2+1)} \textstyle\prod\limits _{n=1}^{q_2} \left(b_n-N\right)\;.
\end{equation}
Since the calculations become rather difficult for higher-order polynomials in the exponential, we will discuss only linear and quadratic entire functions, i.e., ${q_i=0}$  and ${q_i=1}$, respectively.

Let us first discuss the evaluation of $G_{\textrm{IDG}}$. If ${q_2=0}$, we can proceed relatively straightforwardly because
\begin{equation}
    \mathsf{L}^{-1}\left[\left(\left(\lambda+\tfrac{N}{\alpha^2}\right) e^{\jmath^2\lambda}\right)^{-1}\right](s)=e^{-\frac{N }{\alpha ^2}\left(s-\jmath^2\right)}\theta \left(s-\jmath^2\right) \;.
\end{equation}
Upon inserting this formula and the heat kernels $K$ from \eqref{eq:scalarK} into the integral in the second line of \eqref{eq:GK}, we find
\begin{equation}
    G_{\textrm{IDG}}^{0} =
    \begin{dcases}
    \frac{(-1)^{k}}{2^{k+1}\pi^k}\frac{e^{\frac{ (2 k+1)\jmath^2 }{\alpha ^2}}}{2\mathcal{I}\left(0\right)}\left(\frac{1}{\alpha\sinh \tfrac{\rho}{\alpha}} \pp_{\rho}\right)^k \left[\tfrac{e^{\frac{(k+1) \rho }{\alpha }}}{\frac{k+1}{\alpha }} \erfc\left(\tfrac{\rho }{2 \jmath}{+}\tfrac{ (k+1)\jmath}{\alpha }\right){+}\tfrac{e^{-\frac{(k+1) \rho }{\alpha }}}{\frac{k+1}{\alpha }}\erfc\left(\tfrac{ (k+1)\jmath}{\alpha }{-}\tfrac{\rho }{2 \jmath}\right)\right]\;, & N=2k+1\;,
    \\
    \frac{(-1)^k}{2^{k+\frac{3}{2}} \pi ^{k+1}}\frac{e^{\frac{ 2 (k+1)\jmath^2 }{\alpha ^2}}}{2\mathcal{I}\left(0\right)} \left(\frac{1}{\alpha\sinh \tfrac{\rho}{\alpha}} \pp_{\rho}\right)^k \!\!\int_{\frac{\rho}{\alpha}}^{\infty}\!\!\!d u \,\tfrac{e^{-\left(k+\frac{3}{2}\right) u} \erfc\left(\left(k+\frac{3}{2}\right)\frac{\jmath}{\alpha }-\frac{\alpha  u}{2\jmath }\right)-e^{\left(k+\frac{3}{2}\right) u} \erfc\left(\left(k+\frac{3}{2}\right)\frac{\jmath}{\alpha }+\frac{\alpha  u}{2 \jmath}\right)}{\sqrt{\cosh u-\cosh \tfrac{\rho }{\alpha}}}\;, & N=2k+2\;,
    \end{dcases}
\end{equation}
where the superscript $0$ stands for $q_2=0$. For example, in $N=3,4,5$, we get
\begin{equation}
    G_{\textrm{IDG}}^{0} =
    \begin{dcases}
    \frac{e^{\frac{3 \jmath^2}{\alpha ^2}}\csch\tfrac{\rho }{\alpha }}{8 \pi  \alpha  \mathcal{I}(0)}  \left[e^{-\frac{2 \rho }{\alpha }}\erfc\left(\tfrac{2 \jmath}{\alpha }-\tfrac{\rho }{2 \jmath}\right)-e^{\frac{2 \rho }{\alpha }} \erfc\left(\tfrac{2 \jmath}{\alpha }+\tfrac{\rho }{2 \jmath}\right)\right]\;, & N=3\;,
    \\
    -\frac{e^{\frac{ 4\jmath^2 }{\alpha ^2}}}{2^{\frac{7}{2}} \pi^2 \mathcal{I}\left(0\right)} \left(\frac{1}{\alpha\sinh \tfrac{\rho}{\alpha}} \pp_{\rho}\right) \!\!\int_{\frac{\rho}{\alpha}}^{\infty}\!\!\!d u \,\tfrac{e^{-\frac{5}{2} u} \erfc\left(\frac{5\jmath}{2\alpha }-\frac{\alpha  u}{2\jmath }\right)-e^{\frac{5}{2} u} \erfc\left(\frac{5\jmath}{2\alpha }+\frac{\alpha  u}{2 \jmath}\right)}{\sqrt{\cosh u-\cosh \tfrac{\rho }{\alpha}}}\;, & N=4\;,
    \\
    \frac{e^{\frac{5 \jmath^2}{\alpha ^2}}\csch^2\tfrac{\rho }{\alpha }}{16 \pi^2  \alpha^3  \mathcal{I}(0)}  \left[(\coth\tfrac{\rho}{\alpha} + 3)e^{-\frac{3 \rho }{\alpha }}\erfc\left(\tfrac{3 \jmath}{\alpha }-\tfrac{\rho }{2 \jmath}\right) - (\coth\tfrac{\rho}{\alpha} - 3)e^{\frac{3 \rho }{\alpha }} \erfc\left(\tfrac{3 \jmath}{\alpha }+\tfrac{\rho }{2 \jmath}\right) - \tfrac{2\alpha}{\jmath \sqrt{\pi}} e^{-\frac{9 \jmath^2}{\alpha^2}-\frac{\rho^2}{4 \jmath^2}} \right]\;, & N=5\;.
    \end{dcases}
\end{equation}
On the other hand, if ${q_2=1}$, the inverse Laplace transform of $\left(\left(\lambda+{N}/{\alpha^2}\right)\mathcal{I}(-\lambda)\right){}^{-1}$ does not exist (it may exist for some ${q_2>1}$). Nevertheless, the inverse Fourier transform of this expression does exist and it reads
\begin{equation}
     \mathsf{F}^{-1}\left[\left(\left(\lambda+\tfrac{N}{\alpha^2}\right) e^{\jmath^4\lambda\left(\lambda+\frac{b}{\alpha^2}\right)}\right)^{-1}\right](\sigma)=\tfrac{i}{2}  e^{\frac{N (b-N)\jmath^4 }{\alpha ^4}-\frac{i N \sigma }{\alpha ^2}} \erf\left(\tfrac{\sigma }{2 \jmath^2}+\tfrac{i  (b-2N)\jmath^2}{2 \alpha ^2}\right) \;.
\end{equation}
Hence, we can write $G_{\textrm{IDG}}^{1}$ not just by means of the spherical transform in the first line of \eqref{eq:GK} but also using the Schr\"odinger-type version of the second line in \eqref{eq:GK},
\begin{equation}\label{eq:GIDG1}
\begin{aligned}
    G_{\textrm{IDG}}^{1} &=\frac{1}{\mathcal{I}(0)}\int_{\lambda_{\textrm{min}}}^{\infty} d\lambda\, \frac{\mathsf{L}^{-1}\left[K(s)\right](\lambda)}{\left(\lambda+\tfrac{N}{\alpha^2}\right) e^{\jmath^4\lambda\left(\lambda+\frac{b}{\alpha^2}\right)}}=\frac{i e^{\frac{ N (b-N)\jmath^4}{\alpha ^4}}}{2\mathcal{I}(0)}\int_{-\infty-i0^+}^{\infty-i0^+} d\sigma\,  e^{-\frac{i N \sigma }{\alpha ^2}} \erf\left(\tfrac{\sigma }{2 \jmath^2}+\tfrac{i  (b-2N)\jmath^2}{2 \alpha ^2}\right)K(i\sigma)\;, \quad N\geq3\;,
\end{aligned}
\end{equation}
where $\mathsf{L}^{-1}\left[K(s)\right]$ is given by \eqref{eq:inverseofSK} and $K(i\sigma)$ is the analytic continuation of \eqref{eq:scalarK}. Although both integrals are oscillatory, they can be evaluated numerically without any difficulties (at least for odd $N$). The former, however, seems to be more convenient for numerical calculations.

Moving on to $\mathring{G}_{\textrm{IDG}}$, we first consider ${q_1=0}$ and notice that
\begin{equation}\label{eq:laplaceIDGE}
    \mathsf{L}^{-1}\left[\left(\lambda e^{\ell^2\lambda}\right)^{-1}\right](s)=\theta \left(s -\ell^2\right)\;,
\end{equation}
which implies that $\mathring{G}_{\textrm{IDG}}^{0}$ is just an integral of $\mathring{K}$ with the bottom limit being shifted to ${s=\ell^2}$ (when compared to GR). Surprisingly, if ${N=3}$, the resulting integral can still be found in a closed form due to the simple form of $\mathring{K}$ in \eqref{eq:Kring3},
\begin{equation}\label{eq:GIDGring3}
\begin{aligned}
    \mathring{G}_{\textrm{IDG}}^{0} &= \frac{3 \alpha^3 e^{-\frac{\ell ^2}{\alpha ^2}}}{8 \pi \mathcal{E}\left(-\frac{2}{\alpha^2}\right)}\left(\coth \tfrac{\rho}{\alpha}-1\right)  \bigg[e^{\tfrac{\ell ^2}{\alpha ^2}} \left(\left(  2-\tfrac{2 \ell ^2}{\alpha^2} -\tfrac{\rho}{\alpha}\right) e^{\tfrac{2 \rho }{\alpha }} \erfc\left(\tfrac{\ell }{\alpha }+\tfrac{\rho }{2 \ell }\right)-\left(  2-\tfrac{2 \ell ^2}{\alpha^2} +\tfrac{\rho}{\alpha}\right) \erfc\left(\tfrac{\ell }{\alpha }-\tfrac{\rho }{2 \ell }\right)\right)
\\ 
&\feq+4 e^{\rho /\alpha } \erf\tfrac{\rho }{2 \ell }\bigg] \;, \quad N=3\;.
\end{aligned}
\end{equation}
In general ${N\geq3}$, we may calculate $\check{G}_{\textrm{IDG}}^{0}$. Thanks to the second line of \eqref{eq:GKcheck} together with \eqref{eq:laplaceIDGE}, it takes the form of the integral of $K$ from \eqref{eq:scalarK} with the bottom limit ${s=\ell^2}$, which evaluates to
\begin{equation}
    \check{G}_{\textrm{IDG}}^{0}=
    \begin{dcases}
        \frac{(-1)^{k}}{2^{k+1}\pi^k}\frac{1}{2\mathcal{E}\left(-\frac{2}{\alpha^2}\right)}\left(\frac{1}{\alpha\sinh \tfrac{\rho}{\alpha}} \pp_{\rho}\right)^k \left[\tfrac{e^{\frac{k \rho }{\alpha }}}{\frac{k}{\alpha }} \erfc\left(\tfrac{\rho }{2 \ell}{+}\tfrac{ k\ell}{\alpha }\right){+}\tfrac{e^{-\frac{k \rho }{\alpha }}}{\frac{k}{\alpha }}\erfc\left(\tfrac{ k\ell}{\alpha }{-}\tfrac{\rho }{2 \ell}\right)\right]\;, & N=2k+1\;,
        \\
        \frac{(-1)^k}{2^{k+\frac{3}{2}} \pi ^{k+1}}\frac{1}{2\mathcal{E}\left(-\frac{2}{\alpha^2}\right)} \left(\frac{1}{\alpha\sinh \tfrac{\rho}{\alpha}} \pp_{\rho}\right)^k \!\!\int_{\frac{\rho}{\alpha}}^{\infty}\!\!d u \,\tfrac{e^{-\left(k{+}\frac{1}{2}\right) u} \erfc\left(\left(k+\frac{1}{2}\right)\frac{\ell}{\alpha}-\frac{\alpha  u}{2\ell }\right)-e^{\left(k{+}\frac{1}{2}\right) u} \erfc\left(\left(k{+}\frac{1}{2}\right)\frac{\ell}{\alpha}+\frac{\alpha  u}{2 \ell}\right)}{\sqrt{\cosh u-\cosh \frac{\rho }{\alpha}}}\;, & N=2k+2\;.
    \end{dcases}
\end{equation}
Particularly, in $N=3,4,5$, we find
\begin{equation}
    \check{G}_{\textrm{IDG}}^{0} =
    \begin{dcases}
        \frac{1}{8\pi\alpha \mathcal{E}\left(-\frac{2}{\alpha^2}\right)}   \left[(\coth\tfrac{\rho}{\alpha} - 1)  \erfc\left( \tfrac{\ell}{\alpha}{-}\tfrac{\rho }{2 \ell} \right) - (\coth\tfrac{\rho}{\alpha} + 1) \erfc\left(\tfrac{\ell}{\alpha}{+}\tfrac{\rho }{2 \ell}\right)\right]\;, & N=3\;,
        \\
        -\frac{1}{2^{\frac{7}{2}} \pi^2 \mathcal{E}\left(-\frac{2}{\alpha^2}\right)} \left(\frac{1}{\alpha\sinh \tfrac{\rho}{\alpha}} \pp_{\rho}\right) \int_{\frac{\rho}{\alpha}}^{\infty}\!d u \,\tfrac{e^{-\frac{3}{2} u} \erfc\left(\frac{3\ell}{2\alpha}-\frac{\alpha  u}{2\ell }\right)-e^{\frac{3}{2} u} \erfc\left(\frac{3\ell}{2\alpha}+\frac{\alpha  u}{2 \ell}\right)}{\sqrt{\cosh u-\cosh \tfrac{\rho }{\alpha}}}\;, & N=4\;,
        \\
        \frac{1}{16\pi^2\alpha^3 \mathcal{E}\left(-\frac{2}{\alpha^2}\right)}   \bigg[(\coth\tfrac{\rho}{\alpha} - 1)^2 (\coth\tfrac{\rho}{\alpha} + 2) \erfc\left( \tfrac{2\ell}{\alpha}{-}\tfrac{\rho }{2 \ell} \right) - (\coth\tfrac{\rho}{\alpha} + 1)^2 (\coth\tfrac{\rho}{\alpha} - 2) \erfc\left(\tfrac{2\ell}{\alpha}{+}\tfrac{\rho }{2 \ell}\right) \\- \tfrac{2\alpha}{l \sqrt{\pi}} \csch^2\tfrac{\rho}{\alpha} e^{-\frac{4 l^2}{\alpha^2}-\frac{\rho^2}{4 l^2}} \bigg]\;, & N=5\;.
    \end{dcases}
\end{equation}
Again, $\mathring{G}_{\textrm{IDG}}^{0}$ can be found by calculating the 2-dimensional convolutional integral \eqref{eq:Gcheckconv} in arbitrary $N\geq3$ and we recover \eqref{eq:GIDGring3} for ${N=3}$. Let us now focus on ${q_1=1}$. As in the case of $G_{\textrm{IDG}}^{1}$, the inverse Laplace transform of $\left({\lambda\,\mathcal{E}\big(-\lambda-{2}/{\alpha^2}\big)}\right){}^{-1}$ does not exist but its inverse Fourier transform does,
\begin{equation}
    \mathsf{F}^{-1}\left[\left(\lambda e^{\ell^4\lambda\left(\lambda+\frac{a}{\alpha^2}\right)}\right)^{-1}\right](\sigma)=\tfrac{i}{2}  \erf\left(\tfrac{\sigma }{2 \ell^2 }+\tfrac{i  \ell^2 a}{2 \alpha ^2}\right)\;.
\end{equation}
Similar to \eqref{eq:GIDG1}, we can write the integral not only using the TT bi-tensor version of the spherical transform in the first line of \eqref{eq:GKring} but also using the Schr\"odinger-type version of the second line in \eqref{eq:GKring},
\begin{equation}\label{eq:GringIDG1}
\begin{aligned}
    \mathring{G}_{\textrm{IDG}}^{1} &=\frac{1}{\mathcal{E}\left(-\frac{2}{\alpha^2}\right)}\int_{\lambda_{\textrm{min}}}^{\infty} d\lambda\, \frac{\mathsf{L}^{-1}\left[\mathring{K}(s)\right](\lambda)}{\lambda e^{\ell^4\lambda\left(\lambda+\frac{a}{\alpha^2}\right)}}=\frac{i}{2\mathcal{E}\left(-\frac{2}{\alpha^2}\right)}\int_{-\infty-i0^+}^{\infty-i0^+} d\sigma\,  \erf\left(\tfrac{\sigma }{2 \ell^2 }+\tfrac{i  \ell^2 a}{2 \alpha ^2}\right)\mathring{K}(i\sigma)\;, \quad N\geq3\;.
\end{aligned}
\end{equation}
These formulas are especially useful in ${N=3}$, for which $\mathsf{L}^{-1}\big[\mathring{K}(s)\big]$ and $\mathring{K}(i\sigma)$ are given by the closed-form expression \eqref{eq:inverseofSKring3} and \eqref{eq:Kring3}, respectively. Although both integrals are oscillatory, the former can be evaluated without any difficulties while the latter decays too slowly for standard numerical treatments. For the sake of completeness, we also add the formulas for $\check{G}_{\textrm{IDG}}^1$ following from \eqref{eq:GKcheck} in a similar manner to ${G}_{\textrm{IDG}}$. Hence, we have
\begin{equation}
    \check{G}_{\textrm{IDG}}^1=
    \frac{1}{\mathcal{E}\left(-\frac{2}{\alpha^2}\right)}\int_{\lambda_{\textrm{min}}}^{\infty} d\lambda\, \frac{\mathsf{L}^{-1}\left[{K}(s)\right](\lambda)}{\lambda e^{\ell^4\lambda\left(\lambda+\frac{a}{\alpha^2}\right)}}=\frac{i}{2\mathcal{E}\left(-\frac{2}{\alpha^2}\right)}\int_{-\infty-i0^+}^{\infty-i0^+} d\sigma\,  \erf\left(\tfrac{\sigma }{2 \ell^2 }+\tfrac{i  \ell^2 a}{2 \alpha ^2}\right){K}(i\sigma)\;, \quad N\geq3\;,
\end{equation}
from which $\mathring{G}_{\textrm{IDG}}^{1}$ can be calculated via the 2-dimensional convolutional integral \eqref{eq:Gcheckconv}. Both of these oscillatory integrals can be evaluated numerically as in the case of ${G}_{\textrm{IDG}}$.

Again, the investigation of the closed-form expressions for ${N=3}$ in the case ${q_1=0}$ and ${q_2=0}$ shows that both parts of the propagator are regular for ${\rho\to0}$,
\begin{equation}
\begin{aligned}
    G_{\textrm{IDG}}^{0} &\approx\frac{1}{4 \pi \mathcal{I}(0)} \left[\tfrac{1}{\sqrt{\pi}j} e^{-\frac{j^2}{\alpha^2}} - \tfrac{2}{\alpha} e^{\frac{3j^2}{\alpha^2}} \erfc{\tfrac{2j}{\alpha}} \right] \;, 
    \\
    \bar{G}_{\textrm{IDG}}^{0} &\approx\frac{1}{10 \pi \alpha \mathcal{E}\left(-\tfrac{2}{\alpha^2}\right)} \left[\frac{\alpha}{\sqrt{\pi}\ell}  e^{-\frac{\ell^2}{\alpha^2}} + 3 \erfc{\frac{\ell}{\alpha}}\right]\;, \quad {\rho\to0}\;, \quad N=3\;,
\end{aligned}
\end{equation}
and decay exponentially for ${\rho\to\infty}$,
\begin{equation}
    G_{\textrm{IDG}}^{0} \approx \frac{e^{\frac{3j^2}{\alpha^2}}}{2 \pi \alpha \mathcal{I}(0)} e^{-\frac{3\rho}{\alpha}}\;, 
    \quad
    \bar{G}_{\textrm{IDG}}^{0} \approx \frac{24 e^{- \frac{\ell^2}{\alpha^2}}}{\pi \alpha \mathcal{E} \left(-\tfrac{2}{\alpha^2}\right)} e^{-\frac{3\rho}{\alpha}}\;, \quad\rho\to\infty\;, \quad N=3\;.
\end{equation}
Here, only the overall factor depends on the constants of the theory but the damping constant is the same as in GR.

\section{Conclusions}\label{sec:conclusions}

We presented a new covariant method of construction of propagators $\bs{G}$ in the $N$-dimensional Euclidean AdS for a very generic class of gravitational theories (generalizing \cite{DHoker:1999bve} well beyond GR), which makes use of the heat kernel approach inspired by \cite{Frolov:2015bta,Frolov:2015usa,Frolov:2015bia}. First, we derived the equations for propagators \eqref{eq:GF} for arbitrary theory with the Lagrangian that is given by an analytic expression in the metric, curvature, and covariant derivative. In the process, we also corrected a mistake in the original derivation of the equivalent action \cite{Biswas:2016egy,Biswas:2016etb} by presenting a new derivation and generalized the results for the quadratic action to arbitrary dimension ${N\geq3}$, see Appx.~\ref{apx:equivaction}--\ref{apx:quadraticaction}. While working in the Landau gauge, for which only the purely scalar part $G$ and TT tensor part $\bs{G}^{\perp}$ are non-zero, we showed that the propagators can be expressed using the heat kernels for scalars $K$ and heat kernels for symmetric TT rank-2 tensors $\bs{K}^{\perp}$ satisfying the heat equations \eqref{eq:heatequations}, see \eqref{eq:GK} and \eqref{eq:GKbs}. By recalling known properties of MS TT bi-tensors and analyzing the action of the Laplace operator $\Box$, we realized that the MS TT bi-tensors $\bs{G}^{\perp}$ and $\bs{K}^{\perp}$ can be very conveniently represented by the MS bi-scalars $\mathring{G}$ and $\mathring{K}$. We found explicit expressions for $\mathring{K}$, which turned out to be related to the scalar heat kernel $K$ via the convolution \eqref{eq:GKring} [a two-dimensional integral of the form \eqref{eq:convolution}]. For ${N=3}$, we managed to evaluate it to the closed-form \eqref{eq:Kring3}. We tested our heat kernel approach on GR in Sec.~\ref{ssc:GR}, where we rederived the known formulas for the propagators \cite{DHoker:1999bve} and wrote them in an equivalent form of $\mathring{G}_{\textrm{GR}}$ (which seems a bit simpler than $\bar{G}_{\textrm{GR}}$ at least in odd dimensions). Finally, we found explicit formulas for the propagators in various examples of HDG and IDG, see Sec.~\ref{ssc:HDG} and Sec.~\ref{ssc:IDG}, respectively. Interestingly, we arrived at the closed-form expressions for $\mathring{G}_{\textrm{HDG}}$ and $\mathring{G}_{\textrm{IDG}}^{0}$ in ${N=3}$, see \eqref{eq:GHDGring3} and \eqref{eq:GIDGring3}. We also commented on the asymptotic behavior ${N=3}$ and showed that, in contrast to GR, the propagators for HDG and IDG are regular for short geodesic distances but still decay exponentially for long geodesic distances, however possibly, with different constants depending on the theory.

The significance of propagators and heat kernels lies in various applications they can be used for. As this depends on ones interests as well as the gravitational theories in question, we will start with the most important calculations that our novel formulas enable and continue with other interesting follow-up projects and extensions of our work:
\begin{enumerate}[topsep=0pt,itemsep=0pt,label=\textit{\arabic*)}]
    \item Since the propagators are the main building blocks of the AdS/CFT correlation functions, the amplitudes that were calculated in GR so far \cite{DHoker:1999kzh,DHoker:1999mqo,Costa:2014kfa},\footnote{The form of the propagator used in \cite{DHoker:1999kzh} is related to the Landau gauge propagator by a change of the MS bi-tensor basis \cite{DHoker:1999bve}.} can now be evaluated also for other gravitational theories, which may shed more light on the AdS/CFT correspondence beyond GR.
    \item As the full field equations of HDG and IDG are often intractable, one typically has to resort to the weak-field approximations, which are captured by convolution of the propagators with the energy-momentum tensor. An example of this calculation was done for GR by considering a point mass source leading to the linearized Schwarzschild-AdS metric \cite{Kleppe:1994ga}. Now it could be performed also for other theories.
    \item Remark that also our new expressions for the heat kernels for TT rank-2 tensors have their own use in the calculations of partition functions \cite{David:2009xg,Gopakumar:2011qs,Lal:2012ax,Shahidi:2018smw,Lal:2012aku,David:2009xg,Hatzinikitas:2017vgl,Giombi:2008vd,Suzuki:2021pyw}.
    \item Note that above applications may be further promoted also to the BTZ-type black-hole geometries that are constructed from 3-dimensional AdS by some identifications. It is well known that the heat kernels as well as propagators on such geometries are easily obtained out of those on AdS \cite{Mann:1996ze,Suzuki:2021pyw}.
    \item Recall that, similar to \cite{Allen:1986tt,Turyn:1988af,Antoniadis:1986sb,DHoker:1999bve,Faizal:2011sa}, our derivation is fully covariant in the sense that it is independent of coordinates in AdS. Nevertheless, we imposed a specific covariant gauge for the perturbations, the Landau gauge, which was known to provide exceptionally simple results in GR. It would be very interesting to extend our study to the general covariant gauge as it was done in \cite{Faizal:2011sa} in the case of GR.
    \item Another interesting direction is to relax the assumption on the invariance under the isometries of the background space. (A similar studies are especially important in de Sitter space, see \cite{Glavan:2019msf,Glavan:2022nrd}.)
    \item Arguably, one of the most interesting lines of research is the study of propagators in the Lorentzian regime either through the analytic extension of our results or by working directly in the Lorentzian signature. Similar to GR, see e.g. \cite{Kleppe:1994ga}, it would be particularly interesting to analyze effects of various boundary conditions, which become even more complex in higher-derivative theories. In this context, one should also analyse the causality violation appearing in nonlocal theories \cite{Buoninfante:2018mre}.
    \item Recently, it turned out that the amplitudes in AdS/CFT take much simpler form when they are expressed in the Mellin space \cite{Fitzpatrick:2011ia,Nandan:2011wc,Bekaert:2014cea}. Following \cite{Balitsky:2011tw}, it would be also interesting to rewrite also the propagators for HDG and IDG we found to the Mellin space representation. Similarly, it would be interesting to see whether the bulk-to-boundary propagators can still be recast into Dobrev's boundary-to-bulk intertwiners as it is in GR \cite{Leonhardt:2003qu,Leonhardt:2003sn,Costa:2014kfa}.
    \item Following the analysis performed in \cite{Kolar:2022kgx}, it would be interesting to derive some estimates for the propagators for HDG and IDG based on the estimates of the heat kernels \cite{DaviesMandouvalos:1988}. This might be useful for extracting some qualitative information in cases where the integration mentioned above is not tractable.
    \item Despite using the assumption of $\mathcal{F}_i$-analyticity, some of our intermediary results might be easily generalized also to the non-analytic cases and applicable to a wider class of theories, for example, those with the logarithmic terms from 1-loop quantum corrections, see, e.g, \cite{Burzilla:2020utr}.
\end{enumerate}

\section*{Acknowledgements}

The authors would like to thank Korumilli Sravan Kumar (Portsmouth, United Kingdom), Vojt\v{e}ch Witzany (Prague, Czechia), and Pavel Krtou\v{s} (Prague, Czechia) for stimulating discussions. I.K. was supported by Netherlands Organization for Scientific research (NWO) grant no. 680-91-119 and Primus grant PRIMUS/23/SCI/005 from Charles University. T.M. acknowledges the support of the Czech Academy of Sciences (RVO 67985840) and the Czech Science Foundation GA\v{C}R grant no. GA19-09659S.

\appendix

\appsection{Variation of metric and curvature}\label{apx:varmetcur}

In this appendix, we provide variations of the metric density and curvature tensors around MS backgrounds employed mainly in Appx.~\ref{apx:quadraticaction} for derivation of the quadratic action \eqref{eq:quadaction}. 
Let us consider an $N$-dimensional Riemannian manifold $(\hat{M},\hat{\bs{g}})$ that is a linear perturbation of MS background $({M},{\bs{g}})$ (not necessarily of negative constant curvature), i.e., $(\hat{M},\hat{\bs{g}})=(M,\bs{g}+\varepsilon\bs{h})$, $\varepsilon>0$.
An arbitrary functional $A[\bs{\hat{g}}]$ can be expanded around $\hat{\bs{g}}=\bs{g}$ as
\begin{equation}
    A[\bs{\hat{g}}]
    = \sum_{n=0}^{\infty} \frac{\varepsilon^n}{n!} \delta^n A[\bs{h}]\;,
\end{equation}
where $\delta^n A[\bs{h}]$ denotes the $n$-th functional differential at $\hat{\bs{g}}=\bs{g}$
\begin{equation}
    \delta^n A[\bs{h}] := \pp_{\varepsilon}^n A[{\bs{g}}+\varepsilon \bs{h}]\big|_{\varepsilon=0}\;.
\end{equation}
Note that the Riemann tensor of an MS background is fully determined by the corresponding constant Ricci scalar 
\begin{equation}\label{eq:RiemMSS}
    R_{abcd} = \tfrac{2}{N(N-1)}R g_{a[c} g_{d]b}\;, \quad R = \text{const}\;,
\end{equation}
and the traceless Ricci tensor $\bs{S}$ and Weyl tensor $\bs{C}$ thus vanish.

The first and second variation of the metric density around an MS background reads
\begin{equation}\label{eq:varg}
    \delta{\hat{\mathfrak{g}}}^{\nicefrac{1}{2}} = \frac12 {\mathfrak{g}}^{\nicefrac{1}{2}} h = \frac12 {\mathfrak{g}}^{\nicefrac{1}{2}} (h^\top + \Box h^{\bowtie})
\end{equation}
and
\begin{equation}\label{eq:var2g}
\begin{aligned}
    \delta^2{\hat{\mathfrak{g}}}^{\nicefrac{1}{2}} &= {\mathfrak{g}}^{\nicefrac{1}{2}} \left(- \tfrac{1}{2} h_{ab} h^{ab} + \tfrac{1}{4} h^2 \right)
    = {\mathfrak{g}}^{\nicefrac{1}{2}} \left( - \tfrac{1}{2} h^\perp_{ab} h^{{\perp}ab} 
    + h^{{\asymp}a} \left( \Box + \tfrac{1}{N}R \right) h^\asymp_a \right. \\
    &\quad - \left. \tfrac{1}{4} h^{\bowtie} \left( \Box + \tfrac{2}{N}R \right) \Box h^{\bowtie}
    + \tfrac{N-2}{4N} {h^\top}^2
    + \tfrac{N-2}{2N} h^\top \Box h^{\bowtie} \right) + \text{div}\;,
\end{aligned}
\end{equation}
where the metric perturbation $\bs{h}$ is decomposed according to \eqref{eq:decomp} into its TT part $\bs{h}^{\perp}$, transverse covector $\bs{h}^{\asymp}$, and two scalars $h^{\bowtie}$ and $h^{\top}$.
The first and second variation of the Ricci scalar is given by
\begin{equation}\label{eq:varR}
    \delta \hat{R} = \nabla_{b}\nabla_{a}h^{ab} - \Box h - \tfrac{1}{N}R h
      = - \tfrac{N - 1}{N} \left( \Box + \tfrac{1}{N - 1}R \right) h^\top
\end{equation}
and
\begin{equation}\label{eq:var2R}
\begin{aligned}
    \delta^2 \hat{R} &= 2 h^{ab} \Box h_{ab}
    + \frac{3}{2} \nabla_{c}h_{ab} \nabla^{c}h^{ab} - \frac{1}{2} \nabla_{b}h \nabla^{b}h + 2 h^{ab} \nabla_{b}\nabla_{a}h - \nabla_{b}h_{ac} \nabla^{c}h^{ab} - 4 h^{ab} \nabla_{b}\nabla_{c}h_{a}{}^{c}
    - 2 \nabla_{a}h^{ab} \nabla_{c}h_{b}{}^{c} \\
    &\quad + 2 \nabla^{b}h \nabla_{c}h_{b}{}^{c} - \tfrac{2}{N(N-1)}R(h_{ab} h^{ab} - h^2)
    = \tfrac12 h^\perp_{ab} \left( \Box + \tfrac{2(N-2)}{N(N-1)}R \right) h^{{\perp}ab}
    - \tfrac{2}{N}R h^{\asymp}_a \left( \Box + \tfrac{2}{N}R \right) h^{\asymp{a}} \\
    &\quad + \tfrac{1}{2N}R h^{\bowtie} \left( \Box + \tfrac{2}{N}R \right) \Box h^{\bowtie}
    + \tfrac{(N-1)(N+2)}{2 N^2} h^\top \left( \Box + \tfrac{4}{(N-1)(N+2)}R \right) h^\top \\
    &\quad 
    + \tfrac{N-1}{N} h^\top \left( \Box + \tfrac{2}{N(N-1)}R \right) \Box h^{\bowtie}
    + \text{div}\;. 
\end{aligned}
\end{equation}
The variation of $\hat{\bs{S}}$ obviously reflects the traceless property 
\begin{equation}\label{eq:varS}
\begin{aligned}
    \delta \hat{S}_{ab} &= -  \tfrac{1}{2} \Box h_{ab} - \tfrac{1}{2} \nabla_{a}\nabla_{b} h + \tfrac{1}{N}g_{ab} \Box h
    + \nabla_{(a|}\nabla_{c}h_{|b)}{}^{c} -  \tfrac{1}{N} g_{ab}\nabla_{c}\nabla_{d}h^{cd}
    + \tfrac{1}{N(N - 1)}R \left( h_{ab} - \tfrac{1}{N}g_{ab} h \right) \\
    &= - \tfrac{1}{2} \left( \Box - \tfrac{2}{N(N - 1)}R \right) h^\perp_{ab}
    - \tfrac{N - 2}{2N} D_{ab} h^\top
\end{aligned}
\end{equation}
with $D_{ab}$ being a traceless operator defined by
\begin{equation}
    D_{ab} := \nabla_{a}\nabla_{b} - \tfrac{1}{N} g_{ab}\Box\;.
\end{equation}
The variation of the Weyl tensor $\hat{\bs{C}}$ can be conveniently expressed as 
\begin{equation}\label{eq:varC}
     \delta \hat{C}_{abcd} = c_{\{abcd\}} := \tfrac12 c_{[ab][cd]} + \tfrac12 c_{[cd][ab]}
\end{equation}
where $c_{abcd}$ is given by
\begin{equation}\label{eq:varCc}
\begin{aligned}
    c_{abcd} &:= - 2 \nabla_{a}\nabla_{c}h_{bd}
    + \tfrac{2 }{N-2}g_{ac} \left( \nabla_{b}\nabla_{d}h - 2 \nabla_{b}\nabla_{e}h_{d}{}^{e} + \Box h_{bd} - \tfrac{1}{N-1}R h_{bd}
    \right) \\
    &\quad + \tfrac{2 }{(N-1)(N-2)}g_{ac} g_{bd} \left(  \nabla_{f}\nabla_{e}h^{ef} - \Box h + \tfrac{1}{N}R h \right)
    = - 2 \left[ \nabla_{a}\nabla_{c} - \tfrac{1}{N-2}g_{ac} \left( \Box - \tfrac{1}{N-1}R \right) \right] h^\perp_{bd}\;.
\end{aligned}
\end{equation}
The divergence of $\delta\hat{\bs{C}}$ yields
\begin{equation}\label{eq:nablavarC}
    \nabla^a \delta \hat{C}_{abcd} = - \tfrac{N-3}{N-2} \left( \Box - \tfrac{1}{N}R \right)
    \nabla_{[c} h^\perp_{d]b}\;.
\end{equation}

\appsection{Commutators}\label{apx:commutators}

Here, we list commutators of the covariant derivative and wave operator acting on various tensors on MS background utilized in the derivation of the quadratic action in Appx.~\ref{apx:quadraticaction}. All the expressions follow as a direct application of the commutator 
\begin{equation}\label{eq:commutator}
\begin{aligned}
    \left[ \nabla_c, \nabla_d \right] \phi^{a_1 a_2 \cdots}{}_{b_1 b_2\cdots} &= R^{a_1}{}_{ecd} \phi^{e a_2 \cdots}{}_{b_1 b_2 \cdots} + R^{a_2}{}_{ecd} \phi^{a_1 e \cdots}{}_{b_1 b_2 \cdots} + \ldots \\
    &\quad - R^{e}{}_{b_1cd} \phi^{a_1 a_2 \cdots}{}_{e b_2 \cdots} - R^{e}{}_{b_2cd} \phi^{a_1 a_2 \cdots}{}_{b_1 e \cdots} - \ldots\;.
\end{aligned}
\end{equation}
Starting with a general vector $\phi^a$, one can show  
\begin{equation}\label{eq:commutrank1}
    \left[\nabla_a, \Box \right] \phi^{a} = \tfrac{1}{N}R \nabla_a \phi^a\;.
\end{equation}
The commutator applied to a symmetric traceless rank-2 tensor $\phi_{ab}$ gives
\begin{equation}\label{eq:commutrank2}
    \left[\nabla^a, \Box \right] \phi_{ab} = \tfrac{N+1}{N(N-1)}R \nabla^a \phi_{ab}\;.
\end{equation}
For a rank-3 tensor $\phi_{abc}$ with the properties ${\phi_{abc} + \phi_{bca} + \phi_{cab} = 0}$ and ${\phi^a{}_{ab} = \phi^a{}_{ba} = 0}$, it follows that
\begin{equation}\label{eq:commutrank3}
    \left[\nabla^a, \Box \right] \phi_{abc} = \tfrac{N-3}{N(N-1)}R \nabla^a \phi_{a(bc)} + \tfrac{N+1}{N(N-1)}R \nabla^a \phi_{a[bc]}\;.
\end{equation}
Finally, in the case of a completely traceless rank-4 tensor $\phi_{abcd}$ with the Riemann tensor symmetries,
we get
\begin{equation}\label{eq:commutrank4}
    \left[\nabla^a, \Box \right] \phi_{abcd} = \tfrac{1}{N}R \nabla^a \phi_{abcd}\;.
\end{equation}
Note also that for a symmetric TT rank-2 tensor $\phi^\perp_{ab}$, one obtains
\begin{equation}\label{eq:divnablahperp}
    \nabla^a \nabla_b \phi^\perp_{ca} = \tfrac{1}{N-1}R \phi^\perp_{bc}\;.
\end{equation}

\appsection{Equivalent action}\label{apx:equivaction}

The authors of \cite{Biswas:2016egy,Biswas:2016etb} demonstrated that the action \eqref{eq:analyticlagrangian} is equivalent to \eqref{eq:IDG} with respect to the second-order variations around MS backgrounds. In the first step, it was shown that \eqref{eq:analyticlagrangian}  
can be equivalently put into the form
\begin{equation}
    L_0(\hat{\bs{g}},\hat{\bs{R}})
    + \hat{\nabla}_e \hat{R}_{abcd} \mathcal{D}^{abcde}{}_{fghi}(\hat{\bs{g}},\hat{\bs{\nabla}}) \hat{R}^{fghi}\;.
\end{equation}
Then, the term $\hat{\nabla}_e \hat{R}_{abcd} \mathcal{D}^{abcde}{}_{fghi}(\hat{\bs{g}},\hat{\bs{\nabla}}) \hat{R}^{fghi}$ was rewritten using the results of \cite{Biswas:2011ar,Biswas:2013kla} obtained for Minkowski background.
Although it was wrongly considered that the covariant derivatives on MS backgrounds commute, as is actually the case in Minkowski background only, it turns out that the final form of the equivalent action is correct.

Here, we present a new derivation of the equivalent action \eqref{eq:IDG}.
The Lagrangian of any gravitational theory \eqref{eq:analyticlagrangian} can be cast to the form 
\begin{equation}\label{eq:genericL}
    L(\hat{\bs{g}},\hat{\bs{R}},\hat{\bs{\nabla}}) = \sum_{l=0}^\infty \sum_{k=0}^\infty L_{k,2l}(\hat{\bs{g}},\hat{\bs{R}},\hat{\bs{\nabla}})\;,
\end{equation}
where $L_{k,2l}(\hat{\bs{g}},\hat{\bs{R}},\hat{\bs{\nabla}})$ is of the $k$-th order in curvature and of the $2l$-th order in derivatives. The even number of derivatives in the Lagrangian follows from the fact that it is a fully contracted scalar made of $\hat{\bs{g}}$, $\hat{\bs{R}}$, $\hat{\bs{\nabla}}$.
All $\hat{\bs{g}}$, $\hat{\bs{R}}$ involved make up an even number of indices and each contraction among them reduces the number of available indices by two.
Then, each covariant derivative can be contracted either in pair with another covariant derivative or with any of the even number of remaining available indices of $\hat{\bs{g}}$, $\hat{\bs{R}}$.
Let us remark that any quantity denoted by a hat, e.g., the tensor density $\hat{\bs{\mathfrak{X}}}_1$, is constructed from $\hat{\bs{g}}$, $\hat{\bs{R}}$ and $\hat{\bs{\nabla}}$, whereas quantities without hat, e.g., the background metric density ${\mathfrak{g}}^{\nicefrac{1}{2}}$, are evaluated on the background.

For two arbitrary tensor densities $\hat{\bs{\mathfrak{X}}}_1$ and $\hat{\bs{\mathfrak{X}}}_2$ we define the equivalence symbol $\equivn{n}$
\begin{equation}\label{eq:equivn}
    \hat{\bs{\mathfrak{X}}}_1 \equivn{n} \hat{\bs{\mathfrak{X}}}_2 \iff \delta^k \hat{\bs{\mathfrak{X}}}_1 = \delta^k \hat{\bs{\mathfrak{X}}}_2\;, \quad k = 0,\ldots,n\;.
\end{equation}
Then for two scalar densities ${\hat{\mathfrak{g}}}^{\nicefrac{1}{2}}\hat{W}_1$ and ${\hat{\mathfrak{g}}}^{\nicefrac{1}{2}}\hat{W}_2$, where $\hat{W}_1$ and $\hat{W}_2$ are scalars, we can show
\begin{equation}\label{eq:equivofdensities}
        \hat{W}_1 \equivn{n} \hat{W}_2 \iff  {\hat{\mathfrak{g}}}^{\nicefrac{1}{2}}\hat{W}_1 \equivn{n}  {\hat{\mathfrak{g}}}^{\nicefrac{1}{2}}\hat{W}_2\;.
\end{equation}
This is a direct consequence of the definition \eqref{eq:equivn} and non-degeneracy of the background metric, ${{\mathfrak{g}}^{\nicefrac{1}{2}} \neq 0}$ because
\begin{equation}
    {\hat{\mathfrak{g}}}^{\nicefrac{1}{2}}{\hat W}_1 \equivn{n} {\hat{\mathfrak{g}}}^{\nicefrac{1}{2}}{\hat W}_2 \iff  \sum_{k=0}^l \tbinom{l}{k} \delta^{l-k} {\hat{\mathfrak{g}}}^{\nicefrac{1}{2}} \delta^k (\hat{W}_1 - \hat{W}_2) = 0\;, \quad l=0,\dots,n\;.
\end{equation}

In what follows, we would like to construct the Lagrangian $L_{\text{equiv}}$ that is equivalent to $L$ up to a divergence term, ${L \simeq L_{\text{equiv}} + \text{div}}$, where, from now on, we denote $\equivn{2}$ as $\simeq$ for brevity. Then, \eqref{eq:equivofdensities} implies the equivalence of the actions ${S^{\textrm{E}}-S_{\textrm{equiv}}^{\textrm{E}}=O(\varepsilon^3)}$. 

First, notice that $\hat{W} \equivn{n} 0$ corresponds to $\hat{W} = O(\varepsilon^{n+1})$ and therefore any terms of the generic Lagrangian \eqref{eq:genericL} for which $L_{k,2l}(\hat{\bs{g}},\hat{\bs{R}},\hat{\bs{\nabla}}) \simeq 0$ are irrelevant to our analysis.
For any tensor $\hat{\bs{X}}_i$ with the property $\bs{X}_i = 0$, it is obvious that $\delta^n(\hat{\bs{X}}_1 \cdots \hat{\bs{X}}_m) = 0$ when $m>n$ since, after the application of the Leibnitz rule, there appears at least one $\hat{\bs{X}}_i$ evaluated on the background in each term. Similarly, $\delta^n(\hat{\bs{X}}_1 \cdots \hat{\bs{X}}_n) = n! \delta\hat{\bs{X}}_1 \cdots \delta\hat{\bs{X}}_n$. As a consequence, one can show that
\begin{equation}\label{eq:YXm>n}
    \hat{\bs{Y}} \hat{\bs{X}}_1 \cdots \hat{\bs{X}}_m \equivn{n} 0\;, \quad m>n\;,
\end{equation}
and 
\begin{equation}\label{eq:YXm=n}
    \hat{\bs{Y}} \hat{\bs{X}}_1 \cdots \hat{\bs{X}}_m \equivn{n} \bs{Y} \hat{\bs{X}}_1 \cdots \hat{\bs{X}}_m, \quad m=n\;,
\end{equation}
where $\hat{\bs{Y}}$ is an arbitrary tensor.
Now, the irreducible decomposition of the Riemann tensor $\hat{\bs{R}}$ in terms of $\hat{R}$, $\hat{\bs{S}}$ and $\hat{\bs{C}}$ proves to be convenient.
On the MS background, $R = \text{const}$ and $\bs{S} = \bs{C} = 0$, therefore $\hat{\bs{\nabla}} \hat{R}$, $\hat{\bs{S}}$, $\hat{\bs{C}}$ with any number of derivatives satisfy the property of $\hat{\bs{X}}_i$ and due to \eqref{eq:YXm>n} only terms at most quadratic in $\hat{\bs{\nabla}} \hat{R}$, $\hat{\bs{S}}$, $\hat{\bs{C}}$ must be taken into account.
With this knowledge in mind, we can go through all the possible relevant terms of the generic Lagrangian \eqref{eq:genericL} and determine their simplest equivalent counterparts with respect to the second variation around MS backgrounds.

Let us start with the possible terms of $L_{k,0}(\hat{\bs{g}},\hat{\bs{R}})$ of the $k$-th order in curvature without any derivatives:

\begin{enumerate}
    \item Terms involving no $\hat{\bs{S}}$ neither $\hat{\bs{C}}$ are just powers of the Ricci scalar $\hat{R}^k$. If we look at the following variations of $\hat{R}^k$,
        \begin{equation}\label{eq:deltaR^k}
            \delta^0(\hat{R}^k) = R^{k}\;,
            \quad \delta(\hat{R}^k) = k R^{k-1} \delta\hat{R}\;,
            \quad \delta^2(\hat{R}^k) = k R^{k-2}( (k-1) \delta\hat{R} \delta\hat{R} + R \delta^2\hat{R})\;,
        \end{equation}
        the presence of $\delta\hat{R}\delta\hat{R}$ indicates that the equivalent term is at least quadratic in $\hat{R}$. To ensure consistency with \eqref{eq:deltaR^k}, two more constants have to be included and we thus assume the simplest equivalent term of the form $\hat{R}^k \simeq c_0 + c_1 \hat{R} + c_2 \hat{R}^2$. Comparing \eqref{eq:deltaR^k} with $\delta^n(c_0 + c_1 \hat{R} + c_2 \hat{R}^2)$ for $n=0,1,2$, one gets
        \begin{equation}
            \hat{R}^k \simeq \tfrac{(k-1)(k-2)}{2} R^k - k(k-2) R^{k-1} \hat{R} + \tfrac{k(k-1)}{2} R^{k-2} \hat{R}^2,
        \end{equation}
        
    \item Terms of $L_{k,0}(\hat{\bs{g}},\hat{\bs{R}})$ with one $\hat{\bs{S}}$ or $\hat{\bs{C}}$ vanish due to the tracelessness of $\hat{\bs{S}}$ and $\hat{\bs{C}}$
        \begin{equation}\label{eq:equivL:linearSC}
            \hat{R}^{k-1} \Rnode{A}{\hat{\bs{S}}} = 0\;,
                \ncbar[linewidth=.5pt,angle=-90,nodesep=1pt,arm=2pt,offsetA=1.1pt,offsetB=1.1pt]{-}{A}{A}
            \quad \hat{R}^{k-1} \Rnode{A}{\hat{\bs{C}}} = 0\;,
                \ncbar[linewidth=.5pt,angle=-90,nodesep=1pt,arm=2pt,offsetA=-3.3pt,offsetB=1.1pt]{-}{A}{A}
                \ncbar[linewidth=.5pt,angle=-90,nodesep=1pt,arm=2pt,offsetA=1.1pt,offsetB=-3.3pt]{-}{A}{A}
        \end{equation}
        where we introduced a graphical notation to indicate the contractions. Notice that our notation does not take into account the order of indices involved and could thus correspond to multiple configurations. For example, ignoring the symmetries of the Weyl tensor, the contractions in the last equation of \eqref{eq:equivL:linearSC} denotes three cases $\hat{C}^a{}_a{}^{b}{}_{b}$, $\hat{C}^{ab}{}_{ab}$ and $\hat{C}^{ab}{}_{ba}$.
         
    \item The possible terms of $L_{k,0}(\hat{\bs{g}},\hat{\bs{R}})$ quadratic in $\hat{\bs{S}}$, $\hat{\bs{C}}$ 
        \begin{equation}\label{eq:equivL:SC}
        \begin{aligned}
            \hat{R}^{k-2} \hat{\bs{S}} \hat{\bs{S}}:&
            \quad \hat{R}^{k-2} \Rnode{A}{\hat{\bs{S}}} \Rnode{B}{\hat{\bs{S}}} = 0\;,
                \ncbar[linewidth=.5pt,angle=-90,nodesep=1pt,arm=2pt,offsetA=1.1pt,offsetB=1.1pt]{-}{A}{A}
                \ncbar[linewidth=.5pt,angle=-90,nodesep=1pt,arm=2pt,offsetA=1.1pt,offsetB=1.1pt]{-}{B}{B}
            \quad \hat{R}^{k-2} \Rnode{A}{\hat{\bs{S}}} \Rnode{B}{\hat{\bs{S}}} \simeq R^{k-2} \hat{S}_{ab} \hat{S}{}^{ab},
                \ncbar[linewidth=.5pt,angle=-90,nodesep=1pt,arm=2pt,offsetA=1.1pt,offsetB=1.1pt]{-}{A}{B}
                \ncbar[linewidth=.5pt,angle=-90,nodesep=1pt,arm=4pt,offsetA=-1.1pt,offsetB=-1.1pt]{-}{A}{B} \\
            \hat{R}^{k-2} \hat{\bs{S}} \hat{\bs{C}}:&
            \quad \hat{R}^{k-2} \Rnode{A}{\hat{\bs{S}}} \Rnode{B}{\hat{\bs{C}}} = 0\;,
                \ncbar[linewidth=.5pt,angle=-90,nodesep=1pt,arm=2pt,offsetA=1.1pt,offsetB=1.1pt]{-}{A}{A}
                \ncbar[linewidth=.5pt,angle=-90,nodesep=1pt,arm=2pt,offsetA=-3.3pt,offsetB=1.1pt]{-}{B}{B}
                \ncbar[linewidth=.5pt,angle=-90,nodesep=1pt,arm=2pt,offsetA=1.1pt,offsetB=-3.3pt]{-}{B}{B}
            \quad \hat{R}^{k-2} \Rnode{A}{\hat{\bs{S}}} \Rnode{B}{\hat{\bs{C}}} = 0\;,
                \ncbar[linewidth=.5pt,angle=-90,nodesep=1pt,arm=2pt,offsetA=1.1pt,offsetB=3.3pt]{-}{A}{B}
                \ncbar[linewidth=.5pt,angle=-90,nodesep=1pt,arm=4pt,offsetA=-1.1pt,offsetB=1.1pt]{-}{A}{B}
                \ncbar[linewidth=.5pt,angle=-90,nodesep=1pt,arm=2pt,offsetA=1.1pt,offsetB=-3.3pt]{-}{B}{B} \\
            \hat{R}^{k-2} \hat{\bs{C}} \hat{\bs{C}}:&
            \quad \hat{R}^{k-2} \Rnode{A}{\hat{\bs{C}}} \Rnode{B}{\hat{\bs{C}}} = 0\;,
                \ncbar[linewidth=.5pt,angle=-90,nodesep=1pt,arm=2pt,offsetA=-3.3pt,offsetB=1.1pt]{-}{A}{A}
                \ncbar[linewidth=.5pt,angle=-90,nodesep=1pt,arm=2pt,offsetA=1.1pt,offsetB=-3.3pt]{-}{A}{A}
                \ncbar[linewidth=.5pt,angle=-90,nodesep=1pt,arm=2pt,offsetA=-3.3pt,offsetB=1.1pt]{-}{B}{B}
                \ncbar[linewidth=.5pt,angle=-90,nodesep=1pt,arm=2pt,offsetA=1.1pt,offsetB=-3.3pt]{-}{B}{B}
            \quad \hat{R}^{k-2} \Rnode{A}{\hat{\bs{C}}} \Rnode{B}{\hat{\bs{C}}} = 0\;,
                \ncbar[linewidth=.5pt,angle=-90,nodesep=1pt,arm=2pt,offsetA=-3.3pt,offsetB=1.1pt]{-}{A}{A}
                \ncbar[linewidth=.5pt,angle=-90,nodesep=1pt,arm=4pt,offsetA=1.1pt,offsetB=1.1pt]{-}{A}{B}
                \ncbar[linewidth=.5pt,angle=-90,nodesep=1pt,arm=2pt,offsetA=3.3pt,offsetB=3.3pt]{-}{A}{B}
                \ncbar[linewidth=.5pt,angle=-90,nodesep=1pt,arm=2pt,offsetA=1.1pt,offsetB=-3.3pt]{-}{B}{B}
            \quad \hat{R}^{k-2} \Rnode{A}{\hat{\bs{C}}} \Rnode{B}{\hat{\bs{C}}} \simeq \pm \tfrac{3\pm 1}{4} R^{k-2} \hat{C}_{abcd} \hat{C}{}^{abcd}\;,
                \ncbar[linewidth=.5pt,angle=-90,nodesep=1pt,arm=8pt,offsetA=-3.3pt,offsetB=-3.3pt]{A}{B}
                \ncbar[linewidth=.5pt,angle=-90,nodesep=1pt,arm=6pt,offsetA=-1.1pt,offsetB=-1.1pt]{A}{B}
                \ncbar[linewidth=.5pt,angle=-90,nodesep=1pt,arm=4pt,offsetA=1.1pt,offsetB=1.1pt]{A}{B}
                \ncbar[linewidth=.5pt,angle=-90,nodesep=1pt,arm=2pt,offsetA=3.3pt,offsetB=3.3pt]{A}{B}
        \end{aligned}
        \end{equation}
        either vanish due to the tracelessness of $\hat{\bs{S}}$ and $\hat{\bs{C}}$, or can be equivalently rewritten using \eqref{eq:YXm=n}. Both possible full contractions of two $\hat{\bs{S}}$ are equal due to symmetries, i.e., $\hat{S}_{ab} \hat{S}^{ab} = \hat{S}_{ab} \hat{S}^{ba}$. After rearrangement of indices, two fully contracted $\hat{\bs{C}}$ give either $\pm \hat{C}_{abcd} \hat{C}^{abcd}$ or $\pm \hat{C}_{abcd} \hat{C}^{abcd}$ and from the cyclic symmetry of $\hat{\bs{C}}$ it follows that $\hat{C}_{abcd} \hat{C}^{acbd} = \tfrac12 \hat{C}_{abcd} \hat{C}^{abcd}$.
\end{enumerate}
Therefore, we can conclude that $L_{k,0}(\hat{\bs{g}},\hat{\bs{R}})$ is equivalent with respect to the second variation around MS backgrounds to
\begin{equation}\label{eq:Lk0}
    L_{k,0}(\hat{\bs{g}},\hat{\bs{R}}) \simeq \text{const} + \text{const}\,\hat{R} + \hat{R} \mathcal{F}_1(0) \hat{R} + \hat{S}_{ab} \mathcal{F}_2(0) \hat{S}{}^{ab} + \hat{C}_{abcd} \mathcal{F}_3(0) \hat{C}{}^{abcd}\;.
\end{equation}
Note that only terms involving Ricci scalar in $L_{k,0}(\hat{\bs{g}},\hat{\bs{R}})$ and its equivalent counterpart contribute to field equations.

Possible terms of $L_{k,2l}(\hat{\bs{g}},\hat{\bs{R}},\hat{\bs{\nabla}})$ of the $k$-th order in curvature and $2l$-th order of $\hat{\bs{\nabla}}$, $l \geq 1$, contributing to the second variation around MS backgrounds are as follows:
\begin{enumerate}
    \item There are no terms with non-zero number of derivatives without $\hat{\bs{\nabla}}\hat{R}$, $\hat{\bs{S}}$ and $\hat{\bs{C}}$.
    \item Terms of $L_{k,2l}(\hat{\bs{g}},\hat{\bs{R}},\hat{\bs{\nabla}})$, $l \geq 1$, with one $\hat{\bs{\nabla}}\hat{R}$, $\hat{\bs{S}}$ or $\hat{\bs{C}}$ can be expressed as
        \begin{equation}\label{eq:equivL:DX}
        \begin{aligned}
            \hat{R}^{k-1} \hat{\bs{\nabla}}{}^{2l} \hat{R}:&
            \quad \hat{R}^{k-1} \hat{\bs{\nabla}}{}^{2l} \hat{R} \simeq (k-1) R^{k-2} \hat{R} \hat\Box^l \hat{R} + P_{k+1,2l-2}(\hat{\bs{g}},\hat{\bs{R}},\hat{\bs{\nabla}}) + \text{div}\;, \\
            \hat{R}^{k-1} \hat{\bs{\nabla}}{}^{2l} \hat{\bs{S}}:&
            \quad \hat{R}^{k-1} \hat{\bs{\nabla}}{}^{2l} \Rnode{A}{\hat{\bs{S}}} = 0\;,
                \ncbar[linewidth=.5pt,angle=-90,nodesep=1pt,arm=2pt,offsetA=1.1pt,offsetB=1.1pt]{-}{A}{A}
            \quad \hat{R}^{k-1} \Rnode{A}{\hat{\bs{\nabla}}}{}^{2l} \Rnode{B}{\hat{\bs{S}}} = \hat{R}^{k-1} \hat{\bs{\nabla}}{}^{2l} \hat{R} + P_{k+1,2l-2}(\hat{\bs{g}},\hat{\bs{R}},\hat{\bs{\nabla}})\;,
                \ncbar[linewidth=.5pt,angle=-90,nodesep=1pt,arm=2pt,offsetA=1.1pt,offsetB=1.1pt]{-}{A}{B}
                \ncbar[linewidth=.5pt,angle=-90,nodesep=1pt,arm=4pt,offsetA=-1.1pt,offsetB=-1.1pt]{-}{A}{B} \\
            \hat{R}^{k-1} \hat{\bs{\nabla}}{}^{2l} \hat{\bs{C}}:&
            \quad \hat{R}^{k-1} \hat{\bs{\nabla}}{}^{2l} \Rnode{A}{\hat{\bs{C}}} = 0\;,
                \ncbar[linewidth=.5pt,angle=-90,nodesep=1pt,arm=2pt,offsetA=-3.3pt,offsetB=1.1pt]{-}{A}{A}
                \ncbar[linewidth=.5pt,angle=-90,nodesep=1pt,arm=2pt,offsetA=1.1pt,offsetB=-3.3pt]{-}{A}{A}
            \quad \hat{R}^{k-1} \Rnode{A}{\hat{\bs{\nabla}}}{}^{2l} \Rnode{B}{\hat{\bs{C}}} = 0\;,
                \ncbar[linewidth=.5pt,angle=-90,nodesep=1pt,arm=2pt,offsetA=1.1pt,offsetB=3.3pt]{-}{A}{B}
                \ncbar[linewidth=.5pt,angle=-90,nodesep=1pt,arm=4pt,offsetA=-1.1pt,offsetB=1.1pt]{-}{A}{B}
                \ncbar[linewidth=.5pt,angle=-90,nodesep=1pt,arm=2pt,offsetA=1.1pt,offsetB=-3.3pt]{-}{B}{B}
            \quad \hat{R}^{k-1} \Rnode{A}{\hat{\bs{\nabla}}}{}^{2l} \Rnode{B}{\hat{\bs{C}}} = 0 + P_{k+1,2l-2}(\hat{\bs{g}},\hat{\bs{R}},\hat{\bs{\nabla}})\;,
                \ncbar[linewidth=.5pt,angle=-90,nodesep=1pt,arm=8pt,offsetA=-3.3pt,offsetB=-3.3pt]{A}{B}
                \ncbar[linewidth=.5pt,angle=-90,nodesep=1pt,arm=6pt,offsetA=-1.1pt,offsetB=-1.1pt]{A}{B}
                \ncbar[linewidth=.5pt,angle=-90,nodesep=1pt,arm=4pt,offsetA=1.1pt,offsetB=1.1pt]{A}{B}
                \ncbar[linewidth=.5pt,angle=-90,nodesep=1pt,arm=2pt,offsetA=3.3pt,offsetB=3.3pt]{A}{B}
        \end{aligned}
        \end{equation}
        and for $k=1$ the non-zero ones are clearly total derivative terms. Note that in the scalar terms involving covariant derivatives not all the contractions are graphically indicated unless they are essential for subsequent manipulations. 

        The equivalent expression for the scalar $\hat{R}^{k-1} \hat{\bs{\nabla}}{}^{2l} \hat{R}$ is derived in the following way. The covariant derivatives can be always rearrange using \eqref{eq:commutator} into the form
        \begin{equation}
            \hat{R}^{k-1} \hat{\bs{\nabla}}{}^{2l} \hat{R} = \hat{R}^{k-1} \hat{\Box}{}^{l} \hat{R} + P_{k+1,2l-2}(\hat{\bs{g}},\hat{\bs{R}},\hat{\bs{\nabla}})\;,
        \end{equation}
        where $P_{k+1,2l-2}(\hat{\bs{g}},\hat{\bs{R}},\hat{\bs{\nabla}})$ collects all the terms with one more curvature tensor and two derivatives less which arise from the use of the commutators. The variations of $\hat{R}^{k-1} \hat{\Box}{}^{l} \hat{R}$ up to the second order read
        \begin{equation}
        \begin{aligned}
            R^{k-1} \Box{}^{l} R &= 0\;, \\
            \delta(\hat{R}^{k-1} \hat{\Box}{}^{l} \hat{R}) &= R^{k-1} \delta(\hat{\Box}{}^{l} \hat{R}) = R^{k-1} \delta(\hat{\Box}{}^l) R + R^{k-1} \Box{}^{l} \delta\hat{R} = \text{div}\;, \\
            \delta^2(\hat{R}^{k-1} \hat{\Box}{}^{l} \hat{R}) &= R^{k-1} \delta^2( \hat{\Box}{}^{l} \hat{R}) + 2(k-1) R^{k-2} \delta\hat{R} \delta(\hat{\Box}{}^{l} \hat{R})\;,
        \end{aligned}
        \end{equation}
        where we employed \eqref{eq:YXm=n} and \eqref{eq:varFR}.
        Therefore, we can look for an equivalent expression of the form $\hat{R}^{k-1} \hat{\Box}{}^{l} \hat{R} \simeq c_1 \hat{\Box}{}^{l} \hat{R} + c_2 \hat{R} \hat{\Box}{}^{l} \hat{R}$ and comparing the second variations of the both sides we get
         \begin{equation}
            \hat{R}^{k-1} \hat{\Box}{}^{l} \hat{R} \simeq -(k-2)R^{k-1} \hat{\Box}{}^{l} \hat{R} + (k-1)R^{k-2} \hat{R} \hat{\Box}{}^{l} \hat{R} = (k-1)R^{k-2} \hat{R} \hat{\Box}{}^{l} \hat{R} + \text{div}\;.
        \end{equation}

        The derivatives in $\hat{R}^{k-1} \Rnode{A}{\hat{\bs{\nabla}}}{}^{2l} \Rnode{B}{\hat{\bs{S}}}$,
        where $\hat{\bs{S}}$ is fully contracted with $\hat{\bs{\nabla}}{}^{2l}$,
        are commuted unless any of the derivatives contracted with $\hat{\bs{S}}$ acts directly on this tensor. Then, one can use the twice-contracted Bianchi identity
        \begin{equation}\label{eq:BianchiIdDS}
            \hat{\nabla}_{b}\hat{S}_{a}{}^{b} = \tfrac{N-2}{2N} \hat{\nabla}_{a}\hat{R}\;.
        \end{equation}

        Similarly, terms $\hat{R}^{k-1} \Rnode{A}{\hat{\bs{\nabla}}}{}^{2l} \Rnode{B}{\hat{\bs{C}}}$,
        where $\hat{\bs{C}}$ is fully contracted with $\hat{\bs{\nabla}}{}^{2l}$,
        are first put into the form involving $\hat{\nabla}_c \hat{\nabla}_d \hat{C}^{abcd}$ since a divergence of the contracted Bianchi identity implies
        \begin{equation}\label{eq:BianchiIdDDC1}
            \hat{\nabla}_c \hat{\nabla}_d \hat{C}^{abcd} = 0\;.
        \end{equation}
        
    \item Terms of $L_{k,2l}(\hat{\bs{g}},\hat{\bs{R}},\hat{\bs{\nabla}})$, $l \geq 1$, quadratic in $\hat{\bs{\nabla}}\hat{R}$, $\hat{\bs{S}}$, $\hat{\bs{C}}$. First, let us notice that these terms do not contribute to the field equations since \eqref{eq:YXm>n} implies $\hat{R}^{k-2} \hat{\bs{\nabla}} \cdots \hat{\bs{\nabla}} \hat{\bs{X}}_1 \hat{\bs{\nabla}} \cdots \hat{\bs{\nabla}} \hat{\bs{X}}_2 \equivn{1} 0$ for $\hat{\bs{X}}_i$ being any of $\hat{\bs{\nabla}}\hat{R}$, $\hat{\bs{S}}$, $\hat{\bs{C}}$. Moreover, $\hat{R}^{k-2} \hat{\bs{\nabla}} \cdots \hat{\bs{\nabla}} \hat{\bs{X}}_1 \hat{\bs{\nabla}} \cdots \hat{\bs{\nabla}} \hat{\bs{X}}_2 \simeq R^{k-2} \hat{\bs{\nabla}} \cdots \hat{\bs{\nabla}} \hat{\bs{X}}_1 \hat{\bs{\nabla}} \cdots \hat{\bs{\nabla}} \hat{\bs{X}}_2$, following from \eqref{eq:YXm=n}, allows us to arbitrarily move derivatives between $\hat{\bs{X}}_1$ and $\hat{\bs{X}}_2$ on the right-hand side of the equivalence with constant $R^{k-2}$ by introducing total derivative terms. For convenience, we introduce indices $m$, $p$ and $q$ with different ranges
        \begin{equation}
            1 \leq m \leq 2l-1\;, \quad 1 \leq p \leq 2l\;, \quad 0 \leq q \leq 2l\;.
        \end{equation}
        Terms quadratic in $\hat{\bs{\nabla}} \hat{R}$ are equivalent to
        \begin{equation}\label{eq:equivL:DRDR}
            \hat{R}^{k-2} \hat{\bs{\nabla}}{}^{m} \hat{R} \hat{\bs{\nabla}}{}^{2l-m} \hat{R}:
            \quad \hat{R}^{k-2} \hat{\bs{\nabla}}{}^{m} \hat{R} \hat{\bs{\nabla}}{}^{2l-m} \hat{R} \simeq (-1)^m R^{k-2} \hat{R} \hat\Box^l \hat{R} + P_{k+1,2l-2}(\hat{\bs{g}},\hat{\bs{R}},\hat{\bs{\nabla}}) + \text{div}\;,
        \end{equation}
        where we have conveniently moved the derivatives to act on a single $\hat{R}$ and subsequently
        changed their order to form $\hat{\Box}^l$.
        
        All possible scalars involving $\hat{\bs{S}}$ take one of the following form
        \begin{equation}\label{eq:equivL:DXDS}
        \begin{aligned}
            \hat{R}^{k-2} \hat{\bs{\nabla}}{}^{p} \hat{R} \hat{\bs{\nabla}}{}^{2l-p} \hat{\bs{S}}:&
            \quad \hat{R}^{k-2} \hat{\bs{\nabla}}{}^{p} \hat{R} \hat{\bs{\nabla}}{}^{2l-p} \Rnode{A}{\hat{\bs{S}}} = 0\;, \\
                \ncbar[linewidth=.5pt,angle=-90,nodesep=1pt,arm=2pt,offsetA=1.1pt,offsetB=1.1pt]{A}{A}
            &\quad \hat{R}^{k-2} \hat{\bs{\nabla}}{}^{m} \hat{R} \Rnode{A}{\hat{\bs{\nabla}}}{}^{2l-m} \Rnode{B}{\hat{\bs{S}}} = \hat{R}^{k-2} \hat{\bs{\nabla}}{}^{m} \hat{R} \hat{\bs{\nabla}}{}^{2l-m} \hat{R} + P_{k+1,2l-2}(\hat{\bs{g}},\hat{\bs{R}},\hat{\bs{\nabla}})\;, \\
                \ncbar[linewidth=.5pt,angle=-90,nodesep=1pt,armA=0pt,armB=2pt,offsetA=1.1pt,offsetB=-1.1pt,,arrowsize=1.8pt,arrowlength=1,arrowinset=0,ArrowFill=false]{>-}{B}{B}
                \ncbar[linewidth=.5pt,angle=-90,nodesep=1pt,arm=2pt,offsetA=0pt,offsetB=1.1pt]{A}{B}
            &\quad \hat{R}^{k-2} \Rnode{A}{\hat{\bs{\nabla}}}{}^{m+1} \hat{R} \hat{\bs{\nabla}}{}^{2l-m-1} \Rnode{B}{\hat{\bs{S}}} \simeq - \hat{R}^{k-2} \hat{\bs{\nabla}}{}^{m} \hat{R} \hat{\bs{\nabla}}{}^{2l-m} \hat{R} + P_{k+1,2l-2}(\hat{\bs{g}},\hat{\bs{R}},\hat{\bs{\nabla}}) + \text{div}\;,
                \ncbar[linewidth=.5pt,angle=-90,nodesep=1pt,arm=4pt,offsetA=-1.1pt,offsetB=-1.1pt]{A}{B}
                \ncbar[linewidth=.5pt,angle=-90,nodesep=1pt,arm=2pt,offsetA=1.1pt,offsetB=1.1pt]{A}{B}
            \\
            \hat{R}^{k-2} \hat{\bs{\nabla}}{}^{q} \hat{\bs{S}} \hat{\bs{\nabla}}{}^{2l-q} \hat{\bs{S}}:&
            \quad \hat{R}^{k-2} \hat{\bs{\nabla}}{}^{q} \Rnode{A}{\hat{\bs{S}}} \hat{\bs{\nabla}}{}^{2l-q} \Rnode{B}{\hat{\bs{S}}} =
                \ncbar[linewidth=.5pt,angle=-90,nodesep=1pt,arm=2pt,offsetA=-1.1pt,offsetB=-1.1pt]{A}{A}
                \ncbar[linewidth=.5pt,angle=-90,nodesep=1pt,arm=2pt,offsetA=-1.1pt,offsetB=-1.1pt]{B}{B}
            \hat{R}^{k-2} \hat{\bs{\nabla}}{}^{q} \Rnode{A}{\hat{\bs{S}}} \hat{\bs{\nabla}}{}^{2l-q} \Rnode{B}{\hat{\bs{S}}} =
                \ncbar[linewidth=.5pt,angle=-90,nodesep=1pt,arm=2pt,offsetA=-1.1pt,offsetB=-1.1pt]{A}{A}
                \ncbar[linewidth=.5pt,angle=-90,nodesep=1pt,armA=0pt,armB=2pt,offsetA=1.1pt,offsetB=-1.1pt,arrowsize=1.8pt,arrowlength=1,arrowinset=0,ArrowFill=false]{>-}{B}{B}
                \ncbar[linewidth=.5pt,angle=-90,nodesep=1pt,armA=0pt,armB=2pt,offsetA=-1.1pt,offsetB=1.1pt,arrowsize=1.8pt,arrowlength=1,arrowinset=0,ArrowFill=false]{>-}{B}{B}
            \hat{R}^{k-2} \hat{\bs{\nabla}}{}^{q} \Rnode{B}{\hat{\bs{S}}} \hat{\bs{\nabla}}{}^{2l-q} \Rnode{A}{\hat{\bs{S}}} = 0\;,
                \ncbar[linewidth=.5pt,angle=-90,nodesep=1pt,arm=2pt,offsetA=-1.1pt,offsetB=-1.1pt]{A}{A}
                \ncbar[linewidth=.5pt,angle=-90,nodesep=1pt,armA=0pt,armB=2pt,offsetA=1.1pt,offsetB=-1.1pt,arrowsize=1.8pt,arrowlength=1,arrowinset=0,ArrowFill=false]{>-}{B}{B}
                \ncbar[linewidth=.5pt,angle=-90,nodesep=1pt,armA=0pt,armB=2pt,offsetA=-1.1pt,offsetB=1.1pt,arrowsize=1.8pt,arrowlength=1,arrowinset=0,ArrowFill=false]{>-}{B}{B}
            \\
            &\quad \hat{R}^{k-2} \Rnode{A}{\hat{\bs{\nabla}}}{}^{p} \Rnode{B}{\hat{\bs{S}}} \Rnode{C}{\hat{\bs{\nabla}}}{}^{2l-p} \Rnode{D}{\hat{\bs{S}}} = \hat{R}^{k-2} \hat{\bs{\nabla}}{}^{p} \hat{R} \hat{\bs{\nabla}}{}^{2l-p} \hat{\bs{S}} + P_{k+1,2l-2}(\hat{\bs{g}},\hat{\bs{R}},\hat{\bs{\nabla}})\;,
                \ncbar[linewidth=.5pt,angle=-90,nodesep=1pt,arm=2pt,offsetA=0pt,offsetB=0pt]{A}{B}
            \\
            &\quad \hat{R}^{k-2} \Rnode{A}{\hat{\bs{\nabla}}}{}^{p-1} \Rnode{B}{\hat{\bs{S}}} \Rnode{C}{\hat{\bs{\nabla}}}{}^{2l-p+1} \Rnode{D}{\hat{\bs{S}}} \simeq - \hat{R}^{k-2} \hat{\bs{\nabla}}{}^{p} \hat{R} \hat{\bs{\nabla}}{}^{2l-p} \hat{\bs{S}} + P_{k+1,2l-2}(\hat{\bs{g}},\hat{\bs{R}},\hat{\bs{\nabla}}) + \text{div}\;,
                \ncbar[linewidth=.5pt,angle=-90,nodesep=1pt,arm=2pt,offsetA=0pt,offsetB=0pt]{C}{B}
            \\
            &\quad \hat{R}^{k-2} \Rnode{A}{\hat{\bs{\nabla}}}{}^{q} \Rnode{B}{\hat{\bs{S}}} \Rnode{C}{\hat{\bs{\nabla}}}{}^{2l-q} \Rnode{D}{\hat{\bs{S}}} \simeq (-1)^q R^{k-2} \hat{S}_{ab} \hat{\Box}{}^l \hat{S}^{ab} + P_{k+1,2l-2}(\hat{\bs{g}},\hat{\bs{R}},\hat{\bs{\nabla}}) + \text{div}\;,
                \ncbar[linewidth=.5pt,angle=-90,nodesep=1pt,arm=4pt,offsetA=-1.1pt,offsetB=-1.1pt]{B}{D}
                \ncbar[linewidth=.5pt,angle=-90,nodesep=1pt,arm=2pt,offsetA=1.1pt,offsetB=1.1pt]{B}{D}
            \\
            \hat{R}^{k-2} \hat{\bs{\nabla}}{}^{q} \hat{\bs{S}} \hat{\bs{\nabla}}{}^{2l-q} \hat{\bs{C}}:&
            \quad \hat{R}^{k-2} \hat{\bs{\nabla}}{}^{q} \Rnode{A}{\hat{\bs{S}}} \hat{\bs{\nabla}}{}^{2l-q} \Rnode{B}{\hat{\bs{C}}} =
                \ncbar[linewidth=.5pt,angle=-90,nodesep=1pt,arm=2pt,offsetA=-1.1pt,offsetB=-1.1pt]{A}{A}
            \hat{R}^{k-2} \Rnode{A}{\hat{\bs{\nabla}}}{}^{q} \Rnode{B}{\hat{\bs{S}}} \Rnode{C}{\hat{\bs{\nabla}}}{}^{2l-q} \Rnode{D}{\hat{\bs{C}}} = 0\;,
                \ncbar[linewidth=.5pt,angle=-90,nodesep=1pt,arm=2pt,offsetA=-1.1pt,offsetB=-1.1pt]{D}{D}
            \\
            &\quad \hat{R}^{k-2} \Rnode{A}{\hat{\bs{\nabla}}}{}^{p} \Rnode{B}{\hat{\bs{S}}} \Rnode{C}{\hat{\bs{\nabla}}}{}^{2l-p} \Rnode{D}{\hat{\bs{C}}} = \hat{R}^{k-2} \hat{\bs{\nabla}}{}^{p} \hat{R} \hat{\bs{\nabla}}{}^{2l-p} \hat{\bs{C}} + P_{k+1,2l-2}(\hat{\bs{g}},\hat{\bs{R}},\hat{\bs{\nabla}})\;,
                \ncbar[linewidth=.5pt,angle=-90,nodesep=1pt,arm=2pt,offsetA=0pt,offsetB=0pt]{A}{B}
            \\
            &\quad \hat{R}^{k-2} \Rnode{A}{\hat{\bs{\nabla}}}{}^{p-1} \Rnode{B}{\hat{\bs{S}}} \Rnode{C}{\hat{\bs{\nabla}}}{}^{2l-p+1} \Rnode{D}{\hat{\bs{C}}} = - \hat{R}^{k-2} \hat{\bs{\nabla}}{}^{p} \hat{R} \hat{\bs{\nabla}}{}^{2l-p} \hat{\bs{C}} + P_{k+1,2l-2}(\hat{\bs{g}},\hat{\bs{R}},\hat{\bs{\nabla}}) + \text{div}\;,
                \ncbar[linewidth=.5pt,angle=-90,nodesep=1pt,arm=2pt,offsetA=0pt,offsetB=0pt]{C}{B}
            \\
            &\quad \hat{R}^{k-2} \Rnode{A}{\hat{\bs{\nabla}}}{}^{q} \Rnode{B}{\hat{\bs{S}}} \Rnode{C}{\hat{\bs{\nabla}}}{}^{2l-q} \Rnode{D}{\hat{\bs{C}}} \simeq (-1)^q R^{k-2} \left( \tfrac{N-3}{N-2} \hat{S}_{ab} \hat{\Box}^l \hat{S}^{ab} - \tfrac{N-3}{N(N-1)} \hat{R} \hat\Box^l \hat{R} \right) \\
            &\quad\quad + P_{k+1,2l-2}(\hat{\bs{g}},\hat{\bs{R}},\hat{\bs{\nabla}}) + \text{div}\;,
                \ncbar[linewidth=.5pt,angle=-90,nodesep=1pt,arm=4pt,offsetA=-1.1pt,offsetB=-1.1pt]{B}{D}
                \ncbar[linewidth=.5pt,angle=-90,nodesep=1pt,arm=2pt,offsetA=1.1pt,offsetB=1.1pt]{B}{D}
        \end{aligned}
        \end{equation}
        where we introduced an arrow to our graphical notation denoting a contraction with any covariant derivative in the expression.
        The covariant derivative contracted with $\hat{\bs{S}}$ in $\Rnode{A}{\hat{\bs{\nabla}}}{}^p \Rnode{B}{\hat{\bs{S}}}$ 
        is moved to act directly on $\hat{\bs{S}}$ which introduces $P_{k+1,2l-2}$ terms and then the contracted Bianchi identity \eqref{eq:BianchiIdDS} is employed. The case when $\hat{\bs{S}}$ is contracted with $\hat{\bs{\nabla}}$ standing in front of $\hat{\bs{X}}$ in  $\hat{\bs{\nabla}}{}^{p} \Rnode{A}{\hat{\bs{S}}} \Rnode{B}{\hat{\bs{\nabla}}}{}^{2l-p} \hat{\bs{X}}$
        can be translated to the previous one by commuting the contracted derivative to be the outermost one of $\hat{\bs{X}}$ and then moving it to act on $\hat{\bs{S}}$. If two $\hat{\bs{S}}$ are fully contracted with each other in $\Rnode{A}{\hat{\bs{\nabla}}}{}^{q} \Rnode{B}{\hat{\bs{S}}} \Rnode{C}{\hat{\bs{\nabla}}}{}^{2l-q} \Rnode{D}{\hat{\bs{S}}}$,
        the derivatives can be swapped between both tensors to act on a single $\hat{\bs{S}}$ and rearranged to form $\hat{\Box}^l$. The contraction of symmetric $\hat{\bs{S}}$ with an anti-symmetric pair of indices of $\hat{\bs{C}}$ vanishes, therefore, the only non-vanishing full contraction of $\hat{\bs{S}}$ with $\hat{\bs{C}}$ in $\Rnode{A}{\hat{\bs{\nabla}}}{}^{q} \Rnode{B}{\hat{\bs{S}}} \Rnode{C}{\hat{\bs{\nabla}}}{}^{2l-q} \Rnode{D}{\hat{\bs{C}}}$
        after some manipulations with derivatives leads to $\hat{S}_{ab} \hat{\Box}^{l-1} \hat{\nabla}_c \hat{\nabla}_d \hat{C}^{acbd}$ which can be rewritten using a divergence of the contracted Bianchi identity
        \begin{equation}\label{eq:BianchiIdDDC2}
            \hat{\nabla}_{c}\hat{\nabla}_{d}\hat{C}_a{}^{c}{}_{b}{}^{d} = \tfrac{N-3}{N-2} \hat{\Box} \hat{S}_{ab}
            + \tfrac{N-3}{2N(N-1)} \hat{g}_{ab} \hat{\Box} \hat{R}
            - \tfrac{N-3}{2(N-1)} \hat{\nabla}_{a}\hat{\nabla}_{b}\hat{R} + P_{2,0}(\hat{\bs{g}}, \hat{\bs{R}})\;.
        \end{equation}

        Finally, the scalars involving $\hat{\bs{C}}$ without $\hat{\bs{S}}$ read
        \begin{equation}\label{eq:equivL:DXDC}
        \begin{aligned}
            \hat{R}^{k-2} \hat{\bs{\nabla}}{}^{p} \hat{R} \hat{\bs{\nabla}}{}^{2l-p} \hat{\bs{C}}:& 
            \quad \hat{R}^{k-2} \hat{\bs{\nabla}}{}^{p} \hat{R} \hat{\bs{\nabla}}{}^{2l-p} \Rnode{A}{\hat{\bs{C}}} = 
                \ncbar[linewidth=.5pt,angle=-90,nodesep=1pt,arm=2pt,offsetA=-3.3pt,offsetB=1.1pt]{-}{A}{A}
                \ncbar[linewidth=.5pt,angle=-90,nodesep=1pt,arm=2pt,offsetA=1.1pt,offsetB=-3.3pt]{-}{A}{A}
            \hat{R}^{k-2} \hat{\bs{\nabla}}{}^{p} \hat{R} \Rnode{A}{\hat{\bs{\nabla}}}{}^{2l-p} \Rnode{B}{\hat{\bs{C}}}
                \ncbar[linewidth=.5pt,angle=-90,nodesep=1pt,armA=0pt,armB=2pt,offsetA=-3.3pt,offsetB=3.3pt,arrowsize=1.8pt,arrowlength=1,arrowinset=0,ArrowFill=false]{>-}{B}{B}
                \ncbar[linewidth=.5pt,angle=-90,nodesep=1pt,armA=0pt,armB=2pt,offsetA=-1.1pt,offsetB=1.1pt,arrowsize=1.8pt,arrowlength=1,arrowinset=0,ArrowFill=false]{>-}{B}{B}
                \ncbar[linewidth=.5pt,angle=-90,nodesep=1pt,arm=2pt,offsetA=1.1pt,offsetB=-3.3pt]{-}{B}{B}
            = 0\;,
            \\
            &\quad \hat{R}^{k-2} \Rnode{A}{\hat{\bs{\nabla}}}{}^{p} \hat{R} \hat{\bs{\nabla}}{}^{2l-p} \Rnode{B}{\hat{\bs{C}}} \simeq 0 + P_{k+1,2l-2}(\hat{\bs{g}},\hat{\bs{R}},\hat{\bs{\nabla}}) + \text{div}\;,
                \ncbar[linewidth=.5pt,angle=-90,nodesep=1pt,armA=0pt,armB=2pt,offsetA=3.3pt,offsetB=-3.3pt,arrowsize=1.8pt,arrowlength=1,arrowinset=0,ArrowFill=false]{>-}{B}{B}
                \ncbar[linewidth=.5pt,angle=-90,nodesep=1pt,armA=0pt,armB=2pt,offsetA=1.1pt,offsetB=-1.1pt,arrowsize=1.8pt,arrowlength=1,arrowinset=0,ArrowFill=false]{>-}{B}{B}
                \ncbar[linewidth=.5pt,angle=-90,nodesep=1pt,armA=0pt,armB=2pt,offsetA=-1.1pt,offsetB=1.1pt,arrowsize=1.8pt,arrowlength=1,arrowinset=0,ArrowFill=false]{>-}{B}{B}
                \ncbar[linewidth=.5pt,angle=-90,nodesep=1pt,armA=0pt,armB=2pt,offsetA=-3.3pt,offsetB=3.3pt,arrowsize=1.8pt,arrowlength=1,arrowinset=0,ArrowFill=false]{>-}{B}{B}
            \\                
            \hat{R}^{k-2} \hat{\bs{\nabla}}{}^{q} \hat{\bs{C}} \hat{\bs{\nabla}}{}^{2l-q} \hat{\bs{C}}:&
            \quad \hat{R}^{k-2} \Rnode{A}{\hat{\bs{\nabla}}}{}^{q} \Rnode{B}{\hat{\bs{C}}} \Rnode{C}{\hat{\bs{\nabla}}}{}^{2l-q} \Rnode{D}{\hat{\bs{C}}} =
                \ncbar[linewidth=.5pt,angle=-90,nodesep=1pt,arm=2pt,offsetA=-1.1pt,offsetB=-1.1pt]{B}{B}
            \hat{R}^{k-2} \Rnode{A}{\hat{\bs{\nabla}}}{}^{q} \Rnode{B}{\hat{\bs{C}}} \Rnode{C}{\hat{\bs{\nabla}}}{}^{2l-q} \Rnode{D}{\hat{\bs{C}}} = 0\;,
                \ncbar[linewidth=.5pt,angle=-90,nodesep=1pt,arm=2pt,offsetA=-1.1pt,offsetB=-1.1pt]{D}{D}
            \\
            &\quad \hat{R}^{k-2} \Rnode{A}{\hat{\bs{\nabla}}}{}^{p} \Rnode{B}{\hat{\bs{C}}} \Rnode{C}{\hat{\bs{\nabla}}}{}^{2l-p} \Rnode{D}{\hat{\bs{C}}} = \pm \tfrac{N-2}{N-2} \hat{R}^{k-2} \hat{\bs{\nabla}}{}^p \hat{\bs{S}} \hat{\bs{\nabla}}{}^{2l-p} \hat{\bs{C}} \pm \tfrac{N-3}{2N(N-1)} \hat{R}^{k-2} \hat{\bs{\nabla}}{}^p \hat{R} \hat{\bs{\nabla}}{}^{2l-p} \hat{\bs{C}} \\
            &\quad\quad + P_{k+1,2l-2}(\hat{\bs{g}},\hat{\bs{R}},\hat{\bs{\nabla}})\;,
                \ncbar[linewidth=.5pt,angle=-90,nodesep=1pt,arm=2pt,offsetA=0pt,offsetB=0pt]{A}{B}
            \\
            &\quad \hat{R}^{k-2} \Rnode{A}{\hat{\bs{\nabla}}}{}^{p-1} \Rnode{B}{\hat{\bs{C}}} \Rnode{C}{\hat{\bs{\nabla}}}{}^{2l-p+1} \Rnode{D}{\hat{\bs{C}}} = \mp \tfrac{N-2}{N-2} \hat{R}^{k-2} \hat{\bs{\nabla}}{}^p \hat{\bs{S}} \hat{\bs{\nabla}}{}^{2l-p} \hat{\bs{C}} \\
            &\quad\quad \mp \tfrac{N-3}{2N(N-1)} \hat{R}^{k-2} \hat{\bs{\nabla}}{}^p \hat{R} \hat{\bs{\nabla}}{}^{2l-p} \hat{\bs{C}} + P_{k+1,2l-2}(\hat{\bs{g}},\hat{\bs{R}},\hat{\bs{\nabla}}) + \text{div}\;,
                \ncbar[linewidth=.5pt,angle=-90,nodesep=1pt,arm=2pt,offsetA=0pt,offsetB=0pt]{C}{B}
            \\
            &\quad \hat{R}^{k-2} \Rnode{A}{\hat{\bs{\nabla}}}{}^{q} \Rnode{B}{\hat{\bs{C}}} \Rnode{C}{\hat{\bs{\nabla}}}{}^{2l-q} \Rnode{D}{\hat{\bs{C}}} \simeq (-1)^q R^{k-2} \hat{C}_{abcd} \hat\Box^l \hat{C}^{abcd} + P_{k+1,2l-2}(\hat{\bs{g}},\hat{\bs{R}},\hat{\bs{\nabla}}) + \text{div}\;,
                \ncbar[linewidth=.5pt,angle=-90,nodesep=1pt,arm=2pt,offsetA=3.3pt,offsetB=3.3pt]{B}{D}
                \ncbar[linewidth=.5pt,angle=-90,nodesep=1pt,arm=4pt,offsetA=1.1pt,offsetB=1.1pt]{B}{D}
                \ncbar[linewidth=.5pt,angle=-90,nodesep=1pt,arm=6pt,offsetA=-1.1pt,offsetB=-1.1pt]{B}{D}
                \ncbar[linewidth=.5pt,angle=-90,nodesep=1pt,arm=8pt,offsetA=-3.3pt,offsetB=-3.3pt]{B}{D}
        \end{aligned}
        \end{equation}
        where the terms $\hat{\bs{\nabla}}{}^p \hat{R} \hat{\bs{\nabla}}{}^{2l-p} \Rnode{A}{\hat{\bs{C}}}$
        with $\hat{\bs{C}}$ fully contracted with derivatives are simplified using \eqref{eq:BianchiIdDDC1}. The derivative contracted with $\hat{\bs{C}}$ in $\Rnode{A}{\hat{\bs{\nabla}}}{}^{p} \Rnode{B}{\hat{\bs{C}}}$
        is moved to act directly on $\hat{\bs{C}}$ and the resulting expression is subsequently rewritten with the help of the contracted Bianchi identity
        \begin{equation}\label{eq:BianchiIdDC}
            \hat{\nabla}^d \hat{C}_{abcd} = - 2 \tfrac{N-3}{N-2} \hat{\nabla}_{[a}\hat{S}_{b]c} - \tfrac{N-3}{N(N-1)} \hat{g}_{c[b} \hat{\nabla}_{a]}\hat{R}\;.
        \end{equation}
        The terms $\Rnode{A}{\hat{\bs{\nabla}}}{}^{p-1} \Rnode{B}{\hat{\bs{C}}} \Rnode{C}{\hat{\bs{\nabla}}}{}^{2l-p+1} \Rnode{D}{\hat{\bs{C}}}$
        where $\hat{\bs{C}}$ is contracted with $\hat{\bs{\nabla}}$ standing in front of the other $\hat{\bs{C}}$ can be rearrange to the form of the previous case. When two $\hat{\bs{C}}$ are fully contracted with each other, the derivatives in 
        $\Rnode{A}{\hat{\bs{\nabla}}}{}^{q} \Rnode{B}{\hat{\bs{C}}} \Rnode{C}{\hat{\bs{\nabla}}}{}^{2l-q} \Rnode{D}{\hat{\bs{C}}}$ 
        are first moved to act on a single $\hat{\bs{C}}$ and form $\hat{\Box}^l$, then the same argument as in \eqref{eq:equivL:SC} is followed.
\end{enumerate}

In other words, we have just shown that any $L_{k,2l}(\hat{\bs{g}},\hat{\bs{R}},\hat{\bs{\nabla}})$
is equivalent to 
\begin{equation}
    L_{k,2l}(\hat{\bs{g}},\hat{\bs{R}},\hat{\bs{\nabla}}) \simeq \text{const}\, \hat{R} \hat\Box^l \hat{R} + \text{const}\, \hat{{S}}_{ab} \hat\Box^l \hat{{S}}{}^{ab} + \text{const}\, \hat{{C}}_{abcd} \hat\Box^l \hat{{C}}{}^{abcd} + P_{k+1,2l-2}(\hat{\bs{g}},\hat{\bs{R}},\hat{\bs{\nabla}}) + \text{div}
\end{equation}
and $P_{k+1,2l-2}(\hat{\bs{g}},\hat{\bs{R}},\hat{\bs{\nabla}})$ can be rewritten again using \eqref{eq:equivL:DX}, \eqref{eq:equivL:DRDR}, \eqref{eq:equivL:DXDS} and \eqref{eq:equivL:DXDC}.
In this way, we iterate over a decreasing number of derivatives until we end up with $P_{k+l,0}(\hat{\bs{g}},\hat{\bs{R}})$ which takes the form \eqref{eq:Lk0}. Hence,
\begin{equation}
    L_{k,2l}(\hat{\bs{g}},\hat{\bs{R}},\hat{\bs{\nabla}}) \simeq  \text{const} + \text{const}\,\hat{R} + \hat{R}  \mathcal{F}_{1}^{k,2l}(\hat\Box) \hat{R} + \hat{{S}}_{ab}  \mathcal{F}_{2}^{k,2l}(\hat\Box) \hat{{S}}{}^{ab} + \hat{{C}}_{abcd} \mathcal{F}_{3}^{k,2l}(\hat\Box)  \hat{{C}}{}^{abcd} + \text{div}\;,
\end{equation}
where $\mathcal{F}_{i}^{k,2l}(\hat\Box)$ is an operator polynomial in $\hat\Box$ of the order $l$.
The action \eqref{eq:analyticlagrangian} with any analytic Lagrangian \eqref{eq:genericL} is thus equivalent at the level of the second-order variations around an MS background to \eqref{eq:IDG} with
\begin{equation}
    \mathcal{F}_{i}(\hat\Box):=\sum_{k,l}\mathcal{F}_{i}^{k,2l}(\hat\Box)\;,
\end{equation}
i.e., these actions differ only by $O(\varepsilon^3)$ terms.

\appsection{Quadratic action} \label{apx:quadraticaction}

In this appendix, we generalize the derivation of the quadratic action \eqref{eq:quadaction} in $N=3,4$ \cite{Biswas:2016egy,Biswas:2016etb,Mazumdar:2018xjz} (see also \cite{Koshelev:2016xqb,SravanKumar:2019eqt}) to an arbitrary dimension.
For convenience, we will write the equivalent action \eqref{eq:IDG} as
\begin{equation}
    S_{\textrm{equiv}}^{\textrm{E}}[\hat{\bs{g}}]=\int_{\hat{M}}\hat{\mathfrak{g}}^{\nicefrac{1}{2}} \left[ L_{\text{EH+$\Lambda$}} + L_{\hat{R}^2} + L_{\hat{S}^2} + L_{\hat{C}^2} \right]
\end{equation}
with the Lagrangian densities given by
\begin{equation}
    L_{\text{EH+$\Lambda$}} = -\tfrac{1}{2\varkappa} (\hat{R} - 2\Lambda)\;, \quad
    L_{\hat{R}^2} = -\tfrac12 \hat{R} \mathcal{F}_1(\hat\Box) \hat{R}\;, \quad
    L_{\hat{S}^2} = -\tfrac12 \hat{S}_{ab} \mathcal{F}_2(\hat\Box) \hat{S}^{ab}\;, \quad 
    L_{\hat{C}^2} = -\tfrac12 \hat{C}_{abcd} \mathcal{F}_3(\hat\Box) \hat{C}^{abcd}\;.
\end{equation}
Below, we show that second variation of the equivalent action around an MS background $\delta^2 S_{\textrm{equiv}}^{\textrm{E}}[\bs{h}]$, provided the background metric $\bs{g}$ satisfies the vacuum field equations (i.e.\ $\delta S_{\textrm{equiv}}^{\textrm{E}}/\delta\hat{\bs{g}} = 0$),
gives \eqref{eq:quadaction} with $\bs{\tau} = 0$ and
\begin{equation}\label{eq:quadactionEI}
\begin{aligned}
    \mathcal{E}({\Box}) &= 1 + \varkappa\big[2R\mathcal{F}_1(0) + \big(\Box - \tfrac{2}{N(N-1)}R\big) \mathcal{F}_2({\Box}) + 4 \tfrac{N-3}{N-2} \big({\Box} - \tfrac{1}{N-1}R\big) \mathcal{F}_3\big({\Box} + \tfrac{2(N-2)}{N(N-1)}R\big)\big]\;, \\
    \mathcal{I}({\Box}) &= 1 + \varkappa\big[2R \mathcal{F}_1(0) - \tfrac{4(N-1)}{N-2} \big({\Box} + \tfrac{1}{N-1}R\big) \mathcal{F}_1({\Box}) - \tfrac{N-2}{N} {\Box} \mathcal{F}_2\big({\Box} + \tfrac{2}{N-1}R\big)\big]\;.
\end{aligned}
\end{equation}

\subsection{First variation of equivalent action}

Variation of the Einstein--Hilbert term with cosmological constant leads to the well-known result which on MS backgrounds gives
\begin{equation}
    2 \varkappa \delta \left({\hat{\mathfrak{g}}}^{\nicefrac{1}{2}} L_\text{EH+$\Lambda$} \right) = {\mathfrak{g}}^{\nicefrac{1}{2}} \left( R_{ab} - \tfrac{1}{2} g_{ab} R + \Lambda g_{ab} \right) h^{ab} + {\mathfrak{g}}^{\nicefrac{1}{2}} \,\text{div} = {\mathfrak{g}}^{\nicefrac{1}{2}} \left( \Lambda - \tfrac{N - 2}{2N} R \right) h + {\mathfrak{g}}^{\nicefrac{1}{2}} \,\text{div}\;.
\end{equation}

In the parts of $S_{\textrm{equiv}}^{\textrm{E}}[\hat{\bs{g}}]$ quadratic in curvature, we have to deal with variation of the form-factors. Starting with $\hat\Box$ acting on a scalar function $f$, one can show
\begin{equation}\label{eq:varBoxf}
    \delta(\hat\Box) f = \delta(\hat\Box \hat{f}) - \Box \delta \hat{f} = - \nabla_a ( h^{ab} \nabla_b f) + \tfrac12 \nabla^a h \nabla_a f\;.
\end{equation}
For a constant background Ricci scalar $R$, we thus obtain $\delta(\hat\Box)R = 0$ and subsequently $\delta(\hat\Box^{n+1})R = \delta(\hat\Box^n) \Box R + \Box^n \delta(\hat\Box) R = 0$. Hence,
\begin{equation}\label{eq:varFR}
    \delta(\mathcal{F}_i(\hat\Box)) R = 0\;.
\end{equation}
Note also that 
\begin{equation}\label{eq:FR}
    \mathcal{F}_i (\Box) R = f_{i,0} R = \mathcal{F}_i(0) R\;.
\end{equation}

Variation of the $\hat{R}^2$ part of $S_{\textrm{equiv}}^{\textrm{E}}[\hat{\bs{g}}]$ reads
\begin{equation}
\begin{aligned}
    -2 \delta \left({\hat{\mathfrak{g}}}^{\nicefrac{1}{2}} L_{\hat{R}^2} \right) &= \delta \left( {\hat{\mathfrak{g}}}^{\nicefrac{1}{2}} \hat{R} \mathcal{F}_1(\hat\Box) \hat{R} \right)
      = \delta{\hat{\mathfrak{g}}}^{\nicefrac{1}{2}} R \mathcal{F}_1(\Box) R + {\mathfrak{g}}^{\nicefrac{1}{2}} \delta\hat{R} \mathcal{F}_1(\Box) R + {\mathfrak{g}}^{\nicefrac{1}{2}} R \delta(\mathcal{F}_1(\hat\Box)) R + {\mathfrak{g}}^{\nicefrac{1}{2}} R \mathcal{F}_1(\Box) \delta\hat{R} \\
      &= \left( \delta{\hat{\mathfrak{g}}}^{\nicefrac{1}{2}} R + 2 {\mathfrak{g}}^{\nicefrac{1}{2}} \delta\hat{R} \right) \mathcal{F}_1(0) R + {\mathfrak{g}}^{\nicefrac{1}{2}}\,\text{div}\;,
\end{aligned}
\end{equation}
where we employed \eqref{eq:varFR} and \eqref{eq:FR}. In the last term, only $\mathcal{F}_1(0) \delta\hat{R}$ survives from $\mathcal{F}_1(\Box) \delta\hat{R}$ since the terms starting with $\Box$ are just total derivative. Plugging the variation of the metric determinant \eqref{eq:varg} and Ricci scalar \eqref{eq:varR}, one gets
\begin{equation}
    -2 \delta \left( {\hat{\mathfrak{g}}}^{\nicefrac{1}{2}} L_{\hat{R}^2} \right) = {\mathfrak{g}}^{\nicefrac{1}{2}} \left( \tfrac{N-4}{2N} \mathcal{F}_1(0) R^2 h + \text{div} \right)\;.
\end{equation}

Variations $\delta ({\hat{\mathfrak{g}}}^{\nicefrac{1}{2}} L_{\hat{S}^2})$ and $\delta ({\hat{\mathfrak{g}}}^{\nicefrac{1}{2}} L_{\hat{C}^2})$ do not contribute at all since there always appears at least one trace-free Ricci tensor $\bs{S}$ or Weyl tensor $\bs{C}$ which is not varied and vanish when evaluated on the background.
The vacuum field equations for the background $\bs{g}$ following from $\delta S_{\textrm{equiv}}^{\textrm{E}}[\bs{h}] = 0$ then read
\begin{equation}\label{eq:bgFEs}
    \Lambda - \tfrac{N-2}{2N} R - \tfrac{N-4}{2N} \varkappa \mathcal{F}_1(0) R^2 = 0\;.
\end{equation}

\subsection{Second variation of equivalent action}

Decomposing the metric perturbation $\bs{h}$ into the TT tensor $\bs{h}^{\perp}$, transverse covector $\bs{h}^{\asymp}$, and two scalars $h^{\bowtie}$ and $h^{\top}$ \eqref{eq:decomp}, the second variation of the Einstein--Hilbert term with cosmological constant can be expressed as
\begin{equation}\label{eq:var2LEH}
\begin{aligned}
    -2 \varkappa \delta^2 \left( {\hat{\mathfrak{g}}}^{\nicefrac{1}{2}} L_\text{EH+$\Lambda$} \right) &= \delta^2 \left( {\hat{\mathfrak{g}}}^{\nicefrac{1}{2}}(\hat{R} - 2\Lambda) \right) = \delta^2 \left( {\hat{\mathfrak{g}}}^{\nicefrac{1}{2}} \hat{R} \right) - 2\Lambda \delta^2 {\hat{\mathfrak{g}}}^{\nicefrac{1}{2}} \\
    &= {\mathfrak{g}}^{\nicefrac{1}{2}} \Big[ \tfrac{1}{2} h^{ab} \Box h_{ab} - \tfrac{1}{2} h \Box h
      + h \nabla_{c}\nabla_{b}h^{bc} + \nabla_{a}h^{ab} \nabla_{c}h_{b}{}^{c}  \\
    &\quad  - \left( \tfrac{N^2 - 3N + 4}{2N(N - 1)} R - \Lambda \right) h{}_{ab} h^{ab}
      + \tfrac12 \left( \tfrac{N^2 - 5N + 8} {2N(N - 1)} R - \Lambda \right) h^2 + \text{div} \Big] \\
    &= {\mathfrak{g}}^{\nicefrac{1}{2}} \left[ \tfrac12 h^\perp_{ab} \left( \Box - \tfrac{2}{N(N - 1)}R + 2 \Xi \right) h^{{\perp}ab}
      - \tfrac{N - 2}{2N} h^\top \left( \tfrac{N - 1}{N} \Box + \tfrac{1}{N}R + \Xi \right) h^\top + \text{div} \right] \\
    &\quad - 2 \Xi {\mathfrak{g}}^{\nicefrac{1}{2}} \left[ h^{\asymp}_a \left( \Box + \tfrac{1}{N}R \right) h^{{\asymp}a}
      - \tfrac14 h^{\bowtie} \left( \Box + \tfrac{2}{N}R \right) \Box h^{\bowtie} + \tfrac{N - 2}{2N} h^\top \Box h^{\bowtie} \right]\;,
\end{aligned}
\end{equation}
where
\begin{equation}
    \Xi := \Lambda - \tfrac{N - 2}{2N} R\;.
\end{equation}
In GR, the second variation of the Einstein--Hilbert action admits only the TT tensor $\bs{h}^\perp$ and scalar $h^\top$ degrees of freedom propagating on an MS background since the vacuum background field equations imply $\Xi = 0$. Although, the Einstein--Hilbert part contributes also to the transverse covector $\bs{h}^{\asymp}$ and scalar $h^{\bowtie}$ modes in the quadratic action $S_{\textrm{quad}}^{\textrm{E}}[\bs{h}]$, these modes cancel out with the contribution from $L_{\hat{R}^2}$ as we will see later.

Similarly as \eqref{eq:varFR}, one can show that 
\begin{equation}\label{eq:var2FR}
    \delta^2(\mathcal{F}_i(\hat\Box)) R = 0\;,
\end{equation}
which follows from the second variation of $\hat\Box$ acting on a scalar function
\begin{equation}
    \delta^2(\hat\Box) f = \delta^2(\hat\Box \hat{f}) - 2 \delta(\hat\Box\delta \hat{f}) + \Box \delta^2 \hat{f}
    = - h^{bc} \nabla^a h_{bc} \nabla_a f - h^{ab} \nabla_{b} h \nabla_a f
    + 2 \nabla_{b}(h^{ac} h^b{}_c \nabla_{a}f)\;,
\end{equation}
when applied for a constant background Ricci scalar $R$, i.e.\ $\delta^2(\hat\Box)R = 0$ and subsequently 
$\delta^2(\hat\Box^{n+1})R = \delta^2(\hat\Box^n) \Box R + 2 \delta(\hat\Box^n) \delta(\Box) R + \Box^n \delta^2(\hat\Box) R = 0$.

Now, we are able to express the second variation of ${\hat{\mathfrak{g}}}^{\nicefrac{1}{2}} L_{\hat{R}^2}$ 
\begin{equation}
\begin{aligned}
    -2 \delta^2 \left( {\hat{\mathfrak{g}}}^{\nicefrac{1}{2}} L_{\hat{R}^2} \right) &= \delta^2 \left( {\hat{\mathfrak{g}}}^{\nicefrac{1}{2}} \hat{R} \mathcal{F}_1(\hat\Box) \hat{R} \right) = \delta^2 \left( {\hat{\mathfrak{g}}}^{\nicefrac{1}{2}} \hat{R} \right) \mathcal{F}_1(\Box) R + {\mathfrak{g}}^{\nicefrac{1}{2}} R \delta^2(\mathcal{F}_1(\hat\Box)) R + {\mathfrak{g}}^{\nicefrac{1}{2}} R \mathcal{F}_1(\Box) \delta^2 \hat{R} \\
    &\quad + 2 \delta \left( {\hat{\mathfrak{g}}}^{\nicefrac{1}{2}} \hat{R} \right) \delta(\mathcal{F}_1(\hat\Box)) R + 2 \delta{\hat{\mathfrak{g}}}^{\nicefrac{1}{2}} R \mathcal{F}_1(\Box) \delta\hat{R} + 2 {\mathfrak{g}}^{\nicefrac{1}{2}} \delta\hat{R} \mathcal{F}_1(\Box) \delta\hat{R} + 2 {\mathfrak{g}}^{\nicefrac{1}{2}} R \delta(\mathcal{F}_1(\hat\Box)) \delta\hat{R}\;,
\end{aligned}
\end{equation}
where the second and fourth term vanishes due to \eqref{eq:var2FR} and \eqref{eq:varFR}, respectively. The third term can be rewritten as $R\mathcal{F}_1(\Box)\delta^2 R = \mathcal{F}_1(\Box)R \delta^2 R + \text{div}$ and using \eqref{eq:FR} we get
\begin{equation}
\begin{aligned}
    -2 \delta^2 \left( {\hat{\mathfrak{g}}}^{\nicefrac{1}{2}} L_{\hat{R}^2} \right) &= \delta^2 \left( {\hat{\mathfrak{g}}}^{\nicefrac{1}{2}} \hat{R} \right) \mathcal{F}_1(0) R + {\mathfrak{g}}^{\nicefrac{1}{2}} R \mathcal{F}_1(0) \delta^2\hat{R} + 2 \delta{\hat{\mathfrak{g}}}^{\nicefrac{1}{2}} R \mathcal{F}_1(\Box) \delta\hat{R} \\
    &\quad + 2 {\mathfrak{g}}^{\nicefrac{1}{2}} \delta\hat{R} \mathcal{F}_1(\Box) \delta\hat{R} + 2 {\mathfrak{g}}^{\nicefrac{1}{2}} R \delta(\mathcal{F}_1(\hat\Box)) \delta\hat{R} + {\mathfrak{g}}^{\nicefrac{1}{2}}\,\text{div}\;.
\end{aligned}
\end{equation}
It is convenient to rewrite the second term as ${\mathfrak{g}}^{\nicefrac{1}{2}} \delta^2 R \mathcal{F}_1(0) R = \delta^2({\hat{\mathfrak{g}}}^{\nicefrac{1}{2}} \hat{R}) \mathcal{F}_1(0) R - \delta^2{\hat{\mathfrak{g}}}^{\nicefrac{1}{2}} R \mathcal{F}_1(0) R - 2\delta{\hat{\mathfrak{g}}}^{\nicefrac{1}{2}} \delta\hat{R} \mathcal{F}_1(0) R$. After rearranging terms one obtains
\begin{equation}
\begin{aligned}
    -2 \delta^2 \left( {\hat{\mathfrak{g}}}^{\nicefrac{1}{2}} L_{\hat{R}^2} \right) &= 2 \delta^2 \left( {\hat{\mathfrak{g}}}^{\nicefrac{1}{2}} \hat{R} \right) \mathcal{F}_1(0) R - \delta^2{\hat{\mathfrak{g}}}^{\nicefrac{1}{2}} R \mathcal{F}_1(0) R + 2 {\mathfrak{g}}^{\nicefrac{1}{2}} \delta\hat{R} \mathcal{F}_1(\Box) \delta\hat{R} \\
    &\quad + 2 \delta{\hat{\mathfrak{g}}}^{\nicefrac{1}{2}} R \left( \mathcal{F}_1(\Box) - \mathcal{F}_1(0) \right) \delta\hat{R} + 2 {\mathfrak{g}}^{\nicefrac{1}{2}} R \delta(\mathcal{F}_1(\hat\Box)) \delta\hat{R} + {\mathfrak{g}}^{\nicefrac{1}{2}}\,\text{div}\;.
\end{aligned}
\end{equation}
The first two terms can be recast using $L_\text{EH+$\Lambda$}$ as
\begin{equation}
    \left( 2 \delta^2 ({\hat{\mathfrak{g}}}^{\nicefrac{1}{2}} \hat{R}) - \delta^2{\hat{\mathfrak{g}}}^{\nicefrac{1}{2}} R \right) \mathcal{F}_1(0) R = - \left[4 \varkappa \delta^2 \left( {\hat{\mathfrak{g}}}^{\nicefrac{1}{2}} L_\text{EH+$\Lambda$} \right) + \delta^2{\hat{\mathfrak{g}}}^{\nicefrac{1}{2}} (R - 4 \Lambda) \right] \mathcal{F}_1(0) R\;,
\end{equation}
and the terms on the second line yield a total derivative since
\begin{equation}
\begin{aligned}
    \delta(\mathcal{F}_1(\hat\Box))\delta\hat{R} &= \delta(\mathcal{F}_1(\hat\Box) - \mathcal{F}_1(0))\delta\hat{R} = \delta(\hat\Box) \tfrac{\mathcal{F}_1(\Box) - \mathcal{F}_1(0)}{\Box} \delta\hat{R} + \text{div}\\
    &= \tfrac12 \nabla^a h \nabla_a \tfrac{\mathcal{F}_1(\Box) - \mathcal{F}_1(0)}{\Box} \delta\hat{R} + \text{div} = - \tfrac12 h (\mathcal{F}_1(\Box) - \mathcal{F}_1(0)) \delta\hat{R} + \text{div}\;,
\end{aligned}
\end{equation}
where we employed $\delta(\hat\Box^n)f = \delta(\hat\Box) \Box^{n-1} f + \Box \delta(\hat\Box^{n-1})f = \delta(\hat\Box) \Box^{n-1} f + \text{div}$ and \eqref{eq:varBoxf}.
We thus end up with
\begin{equation}\label{eq:var2LR2}
    -2 \delta^2 \left( {\hat{\mathfrak{g}}}^{\nicefrac{1}{2}} L_{\hat{R}^2} \right) = - \left[4 \varkappa \delta^2 \left({\hat{\mathfrak{g}}}^{\nicefrac{1}{2}} \mathcal{L}_\text{EH+$\Lambda$} \right) + \delta^2{\hat{\mathfrak{g}}}^{\nicefrac{1}{2}} (R - 4 \Lambda) \right] \mathcal{F}_1(0) R + 2 {\mathfrak{g}}^{\nicefrac{1}{2}} \delta\hat{R} \mathcal{F}_1(\Box) \delta\hat{R} + {\mathfrak{g}}^{\nicefrac{1}{2}}\,\text{div}\;.
\end{equation}
Substituting \eqref{eq:var2LEH}, \eqref{eq:var2g} and \eqref{eq:var2R} to \eqref{eq:var2LR2} immediately yields
\begin{equation}
\begin{aligned}
    -2 \delta^2 \left({\hat{\mathfrak{g}}}^{\nicefrac{1}{2}} L_{\hat{R}^2} \right) &= {\mathfrak{g}}^{\nicefrac{1}{2}} \left[ h^\perp_{ab} \left( \Box - \tfrac{2}{N(N - 1)}R - \tfrac{N-4}{2N} R \right) h^{{\perp}ab}
      - \tfrac{N - 2}{N} h^\top \left( \tfrac{N - 1}{N} \Box + \tfrac{1}{N}R - \tfrac{N-4}{4N} R \right) h^\top \right] \mathcal{F}_1(0) R \\
    &\quad + \tfrac{N-4}{N} {\mathfrak{g}}^{\nicefrac{1}{2}} \left[ h^\asymp_a \left( \Box + \tfrac{1}{N}R \right) h^{{\asymp}a}
      - \tfrac14 h^{\bowtie} \left( \Box + \tfrac{2}{N}R \right) \Box h^{\bowtie} + \tfrac{N - 2}{2N} h^\top \Box h^{\bowtie} \right] \mathcal{F}_1(0) R^2 \\
    &\quad + \tfrac{2(N-1)^2}{N^2} {\mathfrak{g}}^{\nicefrac{1}{2}} h^{\bowtie} \left( \Box + \tfrac{1}{N-1}R \right) \mathcal{F}_1(\Box) \left( \Box + \tfrac{1}{N-1}R \right) h^{\bowtie} + {\mathfrak{g}}^{\nicefrac{1}{2}}\,\text{div}\;.
\end{aligned}
\end{equation}
Obviously, the second variations of the Einstein--Hilbert term \eqref{eq:var2LEH} and $\hat{R}^2$ part $\delta^2({\hat{\mathfrak{g}}}^{\nicefrac{1}{2}} L_{\hat{R}^2})$ give raise to transverse covector $\bs{h}^\asymp$ and scalar modes $h^{\bowtie}$ in the quadratic action $S_{\textrm{quad}}^{\textrm{E}}[\bs{h}]$. However, these modes cancel out if the MS background satisfies the vacuum field equations \eqref{eq:bgFEs} as can be seen explicitly from
\begin{equation}\label{eq:var2LEHR2}
\begin{aligned}
    -\delta^2 \left( {\hat{\mathfrak{g}}}^{\nicefrac{1}{2}} L_{\text{EH+$\Lambda$+$\hat{R}^2$}} \right) &= \tfrac{1}{2} {\mathfrak{g}}^{\nicefrac{1}{2}} \, h^\perp_{ab} \left( \tfrac{1}{2\varkappa} + \mathcal{F}_1(0) R \right) \left( \Box - \tfrac{2}{N(N - 1)}R + \Xi - \tfrac{N-4}{2N} \varkappa \mathcal{F}_1(0) R^2 \right) h^{{\perp}ab} \\
    &\quad - \tfrac{N - 2}{2N^2} {\mathfrak{g}}^{\nicefrac{1}{2}} \, h^\top \left( \tfrac{1}{2\varkappa} + \mathcal{F}_1(0) R \right) \left[ (N - 1) \Box + R + N \left( \Xi - \tfrac{N-4}{2N} \varkappa \mathcal{F}_1(0) R^2 \right) \right] h^\top \\
    &\quad + \tfrac{1}{N^2} {\mathfrak{g}}^{\nicefrac{1}{2}} \, h^\top \left( (N-1) \Box + R \right) \mathcal{F}_1(\Box) \left( (N-1) \Box + R \right) h^\top + {\mathfrak{g}}^{\nicefrac{1}{2}}\,\text{div} \\
    &\quad - \tfrac{1}{\varkappa} \left( \Xi - \tfrac{N-4}{2N} \varkappa \mathcal{F}_1(0) R^2 \right) {\mathfrak{g}}^{\nicefrac{1}{2}} \left[ h^\asymp_a \left( \Box + \tfrac{1}{N}R \right) h^{{\asymp}a}
      - \tfrac14 h^{\bowtie} \left( \Box + \tfrac{2}{N}R \right) \Box h^{\bowtie} + \tfrac{N - 2}{2N} h^\top \bar\Box h^{\bowtie} \right]\;.
\end{aligned}
\end{equation}

Since the trace-free Ricci tensor $\bs{S}$ and Weyl tensor $\bs{C}$ vanish when evaluated on the background, the second variations of ${\hat{\mathfrak{g}}}^{\nicefrac{1}{2}} L_{\hat{S}^2}$ and ${\hat{\mathfrak{g}}}^{\nicefrac{1}{2}} L_{\hat{C}^2}$ involve only terms where both curvature tensors are varied, i.e.,
\begin{equation}\label{eq:var2LS2C2}
\begin{aligned}
    -2 \delta^2 \left({\hat{\mathfrak{g}}}^{\nicefrac{1}{2}} L_{\hat{S}^2}\right) &= \delta^2 \left({\hat{\mathfrak{g}}}^{\nicefrac{1}{2}} \hat{S}_{ab} \mathcal{F}_1(\hat\Box) \hat{S}^{ab} \right) = 2 {\mathfrak{g}}^{\nicefrac{1}{2}} \delta\hat{S}_{ab} \mathcal{F}_1(\Box) \delta\hat{S}^{ab}\;, \\
    -2 \delta^2 \left({\hat{\mathfrak{g}}}^{\nicefrac{1}{2}} L_{\hat{C}^2} \right) &= \delta^2 \left( {\hat{\mathfrak{g}}}^{\nicefrac{1}{2}} \hat{C}_{abcd} \mathcal{F}_1(\hat\Box) \hat{C}^{abcd} \right) = 2 {\mathfrak{g}}^{\nicefrac{1}{2}} \delta\hat{C}_{abcd} \mathcal{F}_1(\Box) \delta\hat{C}^{abcd}\;.
\end{aligned}
\end{equation}
Although $\delta \hat{S}_{ab}$ \eqref{eq:varS} involves both the TT tensor mode $\bs{h}^\perp$ and the scalar mode $h^\top$, it turns out that the mixed terms do not contribute in $\delta\hat{S}_{ab} \mathcal{F}_2(\Box) \delta\hat{S}^{ab}$ since
\begin{equation}
\begin{aligned}
    (\Box + c) h^\perp_{ab} \mathcal{F}_2(\Box) D^{ab} h^\top &=
    (\Box + c) h^\perp_{ab} \mathcal{F}_2(\Box) \left(\nabla^a \nabla^b - \tfrac{1}{N}g^{ab} \Box \right) h^\top = (\Box + c) h^\perp_{ab} \mathcal{F}_2(\Box) \nabla^a \nabla^b h^\top \\
    &= -\nabla^a \mathcal{F}_2(\Box) (\Box + c) h^\perp_{ab} \nabla^b h^\top + \text{div} = \text{div}\;. 
\end{aligned}
\end{equation}
This follows from the tracelessness of $\bs{h}^\perp$ and the fact that $\Box$ on a TT tensor is a TT tensor which is a direct consequence of the commutator \eqref{eq:commutrank2}. The computation of the pure tensor mode $\bs{h}^\perp$ of $\delta\hat{S}_{ab} \mathcal{F}_2(\Box) \delta\hat{S}^{ab}$ is straightforward, on the other hand, for the pure scalar mode $h^\top$ (ommiting the constant factor $\tfrac{(N-2)^2}{4N^2}$) one gets
\begin{equation}
\begin{aligned}
    D_{ab} h^\top  \mathcal{F}_2(\Box) D^{ab}h^\top &= \left( \nabla_a \nabla_b - \tfrac{1}{N}g^{ab} \Box \right) h^\top  \mathcal{F}_2(\Box) D^{ab}h^\top = \nabla_a \nabla_b h^\top  \mathcal{F}_2(\Box) D^{ab}h^\top = h^\top \nabla_b \nabla_a \mathcal{F}_2(\Box) D^{ab}h^\top + \text{div} \\ 
    &= h^\top \nabla_b \mathcal{F}_2\left(\Box + \tfrac{N+1}{N(N-1)}R\right) \nabla_a D^{ab}h^\top + \text{div} \\
    &= h^\top \mathcal{F}_2\left(\Box + \tfrac{N+1}{N(N-1)}R + \tfrac{1}{N}R\right) \nabla_b \left( \Box \nabla^b - \tfrac{1}{N}\nabla^b \Box \right) h^\top + \text{div} \\
    &= h^\top \mathcal{F}_2\left(\Box + \tfrac{2}{N-1}R\right) \left[ \left(\Box + \tfrac{1}{N}R \right) \Box - \tfrac{1}{N}\Box^2 \right] h^\top + \text{div}\;,
\end{aligned}
\end{equation}
where we employed the tracelessness of $D_{ab}$, the commutator for a traceless rank-2 tensor \eqref{eq:commutrank2} to swap the operators $\nabla_a$, $\mathcal{F}_2(\Box)$ and the commutator for a vector \eqref{eq:commutrank1} to move $\nabla_b$ over $\mathcal{F}_2$ and $\Box$. Therefore, the second variation of ${\hat{\mathfrak{g}}}^{\nicefrac{1}{2}} L_{\hat{S}^2}$ \eqref{eq:var2LS2C2} takes the form
\begin{equation}\label{eq:var2LS2}
\begin{aligned}
    -\delta^2 \left( {\hat{\mathfrak{g}}}^{\nicefrac{1}{2}} L_{\hat{S}^2} \right) &= \tfrac{1}{4} {\mathfrak{g}}^{\nicefrac{1}{2}} \, h^\perp_{ab} \left( \Box - \tfrac{2}{N(N-1)}R \right) \mathcal{F}_2(\Box) \left( \Box - \tfrac{2}{N(N - 1)}R \right) h^{{\perp}ab} \\
    &\quad + \tfrac{(N - 2)^2}{4 N^2} {\mathfrak{g}}^{\nicefrac{1}{2}} \, h^\top \, \mathcal{F}_2 \left( \Box + \tfrac{2}{N-1}R \right) \left( \tfrac{N - 1}{N} \Box + \tfrac{1}{N}R \right) \Box h^\top + {\mathfrak{g}}^{\nicefrac{1}{2}}\,\text{div}\;.
\end{aligned}
\end{equation}

Due to the symmetries and tracelessness of $\delta \hat{C}_{abcd}$, \eqref{eq:varC} and \eqref{eq:varCc}, we can write
\begin{equation}
    \delta \hat{C}_{abcd} \mathcal{F}_3(\Box) \delta \hat{C}^{abcd} = c_{\{abcd\}} \mathcal{F}_3(\Box) \delta \hat{C}^{abcd} = c_{abcd} \mathcal{F}_3(\Box) \delta \hat{C}^{abcd} = - 2 \nabla_{a}\nabla_{c} h^\perp_{bd} \mathcal{F}_3(\Box) \delta \hat{C}^{abcd}\;.    
\end{equation}
Subsequently, taking into account the commutator for a traceless rank-4 tensor with the Riemann tensor symmetries \eqref{eq:commutrank4}, expression for $\nabla^{a}\delta \hat{C}_{abcd}$ \eqref{eq:nablavarC} and commutator for a rank-3 tensor with the properties of $\nabla^a \delta\hat{C}_{abcd}$ \eqref{eq:commutrank3}, it follows 
\begin{equation}
\begin{aligned}
    \delta \hat{C}_{abcd} \mathcal{F}_3(\Box) \delta \hat{C}^{abcd} &= - 2 h^\perp_{bd} \nabla_{c}\nabla_{a} \mathcal{F}_3(\Box) \delta \hat{C}^{abcd} + \text{div} = - 2 h^\perp_{bd} \nabla_{c} \mathcal{F}_3\left(\Box + \tfrac{1}{N}R\right) \nabla_{a}\delta \hat{C}^{abcd} + \text{div} \\
    &= 2\tfrac{N-3}{N-2} h^\perp_{bd} \nabla_{c} \mathcal{F}_3\left(\Box + \tfrac{1}{N}R\right) \left( \Box - \tfrac{1}{N}R \right)
    \nabla^{[c} h^{{\perp}d]b} + \text{div} \\
    &= 2\tfrac{N-3}{N-2} h^\perp_{bd} \mathcal{F}_3\left(\Box + \tfrac{1}{N}R + \tfrac{N-3}{N(N-1)}R\right) \left( \Box - \tfrac{1}{N}R + \tfrac{N-3}{N(N-1)}R \right)
    \nabla_{c} \nabla^{[c} h^{{\perp}d]b} + \text{div} \\
    &= \tfrac{N-3}{N-2} h^\perp_{bd} \mathcal{F}_3\left(\Box + \tfrac{2(N-2)}{N(N-1)}R\right) \left( \Box - \tfrac{2}{N(N-1)}R \right)
    \left( \Box h^{{\perp}db} - \nabla_c \nabla^d h^{{\perp}cb} \right)  + \text{div}\;.
\end{aligned}
\end{equation}
Using \eqref{eq:divnablahperp} and \eqref{eq:var2LS2C2}, we arrive at
\begin{equation}\label{eq:var2LC2}
    -\delta^2 \left({\hat{\mathfrak{g}}}^{\nicefrac{1}{2}} L_{\hat{C}^2} \right) = \tfrac{N-3}{N-2} {\mathfrak{g}}^{\nicefrac{1}{2}} \, h^\perp_{ab} \, \mathcal{F}_3 \left( \Box + \tfrac{2(N-2)}{N(N-1)}R \right) \left( \Box - \tfrac{2}{N(N - 1)}R \right) \left( \Box - \tfrac{1}{N - 1}R \right) h^{{\perp}ab} + \text{div}\;.
\end{equation}
Finally, employing \eqref{eq:bgFEs}, equations \eqref{eq:var2LEHR2},  \eqref{eq:var2LS2}, and \eqref{eq:var2LC2} 
give the quadratic action $S_{\textrm{quad}}^{\textrm{E}}[\bs{h}]$ \eqref{eq:quadaction} with \eqref{eq:quadactionEI} and $\bs{\tau} = 0$. Moreover, if one also uses \eqref{eq:simplificationofF}, formulas \eqref{eq:quadactionEI} further simplify to \eqref{eq:EIop}.

\appsection{Action of \texorpdfstring{$\Box$}{box} on MS TT bi-tensors}\label{apx:powersofbox}

Let $({M},{\bs{g}})$ denote an $N$-dimensional hyperbolic space $\mathbb{H}^N$ and $\bs{\Phi}^{\perp}(\mathrm{x},\mathrm{x}')$ be an MS TT bi-tensor. The aim of this appendix is to review the calculation of $\bar{\Pi}(\Box\bs{\Phi}^{\perp})$ from \cite{Allen:1986tt,Turyn:1988af}, which is mediated through the contraction \eqref{eq:extraction}. Let us start by decomposing $\bs{\Phi}^{\perp}$ as
\begin{equation}
     \bs{\Phi}^{\perp}(\mathrm{x},\mathrm{x}') = \bar\Phi(\rho) \bs{U}_1(\mathrm{x},\mathrm{x}') + \breve\Phi(\rho) \bs{U}_2(\mathrm{x},\mathrm{x}') + \Phi_3(\rho) \bs{U}_3(\mathrm{x},\mathrm{x}')\;,
\end{equation}
where the basis of MS traceless bi-tensors $\bs{U}_k$ in terms of the bi-tensors $\bs{O}_k$ reads
\begin{equation}
\begin{aligned}
    \bs{U}_1 &= \tfrac{1}{N(N-1)} (\bs{O}_1 + N^2 \bs{O}_2 - N \bs{O}_4)\;, \\
    \bs{U}_2 &= -4 \bs{O}_2 - \bs{O}_5\;, \\
    \bs{U}_3 &= - \tfrac{2}{N-1} \bs{O}_1 + \tfrac{2(N-2)}{N-1} \bs{O}_2 + \bs{O}_3 + \tfrac{2}{N-1} \bs{O}_4 + \bs{O}_5\;.
\end{aligned}
\end{equation}
Due to the transversality, the MS bi-scalars $\breve\Phi$, $\Phi_3$ are related to $\bar\Phi$ via
\begin{equation}
    \breve\Phi = - \tfrac{1}{2NC} \left( \pp_\rho + NA \right) \bar\Phi\;, \quad
    \Phi_3 = \tfrac{N-1}{2N(N+1)(N-2)C^2} \left( \pp^2_\rho + (2N+1)A \pp_\rho + \tfrac{N+1}{N-1} ((N-1)A^2 - C^2) \right) \bar\Phi\;,
\end{equation}
where we introduced shorthands
\begin{equation}
    A := \tfrac{1}{\alpha} \coth\tfrac{\rho}{\alpha}\;, \quad C := - \tfrac{1}{\alpha} \csch\tfrac{\rho}{\alpha}\;.
\end{equation}
When applying $\Box$ to $\bs{\Phi}^\perp$, one can employ the following identity for an arbitrary bi-scalar $\Phi(\rho)$ \cite{Allen:1986tt,Turyn:1988af}
\begin{equation}\label{eq:Box(PhiU_k)}
    \Box(\Phi \bs{U}_k) = \Box \Phi \bs{U}_k + \Phi \Box\bs{U}_k\;,
\end{equation}
which simplifies the computation and holds since $\bs{\nabla} \Phi = \bs{n} \pp_\rho \Phi$ and $n^a \nabla_a \bs{U}_k = 0$. Then it remains to express the action of $\Box$ on $\Phi$ and $\bs{U}_k$. Specifically, $\Box\Phi$ is given by \eqref{eq:boxonscalars}, while $\Box\bs{U}_k$ satisfy
\begin{equation}\label{eq:BoxU_k}
\begin{aligned}
    \Box\bs{U}_1 &= -2N(A^2 + C^2) \bs{U}_1 - \tfrac{2N}{N-1}AC \bs{U}_2\;, 
    \\
    \Box\bs{U}_2 &= - 8NAC \bs{U}_1 - (N+2)(A^2 + C^2) \bs{U}_2 - 4AC \bs{U}_3\;, 
    \\
    \Box\bs{U}_3 &= - \tfrac{2(N+1)(N-2)}{N-1}AC \bs{U}_2 - 2(A^2 + C^2) \bs{U}_3\;.
\end{aligned}
\end{equation}
These relations follow from the expressions for the derivatives of the gradients of the geodesic distance $\bs{n}$, $\bs{n}'$, parallel transporter $\bs{\delta}_{\parallel}$ and the functions $A$, $C$ \cite{Allen:1986tt,Turyn:1988af}
\begin{equation}
\begin{gathered}
    \nabla_a n_b = A(g_{ab} - n_a n_b)\;, \quad  \nabla_a n'_{b'} = C(g_{ab} + n_a n'_{b'})\;, 
    \\
    \nabla_a \delta_{\parallel}{}_b{}^{c'} = -(A+C)(g_{ab} n'^{c'} + \delta_{\parallel}{}_a{}^{c'} n_b)\;, 
    \\ 
    \pp_{\rho} A = - C^2\;, \quad \pp_{\rho} C = - AC\;.
\end{gathered}
\end{equation}
Notice that only the $\bs{U}_1$ component actually contributes to the contraction with $n^a n^b n_{a'} n_{b'}$ because
\begin{equation}
    \tfrac{N}{N-1} n^a n^b n_{a'} n_{b'} U_{1ab}{}^{a'b'} = 1\;, \quad n^a n^b n_{a'} n_{b'} U_{2ab}{}^{a'b'} =  n^a n^b n_{a'} n_{b'} U_{3ab}{}^{a'b'} = 0\;.
\end{equation}
Putting everything together, we obtain \eqref{eq:box}. 

Finally, let us point out that for arbitrary MS TT bi-tensor $\bs{\Phi}^{\perp}$, $\Box \bs{\Phi}^{\perp}$ is also a MS bi-tensor. This because $\Box \bs{\Phi}^{\perp}$ is expressible in terms of the MS basis bi-tensors due to \eqref{eq:Box(PhiU_k)}, \eqref{eq:BoxU_k}. Moreover, the operator $\Box$ preserves the TT property of $\bs{\Phi}^{\perp}$.



%

\end{document}